\documentclass[a4paper, amsfonts, amssymb, amsmath, reprint, showkeys, nofootinbib, twoside, superscriptaddress, prb]{revtex4-2}

\usepackage[utf8]{inputenc} 
\usepackage[english]{babel} 

\usepackage{color} 
\usepackage{xcolor} 

\usepackage{hyperref} 
\usepackage{url} 
\hypersetup{            
  colorlinks   = true,  
  urlcolor     = blue,  
  linkcolor    = black, 
  citecolor   = blue    
}

\usepackage{amsmath}    
\usepackage{amsfonts}   
\usepackage{amssymb}    
\usepackage{physics}    
\usepackage{mathtools}  
\usepackage{textcomp} 
\usepackage{mathrsfs} 
\usepackage{cancel} 

\usepackage{cleveref} 
\usepackage{verbatim} 

\usepackage{booktabs} 
\usepackage{float} 

\newcommand{\bs}[1]{\boldsymbol{#1}}

\begin{document}


\title{Conservation laws for interacting magnetic nanoparticles at finite temperature}
\author{Frederik L. Durhuus}
\affiliation{NEXMAP, Department of Physics, Technical University of Denmark, 2800 Kgs.\ Lyngby, Denmark}
\author{Marco Beleggia}
\affiliation{Department of Physics, University of Modena and Reggio Emilia, 41125 Modena, Italy}
\affiliation{DTU Nanolab, Technical University of Denmark, 2800 Kgs. Lyngby, Denmark}
\author{Cathrine Frandsen}
\affiliation{NEXMAP, Department of Physics, Technical University of Denmark, 2800 Kgs.\ Lyngby, Denmark}
\email{fraca@fysik.dtu.dk}

\date{\today}

\begin{abstract}

We establish a general Langevin Dynamics model of interacting, single-domain magnetic nanoparticles in liquid suspension at finite temperature. The model couples the LLG equation for the moment dynamics with the mechanical rotation and translation of the particles.
Within this model, we derive expressions for the instantaneous transfer of energy, linear and angular momentum between the particles and with the environment. 
We demonstrate by numerical tests that all conserved quantities are fully accounted for, thus validating the model and the transfer expressions.
The energy transfer expressions derived here are also useful analysis tools to decompose the instantaneous, non-equilibrium power loss at each MNP into different loss channels. To demonstrate the model capabilities, we analyse simulations of MNP collisions and high-frequency hysteresis in terms of power and energy contributions. 

\end{abstract}

\maketitle

\pagenumbering{arabic}

\section{Introduction}

Magnetic nanoparticles (MNPs) are currently under intense study for a broad range of applications including: inductive components for power electronics\cite{sanusi_investigation_2023}, heating catalytic reactions\cite{almind_optimized_2021,yassine_localized_2020}, drug delivery\cite{tietze_magnetic_2015},  hyperthermia treatment of cancer\cite{pankhurst_applications_2003,pankhurst_progress_2009,perigo_fundamentals_2015} and various biosensor techniques\cite{wu_magnetic_2019} such as magnetic particle imaging \cite{panagiotopoulos_magnetic_2015}.

Central to these technologies is the response of MNPs to applied, magnetic fields, in particular their hysteresis loops and resulting heat dissipation. Analytical models for the magnetic response of an MNP ensemble do exist\cite{rosensweig_heating_2002,carrey_simple_2011}, but they generally assume spatial homogeneity, linear response and quasi-equilibrium. However, recent studies have highlighted the importance of the local temperature profile\cite{cazares-cortes_recent_2019}, MNP interactions\cite{haase_role_2012,munoz-menendez_disentangling_2020,torche_thermodynamics_2020,ortega-julia_estimating_2023}, nonequilibrium dynamics and aggregation\cite{coral_effect_2016,guibert_hyperthermia_2015} in addition to coupling between magnetic and mechanical degrees of freedom\cite{naud_cancer_2020,golovin_towards_2015}. 

One of the most advanced and versatile methods for modelling MNPs is Langevin dynamics\cite{coffey_langevin_2017}, which incorporates all of the aforementioned complexities. Here the equations of motion are derived from the total system energy, with the inclusion of stochastic and damping terms to model thermal noise and dissipation, respectively. Some studies have treated special cases or linearised versions of the equations analytically\cite{lyutyy_forced_2017,lyutyy_power_2018,keshtgar_magnetomechanical_2017,denisov_dynamics_2020-1,de_chatel_magnetic_2009}, but usually in their full non-linear form they are solved numerically. 
The most common method is direct timestep integration\cite{shasha_nonequilibrium_2021,leliaert_adaptively_2017} also known as a Langevin Dynamics (LD) simulation, which we use in the present paper. The main alternative is kinetic Monte-Carlo methods\cite{ruta_unified_2015,anand_hysteresis_2021}.

LD simulation studies include aggregation dynamics in zero-field\cite{durhuus_simulated_2021,rozhkov_self-assembly_2018,anderson_simulating_2021,satoh_brownian_1999}, uniform-\cite{sandreu_aggregation_2011,novikau_influence_2020} and gradient fields\cite{xue_self-assembly_2015}, as well as numerous simulations of hysteresis heating in alternating fields; both in liquid suspension\cite{usov_dynamics_2012,usadel_dynamics_2015,usadel_dynamics_2017,helbig_self-consistent_2023,anderson_simulating_2021,lyutyy_energy_2018,cabrera_unraveling_2017} and fixed in space\cite{haase_role_2012,ruta_unified_2015,kim_dynamical_2018,ortega-julia_estimating_2023,leliaert_individual_2021}. 

The standard way to implement Langevin dynamics of MNPs is to assume a single magnetic moment, $\bs{\mu}$, described by the Landau-Lifshitz-Gilbert (LLG) equation\cite{gilbert_LLG_2004}. The moment is coupled to the orientation vector, $\vb{u}$, which tracks the particles mechanical rotation. Recently several authors\cite{keshtgar_magnetomechanical_2017, usadel_dynamics_2015,kustura_stability_2022,usov_magnetic_2015} derived a more complete model of this magnetomechanical coupling, which includes the intrinsic spin and orbital angular momentum of electrons $\vb{S} = -\bs{\mu}/\gamma$, however only for the single-MNP case. Unlike previous work, e.g.\ \cite{usov_dynamics_2012,berkov_langevin_2006,cabrera_unraveling_2017}, this new model obeys conservation of total angular momentum $\vb{J}$. It has since been studied both numerically\cite{usadel_dynamics_2015,usadel_dynamics_2017} and analytically\cite{lyutyy_power_2018,kustura_stability_2022}.

As stated by Usov \& Liubimov\cite{usov_magnetic_2015} $\vb{J}$-conservation is only important in low-friction cases, such as nanomagnets suspended in vacuum. Nevertheless conservation laws serve as an important test of theoretical arguments in all cases, leading to a more solid foundation upon which to impose approximations. Also there are examples of current\cite{umeda_temperature-variable_2023} and proposed\cite{rusconi_quantum_2017,kustura_stability_2022} experiments where $\vb{J}$-conservation plays a starring role.

Another recent branch of inquiry is the effect of temperature on energy transfer. By the fluctuation-dissipation theorem\cite{kubo_fluctuation-dissipation_1966}, each dissipation channel has corresponding thermal fluctuations, which lead to a mutual energy exchange between particles and environment. These thermal power contributions must be derived explicitly to understand the instantaneous energy transfer at the nanoscale\cite{munoz-menendez_disentangling_2020,leliaert_individual_2021}, which is essential to predict the local temperature distribution.

In this paper we derive a generalisation of the $\vb{J}$-conserving model to systems of multiple MNPs at finite-temperature with dipole-dipole interactions. This leads to additional subtleties in accounting for energy and angular momentum, so we present a detailed analysis of the transfer of conserved quantities that includes analytical and numerical demonstrations of conservation up to entirely physical exchange with the environment. 

This paper may be regarded as a generalisation of the work by Helbig et.\ al.\cite{helbig_self-consistent_2023} to multiple MNPs at finite temperature, and of the work of Leliaert et.\ al.\cite{leliaert_individual_2021} to include mechanical motion. 

A common simplification is the Rigid Dipole Approximation (RDA), where $\bs{\mu}$ is assumed locked to $\vb{u}$, effectively removing the LLG part of the model. This circumvents the moment dynamics, which are generally several orders of magnitude faster than mechanical rotation\cite{berkov_langevin_2006}, thus permitting far longer time steps. Another simplification is the overdamped limit, where inertia is neglected, as friction usually dominates for nano- and micro particles in liquid\cite{purcell_life_1977}.
The RDA, the overdamped limit and the micromagnetics of immobilised particles can all be derived as limiting cases of our model.

Besides model validation, the presented transfer formulas are useful analysis tools. Typically, energy dissipation is determined by the area of a hysteresis curve, however this is only possible for the spatially averaged dissipation of periodically driven systems in steady-state\cite{torche_thermodynamics_2020,munoz-menendez_disentangling_2020}. Using our formulas, one can calculate the instantaneous power dissipation at each MNP and decompose into different loss channels. In particular magnetic losses from Gilbert damping heats the MNP itself, while mechanical losses from viscous damping heats the surrounding fluid. This gives a detailed, local perspective, applicable to non-equilibrium dynamics, which we demonstrate by analysing MNP collisions with and without an external, alternating magnetic field. 

The paper is structured as follows. In \cref{sec:Model} we systematically derive a Langevin dynamics model for MNPs in fluid suspension, which has vacuum- and solid suspension as limiting cases. In \cref{sec:conservation_laws} we formally derive expressions for the transfer of conserved quantities and discuss their application. In \cref{sec:Model_approximations} we discuss common model approximations in the context of conservation laws and the many characteristic frequencies of these systems. In \cref{sec:simulations} we present numerical simulations of 1, 2 and 10 particle systems. Finally, in \cref{sec:Model_generalisations} we discuss possible model generalisations, and to what extent the formal results also generalise.

\section{Model \label{sec:Model}}

\begin{figure}[ht]
    \centering
    \includegraphics[width=0.4\textwidth]{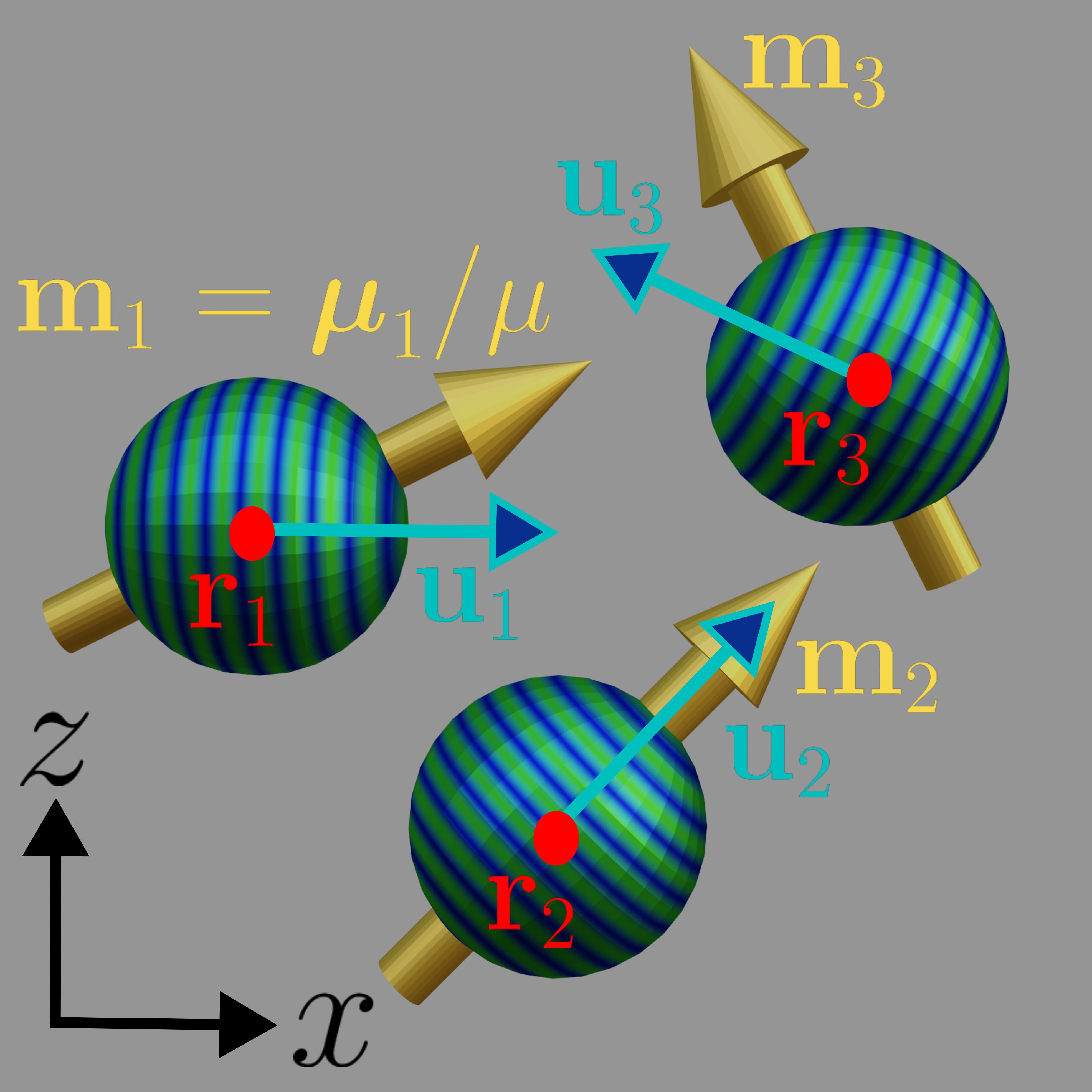}
    \caption{Illustration of system with 3 magnetic nanoparticles (MNPs). Spheres and yellow 3D vectors (magnetic moments) are drawn in Blender\cite{Blender2018} with the same code used in \cref{fig:2MNP_hysteresis_collision,fig:10MNP_collision} to visualise simulations. The center positions $\vb{r}_i$, normalised moments $\vb{m}_i=\boldsymbol{\mu}_i/MV$ and orientation vectors $\vb{u}_i$ are indicated. The anisotropy axes are perpendicular to the blue/green circles and parallel to $\vb{u}_i$, so the anisotropy energy is lowest when $\vb{u}_i \parallel \vb{m}_i$ which is the case for particle 2.}
    \label{fig:system_sketch}
\end{figure}

We consider a collection of identical, ferro- or ferri magnetic, spherical MNPs. Let $\boldsymbol{\mu}_i$ be the net magnetic moment of the $i$'th MNP, $R$ radius, $V$ volume, $\vb{r}_i$ the position of its center and $\vb{v}_i = \vb{\dot{r}}_i$ its velocity. We assume internal exchange coupling is strong enough to ensure uniform magnetisation of constant magnitude, $M$, hence $\bs{\mu}_i = \mu \vb{m}_i$ where $\vb{m}_i$ is a unit vector and $\mu=V M$ is constant. We specify particle orientation by the unit vector $\vb{u}_i$ which is fixed along the chosen anisotropy axis (see \cref{subsec:Uniaxial_anisotropy}). Finally let $\bs{\omega}_i$ denote angular velocity, in the sense that MNP $i$ rotates $\omega_i$ radians per second counterclockwise around the vector $\bs{\omega}_i$. See \cref{fig:system_sketch} for an illustration. This is the model system we analyse in \cref{sec:Model,sec:conservation_laws,sec:Model_approximations,sec:simulations}, and we discuss possible generalisations in \cref{sec:Model_generalisations}.

\subsection{Magnetic interactions \label{sec:Interactions}}

\subsubsection{Magnetic fields\label{subsec:Magnetic_interactions}}

The magnetic interactions of uniformly magnetised spheres are  (up to relativistic corrections) precisely those of point dipoles located at their centers\cite{edwards_interactions_2017}.
That is, each MNP produces a dipole field outside its surface,
and the interaction energy between MNP $i$ and $j$ is $E^\text{int}_{ij} = -\bs{\mu}_i \vdot \vb{B}^\text{dip}_{ji}$ where $\vb{B}^\text{dip}_{ji}$ is the field from MNP $j$ evaluated at $\vb{r}_i$.

The total dipole interaction energy is then
\begin{align*}
    E^\text{int} = \frac{1}{2}\sum_i \sum_{j\neq i} E^\text{int}_{ij} = -\frac{1}{2} \sum_i \bs{\mu}_i \vdot \vb{B}^\text{dip}_i
\end{align*}
with a factor of $1/2$ to avoid double counting. $\vb{B}^\text{dip}_i$ is the field on the $i$'th MNP from all the rest, which is given by\cite{griffiths_EM}
\begin{align}
    \vb{B}^\text{dip}_i =
    \frac{\mu_0}{4\pi} \sum_{\substack{j\neq i}} \frac{1}{r_{ji}^3}\left[3(\bs{\mu}_j \vdot \hat{\vb{r}}_{ji})\hat{\vb{r}}_{ji} - \bs{\mu}_j \right],
    \label{eq:B_dip}
\end{align}
where
\begin{align*}
    \vb{r}_{ji} = \vb{r}_i - \vb{r}_j \qq{,} r_{ji} = \abs{\vb{r}_{ji}}\qq{,} \hat{\vb{r}}_{ji} = \vb{r}_{ji}/r_{ji} 
\end{align*}

We also consider a uniform, but time-varying field $\vb{B}^\text{ext}$ generated by external currents, which gives the Zeeman energy\cite{griffiths_dipoles_1992}
\begin{align}
E^\text{Zee}_i = -\bs{\mu}_i \vdot \vb{B}^\text{ext}.    
\end{align}

\subsubsection{Uniaxial anisotropy \label{subsec:Uniaxial_anisotropy}}
There are a number of energy terms which give the magnetisation a preferred direction relative to the particles crystal lattice; namely shape-, magnetocrystalline-, surface- and strain anisotropy\cite{bedanta_supermagnetism_2015}. The latter two primarily induce non-uniform magnetisation, which we neglect entirely, and shape anisotropy is absent for a sphere. 

Magnetocrystalline anisotropy results from relativistic spin-orbit and spin-spin coupling, which to lowest order yields\cite{landau_vol8_1960} $E_\text{MC} = \frac{1}{2} V \vb{m}^T \mathrm{K}\vb{m}$, where $\mathrm{K}$ is a rank-2 tensor. Because this is a relativistic perturbation, we expect higher order terms to be substantially smaller. We can always pick a local coordinate system such that $\mathrm{K}$ is diagonal. We call the corresponding eigenvalues $K_x, K_y, K_z$. Then, using $m_z^2 = 1 - m_x^2 - m_y^2$, 
\begin{align*}
    \frac{1}{V} E_\text{MC} &= \frac{1}{2} \vb{m}^T\mathrm{K} \vb{m}
    \notag\\
    &= \text{const} + \frac{1}{2} (K_x-K_z)m_x^2 + \frac{1}{2} (K_y - K_z)m_y^2 
\end{align*}
If $K_x,K_y,K_z$ are all different, then this defines a hard-, easy- and intermediate axis, i.e.\ triaxial anisotropy. 

Typically in MNP studies, it is implicitly assumed that there is rotation symmetry about an axis $\vb{u}$, so that two of the eigenvalues are equal. For instance, if $\vb{u} = \vb{e}_x$ then $K_x = K_\parallel$ and $K_y = K_z = K_{\perp}$. The energy may then be written in the coordinate independent form
\begin{align}
    E_\text{MC} = \text{const} - K V (\vb{m} \vdot \vb{u})^2, \label{eq:uniaxial_anisotropy}
\end{align}
where $K = \frac{1}{2}(K_\parallel - K_\perp)$. Evidently \cref{eq:uniaxial_anisotropy} describes uniaxial anisotropy, with either a hard-axis ($K<0$) or an easy-axis ($K>0$).

We refer to the literature for higher order anisotropy terms with lower symmetry\cite{chikazumi1997physics,cullity2011introduction,bedanta_supermagnetism_2015}.

\subsection{Conserved quantities}

\subsubsection{Momenta \label{subsec:momenta}}

The total angular momentum of MNP $i$ in its own rest frame is 
\begin{align}
    & \vb{J}_i = \vb{L}_i + \vb{S}_i \qq{,} \vb{L}_i = I\bs{\omega}_i \qq{,} \vb{S}_i = -\gamma^{-1}\bs{\mu}_i    \label{eq:J_i}
\end{align}
where $I$ is the moment of inertia, $\gamma$ is the gyromagnetic ratio, $\vb{L}_i$ is the mechanical angular momentum and $\vb{S}_i$ is the angular momentum associated with the magnetic moment. $\vb{S}_i$ comprises electron spin and orbital angular momentum\cite{grifftihs_QM}.

The angular momentum of the whole MNP system is
\begin{align}
    &\vb{J} = \vb{L} + \vb{S} 
    \notag \\ 
    \vb{L} = \sum_i (\vb{L}_i + &\vb{r}_i \cross \vb{p}_i) \qq{,} \vb{S} = \sum_i \vb{S}_i  \label{eq:J_multi_MNP}
\end{align}
where $\vb{p}_i$ is the linear momentum. Beside the sum of \textit{intrinsic} angular momenta, $\vb{J}_i$, there is an \textit{extrinsic} mechanical contribution from motion relative to the rest of the system, i.e.\ $\sum_i \vb{r}_i \cross \vb{p}_i$. 

The system linear momentum is all mechanical and given simply by 
\begin{align}
    \vb{p} = \sum_i \vb{p}_i \label{eq:p_multi_MNP}
\end{align}
Note that the present notation is not to be confused with the $\vb{J}, \vb{L}, \vb{S}$ quantum operators used in atomic physics\cite{foot2004atomic,grifftihs_QM}.
We discuss the transfer of linear and angular momenta in \cref{sec:conservation_laws}, in particular proving that the model obeys physical conservation laws.

\subsubsection{Energy}

The system energy is, up to a constant,
\begin{align}
    E &= \frac{1}{2} I \sum_i \omega^2_i + \frac{1}{2} \mathfrak{m} \sum_i v_i^2 - \frac{1}{2} \sum_i \bs{\mu}_i \vdot \vb{B}^\text{dip}_i
    \notag\\
    &\quad - \sum_i \bs{\mu}_i \vdot \vb{B}^\text{ext} - K V \sum_i (\vb{m}_i \vdot \vb{u}_i)^2. \label{eq:E_multi_MNP}
\end{align}

\noindent where $\mathfrak{m}$ is the mass of each particle. The first 2 terms are rotational and translational kinetic energy, while the last 3 are explained in \cref{subsec:Magnetic_interactions,subsec:Uniaxial_anisotropy}.

\subsection{Energy conserving equations of motion \label{subsec:energy_conserving_equations_of_motion}}

The torque $\bs{\tau}$ on a rigid body is defined as the energy gain under an infinitesimal rotation $\delta \bs{\phi}$. That is, when $\vb{u} \xrightarrow{} \vb{u} + \delta \bs{\phi} \cross \vb{u}$ we have by definition
\begin{align*}
    &\Delta E[\delta\bs{\phi}] = -\bs{\tau} \vdot \delta \bs{\phi} + \order{\delta \phi^2}
\end{align*}
To first order in $\delta \phi$,
\begin{align*}
    E[\vb{u} + \delta \bs{\phi} \cross \vb{u}] - E[\vb{u}] &= \pdv{E}{\vb{u}} \vdot (\delta \bs{\phi} \cross \vb{u})
    \\
    &= \left(\vb{u} \cross \pdv{E}{\vb{u}}\right) \vdot \delta \bs{\phi}
\end{align*}
thus the time-derivative of $\vb{L}_i$ is
\begin{align}
    \vb{\dot{L}}_i = \bs{\tau}_i = \pdv{E}{\vb{u}_i} \cross \vb{u}_i \label{eq:L_dot_calculation}
\end{align}

Similarly\cite{keshtgar_magnetomechanical_2017} $\vb{\dot{S}}_i = \partial_{\vb{m}_i} E \cross \vb{m}_i$ for the variation of magnetic angular momentum, and from an infinitesimal displacement, one finds $\vb{\dot{p}}_i = -\partial_{\vb{r}_i}E$ for the variation of linear momentum.

It then follows from \cref{eq:E_multi_MNP,eq:J_i} after some algebra that
\begin{align}
    \vb{\dot{m}}_i &= - \gamma \vb{m}_i \cross (\vb{B}^\text{dip}_i + \vb{B}^\text{ext}_i + \vb{B}^\text{ani}_i) \label{eq:mu_dot_no_damping}
    \\
    I\bs{\dot{\omega}}_i &
    = - \mu_i\vb{m}_i \cross \vb{B}^\text{ani}_i
    \label{eq:omega_dot_no_damping}
    \\
    \mathfrak{m}\vb{\dot{v}}_i &= \vb{F}^\text{dip}_i \label{eq:r_ddot_no_damping}
\end{align}
where
\begin{align}
     \vb{B}^\text{ani}_i &= 2\frac{K_i}{M_i} (\vb{m}_i \vdot \vb{u}_i) \vb{u}_i \label{eq:B_ani}
\end{align}
is an effective field from uniaxial anisotropy and
\begin{align}
    \vb{F}^\text{dip}_i &= \frac{3\mu_0}{4\pi} \sum_{\substack{j\neq i}} \frac{1}{r_{ji}^4}[(\bs{\mu}_i \vdot \hat{\vb{r}}_{ji}) \bs{\mu}_j + (\bs{\mu}_j \vdot \hat{\vb{r}}_{ji}) \bs{\mu}_i
    \notag\\
    &\quad +(\bs{\mu}_i \vdot \bs{\mu}_j)\hat{\vb{r}}_{ji}  -5(\bs{\mu}_i\vdot \hat{\vb{r}}_{ji})(\bs{\mu}_j \vdot \hat{\vb{r}}_{ji}) \hat{\vb{r}}_{ji}]   \label{eq:F_dip}
\end{align}
When differentiating the dipole-dipole interaction, note that $\vb{B}_j^\text{dip}$ contains $\bs{\mu}_i$. 
We observe that the forces in \cref{eq:F_dip}, and the torques $\bs{\mu}_i \cross \vb{B}^\text{dip}_i$, are precisely those between point dipoles\cite{landecker1970analytic}.

\Cref{eq:mu_dot_no_damping,eq:omega_dot_no_damping,eq:r_ddot_no_damping} constitute a molecular dynamics model, which amounts to zero-temperature dynamics in vacuum. In Langevin dynamics, there are also fluctuation and dissipation terms to describe coupling with the degrees of freedom that are not yet in the model.

\subsection{Fluctuations \& dissipation \label{sec:dissipation_and_fluctuations}}

\subsubsection{Gilbert damping \label{sec:Gilbert_damping}}

To model dissipative coupling between magnetic moments and internal degrees of freedom, such as electrons and phonons, we use the phenomenological Gilbert damping torque\cite{gilbert_LLG_2004}, which converts magnetic angular momentum to mechanical:
\begin{align}
    &\vb{\dot{S}}_i = \eval{\vb{\dot{S}}_i}_{\alpha = 0} + \bs{\tau}^{\alpha}_i \qq{and} \vb{\dot{L}}_i = \eval{\vb{\dot{L}}_i}_{\alpha = 0} - \bs{\tau}^{\alpha}_i \label{eq:Gilbert_torque_1}
\end{align}
where at zero temperature
\begin{align}
    \eval{\bs{\tau}^{\alpha}_i}_{T=0} = -\alpha \mu \gamma^{-1} \left[\vb{m}_i \cross \vb{\dot{m}}_i + \vb{m}_i \cross (\vb{m}_i \cross \bs{\omega}_i)\right],   \label{eq:Gilbert_torque_2}
\end{align}
and $\alpha$ is the material-specific Gilbert damping constant. See \cref{sec:fluctuations} for the effect of temperature.

The first term is the standard Gilbert torque in the rest-frame of the particle; the second an additional torque from transforming to the laboratory frame\cite{keshtgar_magnetomechanical_2017,usadel_dynamics_2015}, which has been argued to produce the Barnett effect (change in magnetisation from mechanical rotation)\cite{keshtgar_magnetomechanical_2017,barnett_magnetization_1915}. Indeed the second term is proportional to the Barnett field\cite{keshtgar_magnetomechanical_2017,rusconi_quantum_2017}
\begin{align}
    \vb{B}^\text{bar}_i = -\gamma^{-1} \bs{\omega}_i. \label{eq:B_bar}
\end{align}

\subsubsection{Viscous damping \label{sec:Viscous_damping}}

Often MNPs are studied in fluid suspension, in which case there is
viscous damping from MNPs ploughing through the fluid particles, e.g.\ water molecules. We neglect hydrodynamic interactions between the MNPs (see \cref{sec:Model_generalisations} for discussion and references). For an isolated sphere at low Reynolds number (Stokes flow regime) in a stationary fluid, the exact expressions for the damping force and torque are\cite{rubinow_transverse_1961}
\begin{align*}
    \vb{F}^\text{visc} = - \zeta^\text{t} \vb{v} \qq{,} \bs{\tau}^\text{visc} = -\zeta^\text{r} \bs{\omega}
\end{align*}
where
\begin{align}
    \zeta^\text{t} = 6\pi \eta R \qq{,} \zeta^\text{r} = 8\pi \eta R^3, \label{eq:zeta_sphere}
\end{align}
and $\eta$ is dynamic viscosity.

\subsubsection{Fluctuations \label{sec:fluctuations}}

By the fluctuation-dissipation theorem\cite{kubo_fluctuation-dissipation_1966}; for each dissipation mechanism there are corresponding thermal fluctuations. 
For the viscous damping, random collisions between fluid particles and MNPs lead to Brownian motion, described by a thermal force $\vb{F}^\text{th}$ and torque $\bs{\tau}^\text{th}$. Likewise, to the Gilbert damping there corresponds an effective thermal magnetic field\cite{brown_thermal_1963} $\vb{B}^\text{th}$, which modifies the Gilbert torque by
\begin{align}
    \bs{\tau}^\alpha_i = \eval{\bs{\tau}^\alpha_i}_{T=0} + \bs{\mu}_i \cross \vb{B}^\text{th}_i.  \label{eq:tau_alpha}
\end{align}

These are stochastic vectors with zero mean, i.e.\
\begin{align}
    \expval{\vb{B}^\text{th}_i} = \expval{\vb{F}^\text{th}_i} = \expval{\bs{\tau}^\text{th}_i} = \vb{0} \label{eq:stochastic_vectors}
\end{align}
where $\expval{}$ is an ensemble average in thermal equilibrium. We assume no correlation in the fluctuations of different MNPs, e.g.\ $\expval{B^\text{th}_{i,\alpha}B^\text{th}_{j,\beta}} \sim \delta_{ij}$ where $\delta_{ij}$ is the Kronecker delta, and $\alpha,\beta \in \{x,y,z\}$ denote vector components. We furthermore assume that each vector component is described by a Gaussian distribution with no correlation from one point in time to another (no autocorrelation). This is justified for $\vb{B}^\text{th}$ in general\cite{brown_thermal_1963}, and for $\vb{F}^\text{th},\bs{\tau}^\text{th}$ whenever the MNPs are much larger and slower than the fluid particles\cite{kubo_fluctuation-dissipation_1966}. That is because on the characteristic timescales of MNP motion, there are enough random collisions with fluid particles that the central limit theorem applies (Gaussian distribution) and the time between consecutive impacts is negligible (again, no autocorrelation).
It may then be shown that for a given MNP\cite{brown_thermal_1963}
\begin{align}
 \expval{B^\text{th}_{\alpha}(t)B^\text{th}_{\beta}(t')} &= \frac{2k_BT\alpha}{\gamma \mu} \delta_{\alpha\beta} \delta(t-t')  \label{eq:B_th}
\end{align}
and\cite{blundell2010concepts,kamenev_field_2011}
\begin{align}
\expval{F^\text{th}_{\alpha}(t)F^\text{th}_{\beta}(t')} &= 2k_BT\mathrm{\zeta}^\text{t} \delta_{\alpha \beta}\delta(t-t')   \label{eq:F_th}
    \\ 
\expval{\tau^\text{th}_{\alpha}(t)\tau^\text{th}_{\beta}(t')} &= 2k_BT\mathrm{\zeta}^\text{r} \delta_{\alpha\beta} \delta(t-t')  \label{eq:tau_th}
\end{align}
where $k_B$ is Boltzmann's constant. $\delta_{\alpha\beta}$ signifies that different vector components are uncorrelated, which is true in \cref{eq:F_th,eq:tau_th} only for spherical particles\cite{han_brownian_2006,happel_low_1981}.

\subsection{Full Langevin dynamics \label{sec:Full_Langevin_dynamics}}

Here we present the general equations of motion for our Langevin dynamics model of interacting MNPs. 

Inserting \cref{eq:tau_alpha,eq:Gilbert_torque_2,eq:J_i,eq:mu_dot_no_damping} in \cref{eq:Gilbert_torque_1} yields the LLG equation
\begin{align}
    \vb{\dot{m}}_i = -\gamma \vb{m}_i \cross \vb{B}^\text{eff}_i + \alpha \vb{m}_i \cross \vb{\dot{m}}_i,  \label{eq:mu_dot_mid}
\end{align}
where
\begin{align}
    \vb{B}^\text{eff}_i = \vb{B}^\text{ext}_i + \vb{B}^\text{dip}_i + \vb{B}^\text{ani}_i + \vb{B}^\text{th}_i - \alpha\vb{m}_i \cross \vb{B}^\text{bar}_i,   \label{eq:B_eff}
\end{align}
is an effective magnetic field. $\vb{B}^\text{ext}$ is a uniform applied field and the other component fields are given in \cref{eq:B_dip,eq:B_ani,eq:B_bar,eq:B_th}.
We emphasise that only $\vb{B}^\text{dip}$ and $\vb{B}^\text{ext}$ are actual magnetic fields governed by Maxwells equations. In \cref{appsec:rewriting_LLG_equation} we isolate $\vb{\dot{m}}_i$ in \cref{eq:mu_dot_mid}, which yields
\begin{align}
    \vb{\dot{m}}_i &= -\gamma' \vb{m}_i \cross \vb{B}^\text{eff}_i - \alpha \gamma' \vb{m}_i \cross \left[\vb{m}_i \cross \vb{B}^\text{eff}_i\right] \label{eq:mu_dot}
\end{align}
where $\gamma' = \gamma / (1+\alpha^2)$.

The mechanical equations of motion are (cf. \cref{appsec:mechanical_equations_of_motion} for intermediate steps)
\begin{align}
    I\bs{\dot{\omega}}_i &= \mu \gamma^{-1} \vb{\dot{m}}_i + \mu\vb{m}_i \cross \vb{B}_i
    - \zeta^\text{r} \bs{\omega}_i + \bs{\tau}^\text{th}_i \label{eq:omega_dot}
    \\
    \mathfrak{m} \vb{\dot{v}}_i &= \vb{F}^\text{dip}_i - \zeta^\text{t} \vb{v}_i + \vb{F}^\text{th}_i \label{eq:r_ddot}
    \\
    \vb{\dot{u}}_i &= \bs{\omega}_i \cross \vb{u}_i \qq{,} \vb{\dot{r}}_i = \vb{v}_i \label{eq:u_dot}
\end{align}
where $\vb{F}^\text{dip}$ is given in \cref{eq:F_dip}, $\vb{B}_i = \vb{B}^\text{ext} + \vb{B}^\text{dip}_i$ is the magnetic field from all sources outside MNP $i$ and $\vb{m} = \bs{\mu} /\mu$ is the normalised magnetic moment. 
For a given initial state specified by $\vb{m}_i, \vb{u}_i, \bs{\omega}_i, \vb{r}_i,\vb{v}_i$, the time-evolution of the system is fully described by \cref{eq:mu_dot,eq:omega_dot,eq:r_ddot,eq:u_dot}.

\Cref{eq:mu_dot} can alternatively be written
\begin{align}
    \vb{\dot{m}}_i = \bs{\Omega}_i \cross \vb{m}_i \qq{,} \bs{\Omega}_i = \gamma' \vb{B}_i^\text{eff} + \alpha \gamma' \vb{m}_i \cross \vb{B}_i^\text{eff},     \label{eq:Omega}
\end{align}
where $\bs{\Omega}$ is the angular velocity of the magnetic moments. This non-standard form highlights the similarity between the LLG and mechanical rotation, making comparison easier. We note that the magnetic moments are in steady-state, i.e.\ $\vb{\dot{m}}_i = 0$, precisely when $\vb{m}_i \parallel \vb{B}^\text{eff}_i$.

\section{Conservation laws \label{sec:conservation_laws}}

\subsection{Angular momentum transfer \label{subsec:Angular_momentum_conservation}}

We consider the change in angular momentum, see \cref{subsec:momenta}, given the equations of motion \cref{eq:mu_dot,eq:omega_dot,eq:r_ddot,eq:u_dot}.

For the intrinsic angular momentum of a single MNP, we find
\begin{align}
    \vb{\dot{S}}_i &= -\gamma^{-1} \bs{\dot{\mu}}_i \label{eq:Si_dot}
    \\
    \vb{\dot{L}}_i &= \gamma^{-1} \bs{\dot{\mu}}_i - \zeta^\text{r} \bs{\omega}_i + \bs{\mu}_i \cross \vb{B}_i + \bs{\tau}^\text{th}_i   \label{eq:Li_dot}
    \\
    \vb{\dot{J}}_i &= - \zeta^\text{r} \bs{\omega}_i + \bs{\mu}_i \cross \vb{B}_i + \bs{\tau}^\text{th}_i   \label{eq:Ji_dot}
\end{align}
Note that regardless of the exact dynamics of the magnetic moment, described by $\bs{\dot{\mu}}_i$, the change in $\vb{S}_i$ is compensated by the torque
\begin{align}
    \bs{\tau}^\text{EdH}_i = -\vb{\dot{S}}_i = \gamma^{-1} \bs{\dot{\mu}}_i \label{eq:tau_EdH}.
\end{align}
We dub it the Einstein--de-Haas torque because it produces the Einstein--de-Haas effect, i.e.\ changes in magnetisation induce rotation. $\bs{\tau}^\text{EdH}$ bears a striking resemblance to the Barnett field \cref{eq:B_bar}, because they are two sides of the same phenomenon; namely that angular momentum conservation necessitates a direct coupling between magnetic and mechanical rotation. Thus the only ways $\vb{J}_i$ can change are by viscous friction or coupling to external fields. The effective fields $\vb{B}^\text{ani}_i, \vb{B}^\text{th}$ and $\vb{B}^\text{bar}_i$ can only convert between magnetic- and mechanical angular momentum within the $i$'th MNP, so their exact form, and whether they are included in the model, is irrelevant for $\vb{J}$-conservation.

The change in total angular momentum is
\begin{align*}
    \vb{\dot{J}} &= \sum_i \left(\vb{\dot{J}}_i 
    + \vb{r}_i \cross \vb{\dot{p}}_i \right) 
    \\
    &= \sum_i \bigg(\vb{r}_i \cross \vb{F}^\text{dip}_i + \bs{\mu}_i \cross \vb{B}^\text{dip}_i + \bs{\mu}_i \cross \vb{B}^\text{ext}_i
    \\
    &\hspace{0.92cm} - \zeta^\text{t}\vb{r}_i \cross \vb{v}_i - \zeta^\text{r} \bs{\omega}_i + \bs{\tau}_i^\text{th} + \vb{r}_i \cross \vb{F}^\text{th}_i\bigg)
\end{align*}
The first two terms are identical to the case of magnetic point dipoles alone in vacuum. This can be verified by repeating the above calculation using \cref{eq:r_ddot_no_damping,eq:omega_dot_no_damping,eq:mu_dot_no_damping} at $B^\text{ext}=0$. We prove in \cref{appsec:conservation_laws_in_magnetostatics} that this contribution is 0. 
Thus
\begin{align}
    \vb{\dot{J}} &= \sum_i \bigg(\bs{\mu}_i \cross \vb{B}^\text{ext}_i
    -\zeta^\text{r} \bs{\omega}_i - \zeta^\text{t} \vb{r}_i \cross \vb{v}_i
    \notag \\
    &\hspace{0.92cm} + \bs{\tau}_i^\text{th} + \vb{r}_i \cross \vb{F}^\text{th}_i \bigg) \label{eq:J_dot}
\end{align}
The first term describes coupling to external $\vb{B}$-fields, which is in general non-zero. Consider for instance a single dipole in a uniform applied field. The next two describe losses to viscous friction and the last two thermal transfer between MNPs and fluid particles. In conclusion, the model is indeed angular momentum conserving, except for entirely physical transfer to and from the environment.

We note that the argument relies on the cancellation of $\sum_i \vb{m}_i \cross \vb{B}_i^\text{dip}$ and $\sum_i \vb{r}_i \cross \vb{F}^\text{dip}_i$. 
The former is the change in intrinsic angular momentum of individual MNPs from dipole torques. The latter is the change in extrinsic angular momentum from dipole forces, i.e.\ the contribution from the MNPs velocity relative to a shared center-of-mass. Thus the dynamics of interacting dipoles conserve angular momentum when and only when one considers both rotational and translational degrees of freedom. 

\subsection{Absence of field momentum}

In \cref{subsec:Angular_momentum_conservation}, we proved that for MNPs isolated in vacuum, our model conserves the vector $\vb{J}$, which is the sum of mechanical and magnetic angular momentum (see \cref{subsec:momenta}). 
However, in the context of electrodynamics, linear- and angular momentum are conserved only when also accounting for momentum in the electromagnetic fields and the effect of "hidden momentum" \cite{babson_hidden_2009, romer_angular_1966,hnizdo_conservation_1992}. The former is given by $\vb{J}^\text{EM} = \int \vb{E} \cross \vb{B} \dd\vb{r}$ where $\vb{E}$ is electric field\cite{griffiths_EM}.

As the MNPs move, their dipole fields change, inducing an $\vb{E}$-field; so why did we not need to account for field momentum? The answer is that we calculate all magnetic interactions using magneto\textit{statics}.
That is we both neglect relativistic effects like hidden momentum and induced $\vb{E}$-fields. Christiansen et.\ al.\cite{christiansen_theoretical_2022} found that induced $\vb{E}$-fields from moment precession in single-domain MNPs are indeed weak compared to other nanoscale induction phenomena, justifying the second assumption.

For a formal proof of angular momentum conservation in electrodynamics, we refer to \cite[problem 6.10]{jackson_classical_1999}. In \cref{appsec:conservation_laws_in_magnetostatics} we modify the argument in \cite{jackson_classical_1999} to prove that linear- and angular momentum are conserved for any steady current distribution when $\vb{E}=0$; without any momenta in the fields. Magnetic dipoles are equivalent to infinitesimal current loops, so the proof applies to MNPs.

This begs a second question: if there is no momentum in the external field, $\vb{B}^\text{ext}$, how can it exert forces and torques on the MNPs? 
The answer is that $\vb{B}^\text{ext}$ is itself generated by a current distribution, typically in a coil, and the dipole fields from the MNPs exert corresponding forces and torques on this current distribution, ensuring momentum balance.

Finally, we note that if the MNPs have ionisable surface groups or are dispersed in polar solvent, they may carry a surface charge, in which case even the static EM fields carry momentum\cite{sharma_field_1988}. Then $\vb{L}^\text{EM}$ changes as electric charges and magnetic dipoles rearrange, which is exactly balanced by mechanical torques. Whether these torques are significant is beyond the scope of the present study.

\subsection{Linear momentum transfer}

The change in system linear momentum is
\begin{align*}
    \vb{\dot{p}} = \sum_i \vb{\dot{p}}_i = \sum_i \mathfrak{m}_i \vb{\dot{v}}_i
\end{align*}
which is given in \cref{eq:r_ddot}. 
$\vb{F}^\text{dip}$ is given by a pairwise, velocity-independent potential, hence it obeys Newtons third law.
It follows that
\begin{align}
    \vb{\dot{p}} = \sum_i \left(-\zeta^\text{t} \vb{v}_i + \vb{F}^\text{th}_i\right) \label{eq:p_dot}
\end{align}
That is, linear momentum is conserved except for viscous friction and Brownian forces, i.e.\ momentum exchange with the fluid particles.

\subsection{Energy transfer}

The equations of motion \cref{eq:mu_dot_no_damping,eq:omega_dot_no_damping,eq:r_ddot_no_damping,eq:u_dot} by construction conserve the system energy (\cref{eq:E_multi_MNP}).
For the complete model with damping and fluctuations (\cref{eq:mu_dot,eq:omega_dot,eq:r_ddot,eq:u_dot}), we find after considerable algebra (See \cref{appsec:E_dot})
\begin{align}
    \dot{E} = \sum_i (P^\text{hyst}_i + P^\text{mag}_i + P^\text{rot}_i + P^\text{trans}_i) \label{eq:E_dot}
\end{align}
where 
\begin{align}
 P_i^\text{hyst} &= -\mu \vb{m}_i \vdot \vb{\dot{B}}^\text{ext}       \label{eq:P_in}
 \\
 P_i^\text{mag} &= -\alpha \mu \gamma^{-1}([\bs{\Omega}_i - \bs{\omega}_i] \cross \vb{m}_i)^2 
 \notag\\
 &\hspace{0.75cm} + \mu [(\bs{\Omega}_i - \bs{\omega}_i) \cross \vb{m}_i] \vdot \vb{B}^\text{th}_i  \label{eq:P_mag}
 \\
 P_i^\text{rot} &= -\zeta^\text{r} \omega^2_i + \bs{\omega}_i \vdot \bs{\tau}^\text{th}_i   \label{eq:P_rot}
 \\
 P_i^\text{trans} &= -\zeta^\text{t} v^2_i + \vb{v}_i \vdot \vb{F}^\text{th}_i      \label{eq:P_trans}
\end{align}

$\dot{E}$ is the net energy transfer between the MNPs and their environment. By environment we mean both the liquid or solid medium, the external currents generating $\vb{B}^\text{ext}$ and internal degrees of freedom in the particles, like electronic states and lattice vibrations. $\dot{E} > 0$ means the MNPs are gaining energy while $\dot{E} < 0$ means they are losing energy to the environment.

The hysteresis power $P^\text{hyst}$ describes the work done by $\vb{B}^\text{ext}$ on the magnetic moments. As shown below, a hysteresis curve measures the time-averaged value of $P^\text{hyst}$ hence the name.
$P^\text{hyst}$ is the main source of energy into the MNP system, while the other terms generally dissipate this energy into the particle surroundings.
$P^\text{rot}, P^\text{trans}$ describe exchange of kinetic energy in a fluid medium, so the resulting viscous energy losses go to heating the surrounding fluid. $P^\text{mag}$ describes Gilbert damping, i.e.\ energy transfer from the single-domain moments to internal degrees of freedom like phonon modes\cite{streib_magnon-phonon_2019}, spin-waves\cite{streib_magnon-phonon_2019} and electronic transitions\cite{simensen_magnon_2020}. Gilbert damping heats the MNPs themselves. 

While there is thermal conduction between MNPs and fluid, a non-uniform temperature profile is possible in driven systems, so the relative magnitude of viscous and magnetic losses will influence the equilibrium temperature distribution. 

For $P^\text{mag}, P^\text{rot}$ and $P^\text{trans}$, the first term is pure damping, while the second describes a 2-way transfer from thermal fluctuations. Over time, thermal fluctuations will typically average to 0, but at a given instant, there may be a net energy transfer from environment to system. The two terms are related by the fluctuation-dissipation theorem\cite{kubo_fluctuation-dissipation_1966}, so if $\alpha, \zeta^\text{r}$ or $\zeta^\text{t}$ is zero, the corresponding fluctuations are absent, in which case $P^\text{mag}, P^\text{rot}$ or $P^\text{trans}$ respectively is 0.

Each damping mechanism is local, so $P_i^\text{mag}$ changes the internal temperature of the $i$'th MNP and $P_i^\text{rot}, P_i^\text{trans}$ heat the fluid in the vicinity of the $i$'th MNP. That said, the distribution of power dissipation in the fluid would require further analysis. Likewise, by tracking the dissipation of momenta, one can infer how much the MNP motion stirs the fluid, but not the exact form of waves and vortices formed.
We observe that $P^\text{mag}$ depends on the difference in angular velocity between mechanical- and moment rotation, so Gilbert damping is given by changes in moment orientation \textit{relative} to the particles atomic lattice. Thus one can distinguish magnetic losses for a single MNP by computing the hysteresis curve in a local, co-rotating coordinate system as done in Ref.\ \cite{helbig_self-consistent_2023}. 

When setting $\omega_i = 0$, \cref{eq:P_mag} is in complete agreement with Ref.\ \cite[eq. 6]{leliaert_individual_2021}, which included the thermal power, but did not consider mechanical motion. 
To show the equivalence, one may use the identity $\abs{\bs{\Omega} \cross \vb{m}}^2 = \gamma \gamma'\abs{\vb{B}^\text{eff} \cross \vb{m}}^2$. 
\Cref{eq:P_rot,eq:P_trans} are consistent with known expressions for rigid-body particles in fluid \cite{hill_extremum_1956,kim_microhydrodynamics_1991}.

Now consider if $\vb{B}^\text{ext}$ is periodic with frequency $f_\text{ext}$. Then the time-averaged input power is
\begin{align*}
    \overline{P^\text{hyst}} &= -f_\text{ext} \oint_\text{cycle} \vb{m}_\text{tot} \vdot \vb{\dot{B}}^\text{ext} \dd t 
    \\
    &= -f_\text{ext} \oint_\text{cycle} \vb{m}_\text{tot} \vdot \dd \vb{B}^\text{ext},
\end{align*}
In words; the average energy gained by the MNPs per field-cycle is given by the area of the hysteresis curve. All of this energy is eventually dissipated as heat, so the hysteresis area is a standard way to estimate heating power\cite{rosensweig_heating_2002,carrey_simple_2011,usadel_dynamics_2015}. 

The hysteresis approach has several limitations. It requires periodic driving and that the system is in steady state, or at least changes slowly relative to $f_\text{ext}$. Also it gives an average power, so $f_\text{ext}$ limits time-resolution, and it gives no information on where and how the energy is dissipated. In simulations, one can compute the hysteresis curve of each MNP, but this includes contributions from the dipolar fields which exchange energy between MNPs, therefore hysteresis curves cannot determine local heating in interacting systems\cite{torche_thermodynamics_2020,munoz-menendez_disentangling_2020}.

Using \cref{eq:P_in,eq:P_mag,eq:P_rot,eq:P_trans} directly gives a local perspective on the instantaneous power, both input and loss channels, which also works for transient and non-equilibrium dynamics of interacting MNP systems. If coupled to a model of heat transfer, one can in principle simulate the entire temperature distribution. Also plotting \cref{eq:P_in,eq:P_mag,eq:P_rot,eq:P_trans} vs.\ time gives an alternative way to visualise and interpret various MNP dynamics, as we demonstrate in \cref{sec:simulations}.

\section{Model approximations \label{sec:Model_approximations}}

\Cref{eq:mu_dot,eq:omega_dot,eq:r_ddot,eq:u_dot} constitute a highly general, numerically solvable model for the dynamics of an MNP collection. It can be used to study aggregation, hysteresis, magnetic susceptibility, Brownian diffusion, moment-dynamics in clusters etc.\ The price is heavy computations, with little hope for analytical solutions, therefore it is common to consider approximate models.
Here we discuss several useful approximations in the context of conservation laws.

\subsection{Parameter ranges \label{subsec:Parameter_ranges}}

In \cref{tab:parameters} we list the model parameters, including typical ranges in MNP studies for particles of Fe, Co, Ni and their oxides.

Spherical MNPs typically cease to be single-domain at \cite{singamaneni_magnetic_2011} $R = 50\: \mathrm{nm}$ or less, with some notable exceptions for rare earth compounds \cite{michels_magnetic_2021}.
Studies are typically conducted at room- or body temperature for biomedical applications, but $T = 950\: \mathrm{C}^{\circ}$ has been tested for catalysis\cite{almind_optimized_2021}.
Water, as one of the least viscous liquids, has $\eta = 1.0\: \mathrm{mPa\cdot s}$ at $T=300\:\mathrm{K}$. 
Typical MNP materials include maghemite ($\mathrm{Fe_3O_4}$) for biomedicine and cobalt for catalysis. The bulk, saturation magnetisations are respectively $M_s^{\mathrm{Fe_3O_4}} = 412 \:\mathrm{kA/m}$ and $M_s^\mathrm{Co} = 1.45\mathrm{MA/m}$, however the magnetisation of nanomagnets is typically smaller than bulk due to disordered surface spins\cite{disch_quantitative_2012}. The same surface effects increase the effective anisotropy constant for very small particles\cite{bodker_surface_1994}; for example\cite{pisane_unusual_2017} $K = 20\:\mathrm{kJ/m^3}$ for bulk maghemite but for $R = 2 \: \mathrm{nm}$ particles experimental results are around $K = 90 \: \mathrm{kJ/m^3}$.

Using ferromagnetic resonance Bhagat \& Lubitz\cite{bhagat_temperature_1974} measured the Gilbert damping in monocrystalline, bulk Fe, Co and Ni as a function of temperature. Reading off their relaxation parameter at $T=300\:\mathrm{K}$ and multiplying by\cite{gilmore_identification_2007} $4\pi/(\mu_0\gamma M_s)$ to convert to $\alpha$, the results are $\alpha_\text{Fe} = 2.2\cdot 10^{-3}$, $\alpha_\text{Co}=3.5\cdot 10^{-3}$ and $\alpha_\text{Ni} = 2.5\cdot 10^{-2}$.
By doping NiFe thin films $\alpha_\mathrm{NiFe} = 0.1$ has been measured at 300 K\cite{bailey_control_2001}.
At the other end of the scale is Yttrium-Iron-Garnet(YIG), which for a monocrystalline, $R=300\:\mathrm{\mu m}$ sphere was found to have\cite{klingler_gilbert_2017} $\alpha_{YIG} = 2.7\cdot 10^{-5}$. 
In summary, $\alpha$ is generally $\ll 1$ with typical values around $10^{-3}$ to $10^{-2}$ and outliers down to less than $10^{-4}$. 
However when simulating phenomena with a weak $\alpha$-dependence, one can speed up magnetic relaxation by using $\alpha$ near unity, thus potentially reducing computation time.

When two MNPs are in surface contact, the field from one on the other is
\begin{align*}
    B^\text{dip}_\text{contact} = \frac{\mu_0 \mu}{2\pi R_h^3} = \frac{2}{3}\mu_0 M
\end{align*}
which ranges from $8\:\mathrm{mT}$ to $800\:\mathrm{mT}$ for the considered magnetisations. So it is highly case specific whether the external field exceeds the dipole field.

For reference the mass, moment of inertia and magnetic moment of a spherical, single-domain magnet are $\mathfrak{m} = \frac{4\pi}{3}\rho R^3$, $I=\frac{8\pi}{15}\rho R^5$ and $\mu=\frac{4\pi}{3}M R^3$ respectively. Friction coefficients are given in \cref{eq:zeta_sphere}.

\begin{table}[H]
    \centering
    \begin{tabular}{|c|c|c|c|}
        \hline
        Symbol & Description & Unit & Values \\
        \hline \hline
        $\gamma$ & Gyromagnetic ratio & $\mathrm{s^{-1} \: T^{-1}}$& $1.76 \cdot 10^{11}$ \\
        \hline
        $R$ & Radius & $\mathrm{nm}$ & $1..50$\\
        \hline
        $T$ & Temperature & $\mathrm{K}$ & $0..1000$ \\
        \hline
        $\eta$ & Dynamic viscosity & $\mathrm{kg / (m \cdot s)}$ & $10^{-3}..10^{-1}$ \\
        \hline 
        $\rho$ & Mass density& $\mathrm{g/cm^3}$ & $1..10$ \\
        \hline
        $M$ & Magnetisation & $\mathrm{A / m}$ & $10^{4}.. 10^{6}$ \\
        \hline
        $B_\text{ext}$ & Applied field magnitude & $\mathrm{T}$ & 0..1 \\
        \hline
        $f_\text{ext}$ & Applied field frequency & $\mathrm{Hz}$ & $0 .. 10^{7}$ \\
        \hline
        $K$ & Anisotropy energy density & $\mathrm{J / m^{3}}$ & $10^{3}..10^{5}$ \\
        \hline
        $\alpha$ & Gilbert damping & $\mathrm{1}$ & $10^{-3} .. 0.5$ \\
        \hline
    \end{tabular}
    \caption{Typical parameter ranges in studies of single-domain MNPs made from Fe, Co, Ni and their oxides.}
    \label{tab:parameters}
\end{table}

\subsection{Overdamped limit \label{subsec:Overdamped_limit}}

For micron scale and smaller objects moving in liquid, inertia tends to be negligible\cite{purcell_life_1977}. The argument is that viscous drag dampens away transient responses so quickly that objects are essentially always moving at terminal velocity. Therefore in Langevin dynamics studies of microswimmers the inertia term is traditionally neglected\cite{lowen_inertial_2020}. For MNPs torques and forces can change on ns time-scales or faster due to moment precession\cite{berkov_langevin_2006}, so it is not immediately clear if inertia can always be neglected. That said, some studies\cite{usadel_dynamics_2015}, including our own in \cref{sec:simulations}, have found no apparent difference between the MNP trajectories for simulations with/without inertia.

In the overdamped limit ($\mathfrak{m}, I \xrightarrow{} 0$) \cref{eq:omega_dot,eq:r_ddot} reduce to
\begin{align}
    \zeta^\text{r} \bs{\omega}_i &= \mu \gamma^{-1} \vb{\dot{m}}_i + \mu \vb{m}_i \cross \vb{B}_i + \bs{\tau}^\text{th}_i   \label{eq:omega_overdamped}
    \\
    \zeta^\text{t} \vb{v}_i &= \vb{F}^\text{dip}_i + \vb{F}^\text{th}_i \label{eq:r_dot_overdamped}
\end{align}
while \cref{eq:u_dot,eq:mu_dot,eq:Omega} remain unchanged. Then linear- and angular velocities ($\vb{v}, \bs{\omega}$) are given by the \textit{immediate} forces and torques, so the system state is specified just by $\vb{m}_i, \vb{u}_i, \vb{r}_i$.

The power expressions \cref{eq:P_in,eq:P_mag,eq:P_rot,eq:P_trans} are formally unchanged, while $\vb{\dot{p}} = \vb{\dot{L}} = 0$. The reason is that the transient, inertial dynamics are approximated as decaying infinitely quickly, so the system is always in force and torque balance. In this case one can still calculate momenta directly from $\vb{v}(t)$ and $\bs{\omega}(t)$, but \cref{eq:Li_dot,eq:Ji_dot,eq:J_dot} are pointless.

\subsubsection{Characteristic frequencies \label{subsec:characteristic_frequencies}}

In the overdamped limit we can easily determine characteristic frequency scales. For numerical estimates, we consider the parameter ranges in \cref{tab:parameters}. Using that $\alpha \ll 1$ we have $\gamma' \approx \gamma$, so $\Omega \sim \gamma B^\text{eff}$ which is the well-known Larmor frequency for moment precession. 
For mechanical rotation, inserting \cref{eq:mu_dot} in \cref{eq:omega_overdamped} yields $\zeta^\text{r}\omega \sim \mu B^\text{ani} + \alpha \mu B^\text{eff}$. If $\alpha B^\text{eff} \lesssim B^\text{ani}$ then $\omega \sim \mu B^\text{ani}/\zeta^\text{r} \sim K/(3\eta)$, which ranges from 30 kHz to 300 MHz. In general $\omega \lesssim \mu B^\text{eff} / \zeta^\text{r}$, so whenever the overdamped limit is applicable
\begin{align*}
    \frac{\omega}{\Omega} \lesssim \frac{\mu}{\gamma \zeta^\text{r}} = \frac{M}{6\gamma \eta} \leq 0.001. 
\end{align*}
We conclude that for any overdamped MNP system in the parameter space of \cref{tab:parameters}, mechanical rotation is at least 3 orders of magnitude slower than moment precession.

The characteristic rate of magnetic relaxation is \cite{berkov_langevin_2006} $\alpha \Omega$ which also exceeds $\omega$ for all parameter combinations in \cref{tab:parameters}. Both magnetic frequency scales are apparent in \cref{fig:1MNP_hysteresis}.

\subsubsection{Neglecting the Barnett field}

For the Barnett field, \cref{eq:B_bar}, we have 
\begin{align*}
    \frac{B^\text{bar}}{B^\text{eff}} \sim \frac{\omega}{\Omega} \leq 0.001.
\end{align*}
Thus for typical MNP simulations in liquid, the Barnett field is a tiny correction to $\vb{B}^\text{eff}$ and, as pointed out in \cref{subsec:Angular_momentum_conservation}, it does not change whether angular momentum is conserved. Similarly, one can to a good approximation set $\omega = 0$ in \cref{eq:P_mag}, i.e.\ calculate $P^\text{mag}$ in laboratory coordinates rather than local coordinates. While this vindicates previous numerical studies, e.g.\cite{helbig_self-consistent_2023,usov_dynamics_2012,usov_dynamics_2019,berkov_langevin_2006,cabrera_unraveling_2017}, we find that using the exact expressions does not significantly increase computational complexity.

For MNPs suspended in vacuum or low-density gas, the mechanical rotation can be much greater, so the above argument breaks down. Indeed in low friction conditions the Barnett field has been theorised to produce additional lines in a ferromagnetic resonance spectrum\cite{keshtgar_magnetomechanical_2017}, and to enable stable levitation of a nanomagnet even in a static applied field\cite{rusconi_quantum_2017,kustura_stability_2022}. Also, by rapidly rotating dried and frozen samples of $\mathrm{Fe_3O_4}$ nanoparticles the Barnett field has been measured with enough accuracy to infer orbital corrections to the gyromagnetic ratio\cite{umeda_temperature-variable_2023}. To simulate near-vacuum conditions, one may use the full inertial model with a low value of the friction coefficient $\eta$.

\subsubsection{Alternative power expressions}

Let $\bs{\tau}^\text{det} = \zeta^\text{r}\bs{\omega}_i - \bs{\tau}^\text{th}$ be the deterministic part of the non-viscous torque, i.e.\ the torque from potential gradients. In the overdamped limit, we can rewrite the rotational power (\cref{eq:P_rot}) as
\begin{align}
    P^\text{rot}_i = - \bs{\omega}_i \vdot \bs{\tau}^\text{det}_i = -\frac{1}{\zeta^\text{r}} [(\tau^\text{det}_i)^2 + \bs{\tau}^\text{det}_i \vdot \bs{\tau}^\text{th}_i]  \label{eq:P_rot_alternative}
\end{align}
and similarly the translational power is
\begin{align}
    P^\text{trans}_i = -\vb{v}_i \vdot \vb{F}_i^\text{det} = - \frac{1}{\zeta^\text{t}}[(F^\text{det}_i)^2 + \vb{F}^\text{det}_i \vdot \vb{F}^\text{th}_i] \label{eq:P_trans_alternative}
\end{align}

Because terms that are second order in the thermal fluctuations have been cancelled by hand, \cref{eq:P_rot_alternative,eq:P_trans_alternative} are slightly more numerically stable than \cref{eq:P_rot,eq:P_trans}.

It is seen that when $\bs{\tau}^\text{det}=0$ ($\vb{F}^\text{det}=0$) the translational (rotational) energy transfer is precisely 0, despite thermal fluctuations. The reason is that if there is no inertia and the potential energy is flat wrt.\ particle positions and orientations, the system cannot store mechanical energy. Likewise, Leliaert et.\ al.\cite{leliaert_individual_2021} showed that when the potential has no dependence on moment direction, such as for an immobilised MNP in zero-field and without uniaxial anisotropy, there is no magnetic energy transfer; not even thermal.

\subsection{Rigid dipole approximation \label{sec:Rigid_dipole_approximation}}

The popular rigid dipole approximation (RDA)\cite{coffey_inertial_1996,durhuus_simulated_2021,usadel_dynamics_2017}, says that the moment is locked to the mechanical rotation of the particle, i.e.\ $\vb{m} = \vb{u}$. Then \cref{eq:mu_dot} drops out, $\alpha$ is no longer a model parameter and \cref{eq:omega_dot} reduces to
\begin{align*}
    I\bs{\dot{\omega}}_i + \zeta^\text{r} \bs{\omega}_i = \mu \gamma^{-1} \vb{\dot{u}}_i + \mu \vb{u}_i \cross \vb{B}_i + \bs{\tau}^\text{th}_i \qq{(RDA)}
\end{align*}
Using that $\vb{\dot{u}} = \boldsymbol{\omega} \cross \vb{u}$ this can be rewritten
\begin{align*}
    I\bs{\dot{\omega}}_i + \zeta^\text{r} \bs{\omega}_i = + \mu \vb{u}_i \cross (\vb{B}_i + \vb{B}^\text{bar}_i) + \bs{\tau}^\text{th}_i \qq{(RDA)}
\end{align*}
Typically the $\vb{\dot{u}}_i$ term is also neglected, which evidently is equivalent to neglecting the Barnett field, hence justified in liquid suspension.
 
Because you avoid the high frequency moment dynamics, the RDA enables a major increase in simulation speed, and facilitates analytical solutions\cite{usadel_dynamics_2017}. The power expressions \cref{eq:E_dot,eq:P_in,eq:P_rot,eq:P_trans} are unchanged but $P^\text{mag} = 0$, i.e.\ Gilbert damping is excluded from the model. 

The RDA relies on 2 assumptions : \textbf{(1)} that transient, moment dynamics are irrelevant for the phenomenon of interest and \textbf{(2)} that $\vb{B}^\text{eff}$ is dominated by $\vb{B}^\text{ani}$ so that $\vb{m} \parallel \vb{u}$ in steady-state. We refer to Ref. \cite{usadel_dynamics_2017} for further analysis of when and why the RDA is applicable. 

In some regimes, even though $\vb{m} \parallel \vb{u}$ is nearly true most of the time, Gilbert damping is the primary loss channel\cite{usov_dynamics_2012,helbig_self-consistent_2023}, so the RDA should be applied with great care when studying energy transfer. An example is seen in \cref{fig:1MNP_hysteresis}. We also present a simulation of zero-field aggregation in \cref{fig:10MNP_collision} where the RDA would have been reasonable.

\subsection{Immobilised particles}

Consider a collection of MNPs immobilised in a solid substrate, i.e.\ no mechanical rotation or translation possible. Then $\vb{B}^\text{bar}, P^\text{rot}, P^\text{trans} = 0$ and \cref{eq:omega_dot,eq:r_ddot,eq:u_dot} drop out of the model, leaving just the LLG part; \cref{eq:mu_dot}. $\vb{r}_i, \vb{u}_i$ are constants which still enter through $\vb{B}^\text{dip}_i$ and $\vb{B}^\text{ani}_i$ respectively, but $\vb{m}_i$ are the only dynamical variables. Then the model is essentially an example of micromagnetics.

\section{Simulations \label{sec:simulations}}

\begin{figure*}[t]
    \centering
    \begin{minipage}[l]{0.45\textwidth}
        \includegraphics[width=\textwidth]{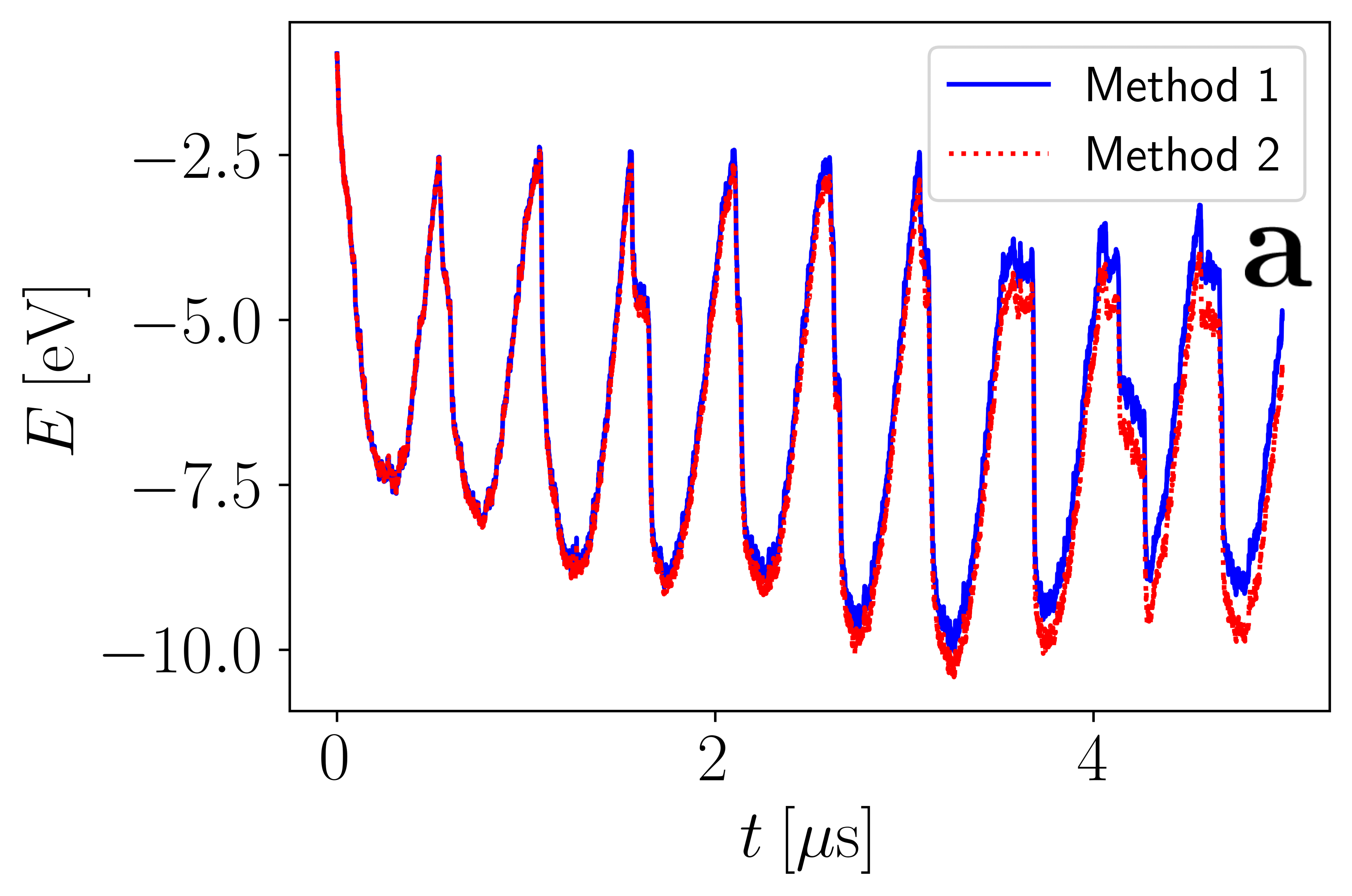}
    \end{minipage}%
    \begin{minipage}[r]{0.45\textwidth}
        \includegraphics[width=\textwidth]{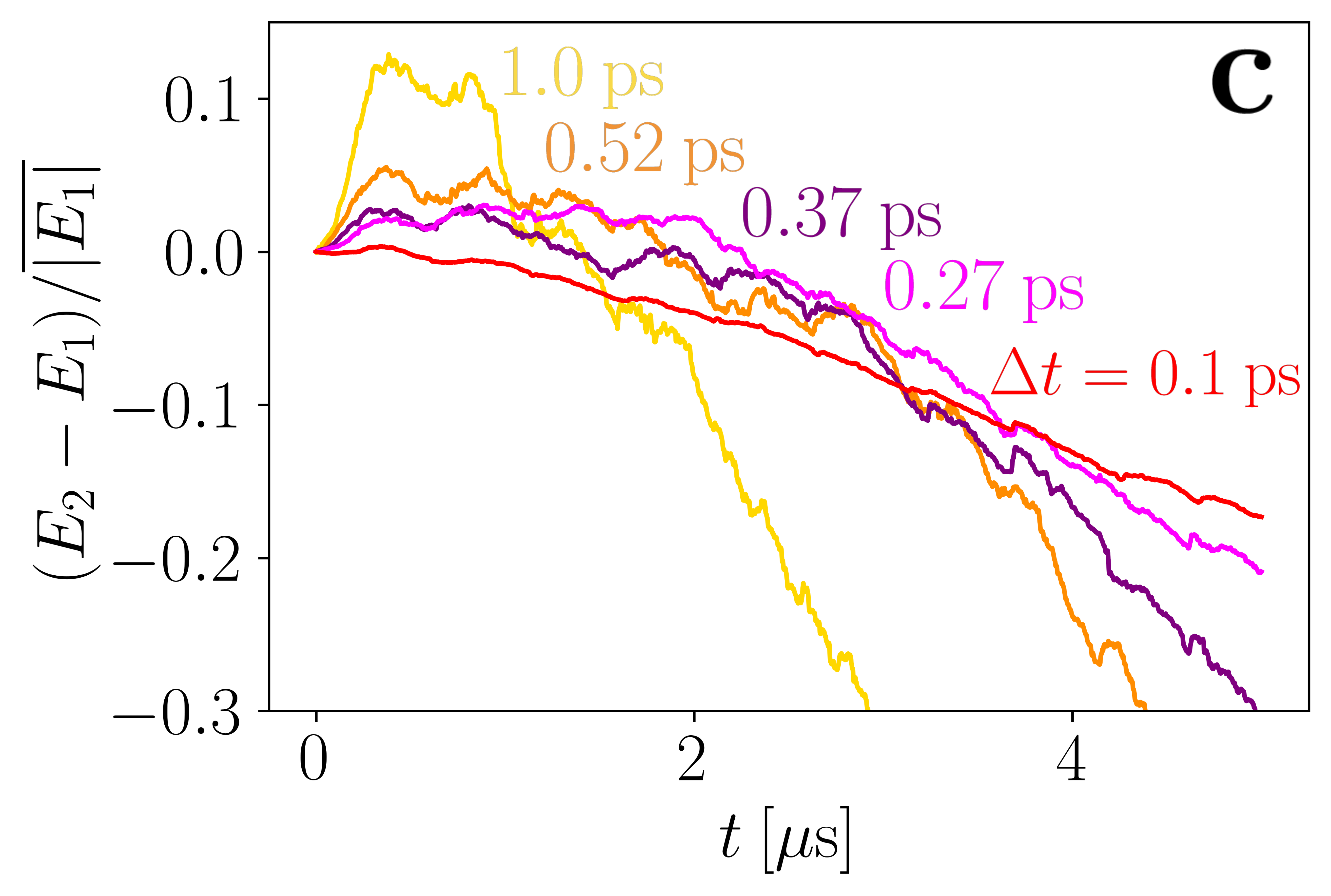}
    \end{minipage}
    \begin{minipage}[l]{0.45\textwidth}
        \includegraphics[width=\textwidth]{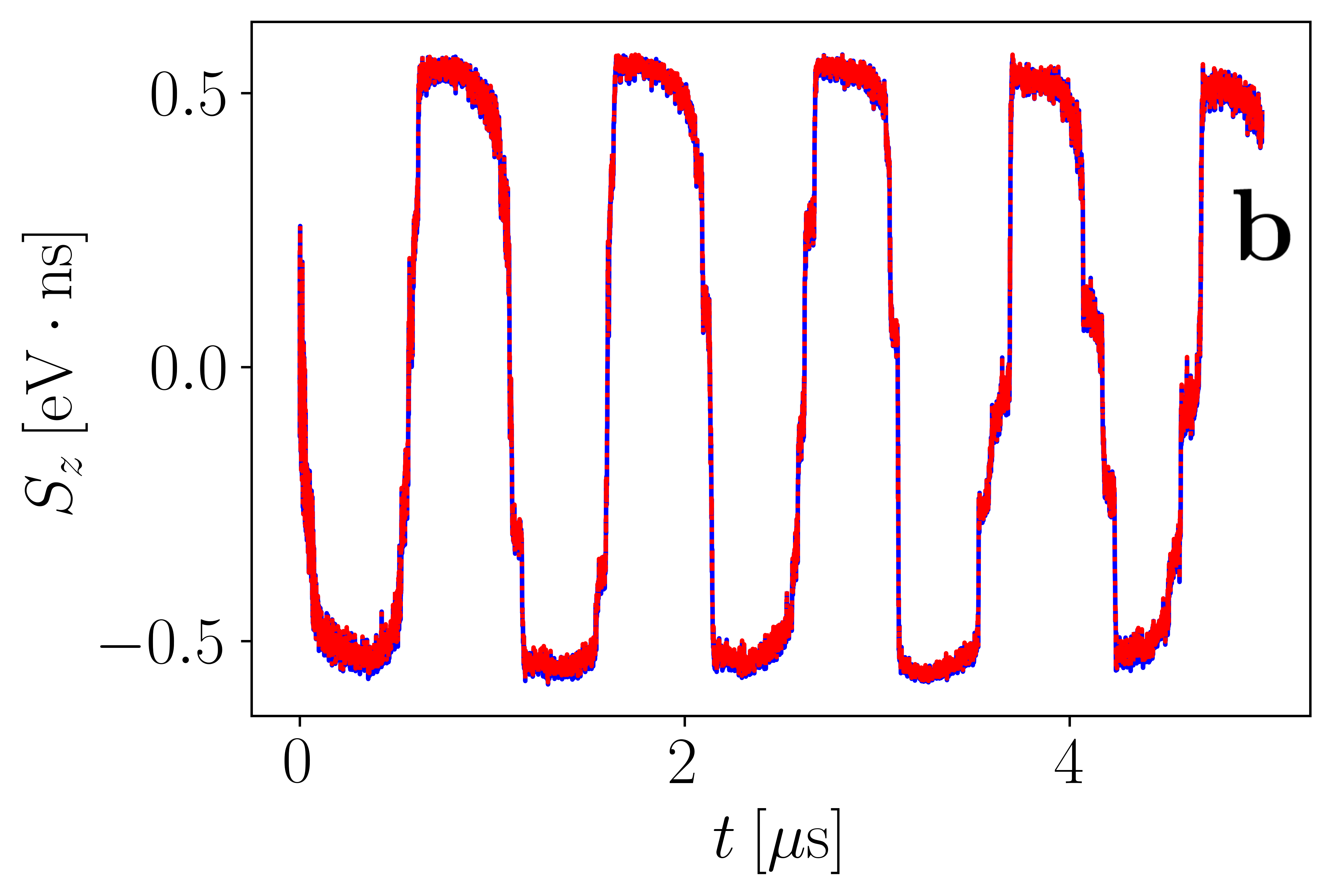}
    \end{minipage}%
    \begin{minipage}[r]{0.45\textwidth}
        \includegraphics[width=\textwidth]{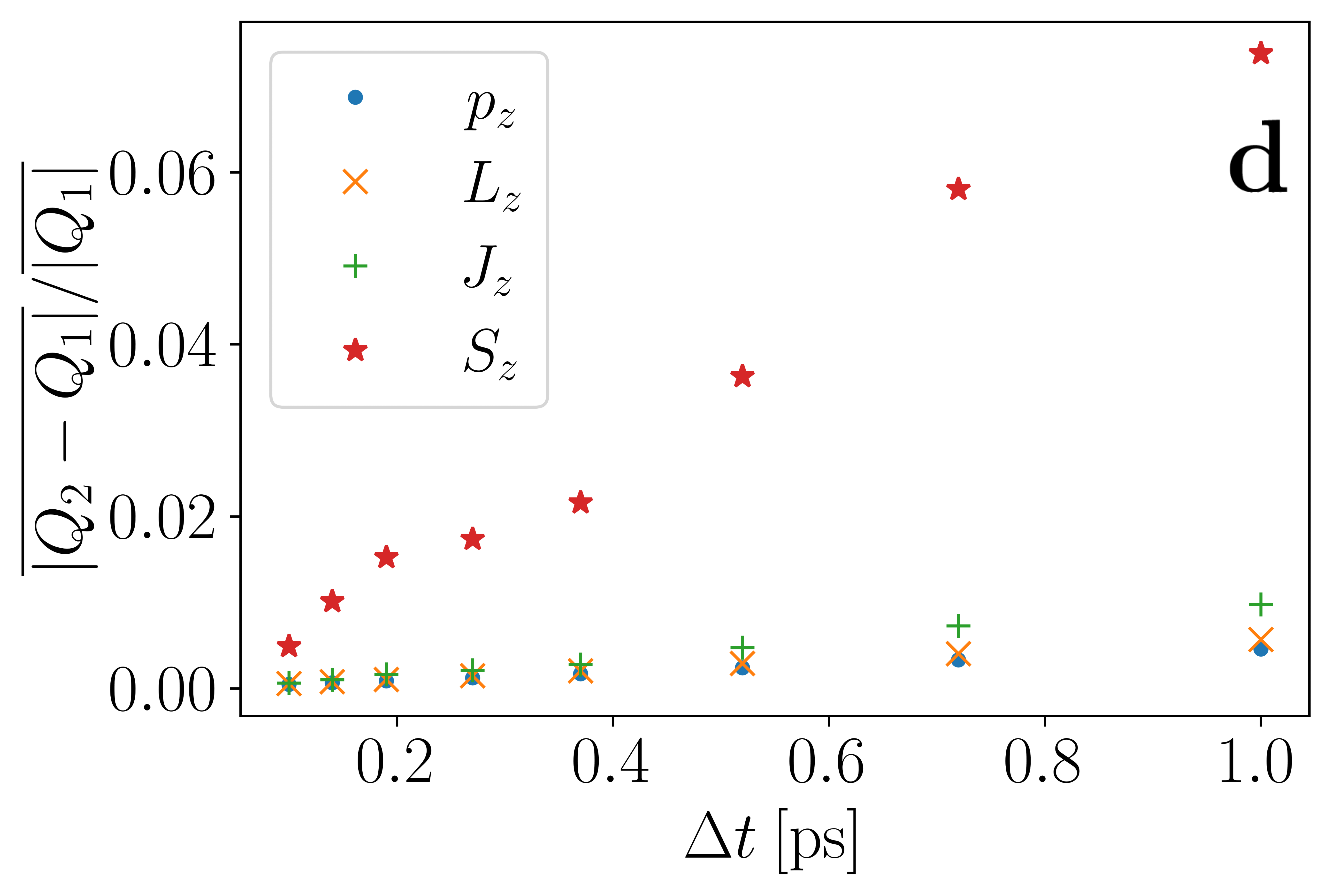}
    \end{minipage}
    \caption{Conservation law tests. The test system is 10 MNPs with $R=10\: \mathrm{nm}$, $M=400 \: \mathrm{kA/m}$, $K=10\:\mathrm{kJ/m^3}$, $\rho=4.9\:\mathrm{g/cm^3}$ and $\alpha=0.01$ dispersed in liquid with $\eta = 1\: \mathrm{mPa\cdot s}$ and $T=300\:\mathrm{K}$. There is an external field of the form $\vb{B}^\text{ext} = B^\text{ext} \sin(2\pi f^\text{ext} t) \vb{e}_z$ where $B^\text{ext}=50\:\mathrm{mT}$ and $f^\text{ext} = 1\:\mathrm{MHz}$. The maximal anisotropy field strength is then $B^\text{ani}_\text{max} = 2K/M = 50\:\mathrm{mT}$. Initial conditions are random with a volume fraction of 0.1, but the same across all simulations. An equivalent initial state is seen in \cref{fig:10MNP_collision}c.
    \textbf{a,b)} System energy and $z$-component of magnetic angular momentum simulated with methods 1 and 2 at $\Delta t = 0.1\: \mathrm{ps}$. \textbf{c)} Relative difference between $E_2$ and $E_1$ vs. time at different timesteps. \textbf{d)} Average error in $z$-components of system momenta vs.\ timestep for method 2 relative to method 1. $\overline{Q}$ refers to a time-average over the whole $5\: \mathrm{\mu s}$ simulation, where $Q$ is a placeholder for $p_z, L_z, J_z$ and $S_z$. \textbf{b,d)} include inertia for tracking momentum but in \textbf{a,c)} the overdamped limit is used and kinetic energy is excluded. 
    For each datapoint, the method 1 and 2 estimates are from the same simulation, so the thermal noise is identical.}
    \label{fig:conservation_law_tests}
\end{figure*}

\subsection{Method \label{subsec:Method}}

\subsubsection{Timestep integration}
To simulate the derived model, our method of choice is timestep integration. The core idea is to discretise time into steps of $\Delta t$ which are short compared to all characteristic timescales in the system. This justifies a number of numerical schemes for updating the configuration from $t$ to $t + \Delta t$.
By iteratively updating the system, its entire time-evolution is evaluated. 

In addition to the model in \cref{sec:Full_Langevin_dynamics} we include the WCA-force:
\begin{align}
    \vb{F}^\text{WCA}_i = \sum_{j\neq i} \begin{cases}
    12 \epsilon_\text{WCA} \left[\frac{(2R)^{12}}{r_{ji}^{13}} -  \frac{(2R)^6}{r_{ji}^7}\right] 
    & r_{ji} \leq 2R
    \\
    0 & r_{ji} > 2R
    \end{cases} 
    \label{eq:F_WCA}
\end{align}
and associated potential. We use $\epsilon_\text{WCA} = 10^{-19}\: \mathrm{J}$ throughout. This short-range, purely-repulsive potential limits particle overlap so the code can handle collisions, but contributes little to the system energy (see \cref{fig:10MNP_collision}b).

We present our algorithms in \cref{appsec:implementation}. The mechanical equations (\cref{eq:omega_dot,eq:r_ddot}) are second order with inertia but first order in the overdamped limit, so two different algorithms are needed. We use the simple Euler method for updating position, velocity etc.\ and the Euler-Rodriguez formula for rotations\cite{cheng_historical_1989} because it is exceedingly good at conserving vector magnitudes. For moment rotations, we also tested a combination of Heuns method and the Euler-Rodriguez formula, but for ease of reproducibility all presented data uses the simple scheme.

One can improve performance by various higher-order integrators and adaptvie-timestepping\cite{leliaert_adaptively_2017} or symplectic algorithms for the second-order equations\cite{young2014leapfrog}. However there are already a number of technical subtleties when integrating stochastic equations\cite{gardiner_handbook_2002}, so we prioritised clarity over efficiency.

\subsubsection{Stochastic integration \label{subsubsec:Stochastic_integration}}

Our model is a set of coupled stochastic differential equations (SDEs), because it contains the stochastic vectors $\vb{B}^\text{th}, \vb{F}^\text{th}, \bs{\tau}^\text{th}$. Since these have zero auto-correlation (see \cref{eq:B_th,eq:F_th,eq:tau_th}), they change significantly on an infinitesimal timescale. Therefore they obey different rules of calculus than the deterministic vectors. See e.g.\cite{sarkka_applied_2019,gardiner_handbook_2002,oksendal_stochastic_1998} for an introduction to SDEs and stochastic calculus.

When discretising time, the closest we can get to zero auto-correlation is to randomise the stochastic vectors every timestep. Formally this amounts to the substitution $\delta(t-t') \xrightarrow{} \frac{1}{\Delta t} \delta_{nn'}$ in \cref{eq:B_th,eq:F_th,eq:tau_th} where $n$ is timestep index. Between updates, the thermal vectors are constant so the time-evolution is deterministic. When integrating an SDE like \cref{eq:E_dot} or \cref{eq:J_dot}, it matters at what point in the deterministic intervals the integrand is evaluated, because $\vb{\dot{S}}, P^\text{mag}, P^\text{rot}$ and $P^\text{trans}$ all contain \textit{multiplicative noise} terms; products of deterministic- and stochastic variables like $\vb{v} \vdot \vb{F}^\text{th}$ or $\bs{\mu} \cross \vb{B}^\text{th}$. If integrated incorrectly, multiplicative noise terms may produce artificial \textit{thermal drift}, which is an error proportional to integration time $t$. Unlike numerical errors from time-discretisation, thermal noise results from a formal error in the interpretation of an SDE, hence it remains even as the timestep approaches zero.

For continuous processes, as are common in physical systems, the correct choice is often midpoint integration, also known as a Stratonovich integral. However there are counterexamples. For example Ref. \cite{pesce_stratonovich--ito_2013} experimentally studies an electric circuit that can be tuned continuously from Stratonovich calculus to Itô (initial point integration) depending on input parameters. We note that the regular rules of differentiation apply only in Stratonovich calculus, so this is a prerequisite for the derivation of $\dot{E}$ in \cref{appsec:E_dot}.

Here it should be mentioned that the LLG itself contains multiplicative noise terms. However it has been shown that the drift is only in the magnitude of magnetic moments\cite{berkov_thermally_2002}, so as long as $m_i$ are held constant, one can use all the tools and algorithms for time-stepping regular ODEs\cite{leliaert_adaptively_2017}. For the energy transfer we are not so lucky.

\subsubsection{Evaluating conserved quantities}

For the instantaneous values of energy and momenta at a given time, we simply insert the simulated system configuration in \cref{eq:E_multi_MNP,eq:J_multi_MNP,eq:p_multi_MNP}. Similarly the transfer rates are given by \cref{eq:p_dot,eq:J_dot,eq:E_dot}. This directly gives a way to visualise and interpret MNP dynamics in terms of energy, as we demonstrate in \cref{subsec:Single-particle_hysteresis,subsec:Hysteresis_during_collision,subsec:Zero-field_aggregation}. One caveat is that we set kinetic energy to 0 in the overdamped limit for self-consistency.

When including thermal noise or for practical applications, it is often more informative to consider time averages. For example for hyperthermia the figure of merit is the average input power. This is typically estimated from the area of the hysteresis curve, but can equivalently be calculated by the time-average $\overline{P^\text{hyst}} = \frac{1}{t} \int_0^t P^\text{hyst}(t') \dd t'$. With the present formalism we can likewise average magnetic- and viscous loss power, as well as forces and torques. 

To test the self-consistency of our analytical results, we consider 2 different methods for evaluating the total energy and momenta numerically. Method 1 is to take the system configuration directly. For example in the overdamped case $E_1(t) = E[\vb{m}(t), \vb{u}(t), \vb{r}(t)]$. Method 2 is to integrate the transfer rates, e.g.\ $E_2(t) = E(t=0) + \int_0^t \dot{E}(t') \dd t'$. The numerical results are presented in \cref{subsec:Conservationa_law_tests} and \cref{fig:conservation_law_tests}. Method 2 involves the same integral as the time-averaged transfer rates, so the consistency of the two methods demonstrates the elimination of thermal noise, and that we can numerically integrate the transfer rates in practice. The comparison can also be used to benchmark a given numerical implementation.

We find that in the overdamped limit, using midpoint integration eliminates all thermal drift, so the Stratonovich interpretation is correct and we can integrate all power contributions at finite temperature. This also justifies using the regular rules of differentiation in deriving \cref{eq:E_dot}. As discussed in \cref{subsec:Overdamped_limit}, $\vb{\dot{p}} = \vb{\dot{J}} = 0$ in the overdamped limit, so we cannot compute changes in the MNP momenta. One can still integrate the individual forces and torques, but our self-consistency test requires inertia.

$\vb{\dot{S}}$ is prone to thermal noise, because unlike the single-particle moments the magnitude of $\vb{S}$ fluctuates, but midpoint integration works. Thus with inertia and midpoint integration specifically for $\vb{\dot{S}}$ we can integrate all momentum components at finite temperature. However when integrating the total power, significant thermal drift is present for both Itô and Stratonovich integration regardless of timestep, and we have not found a suitable alternative. It is possible the correct SDE interpretation depends on the magnitude of the inertia, analogously to\cite{pesce_stratonovich--ito_2013}.
See \cref{tab:integration_methods} for a summary of when we can integrate the transfer rates.

We note that for all simulations presented in \cref{subsec:Conservationa_law_tests,subsec:Zero-field_aggregation,subsec:Hysteresis_during_collision} we simulated both the overdamped limit and full, inertial model, and also compared methods 1 and 2 for energy and momenta to the extent it was possible without thermal drift. There was no apparent difference between inertial and overdamped curves, so MNP mass is negligible for all presented simulations.

\begin{table}[]
    \centering
    \begin{tabular}{|c|c|c|c|} 
    \hline
    & &Energy & Momenta \\
    \hline
   Overdamped  & $T=0$ & \checkmark & $\cross$  \\
   \hline
   Overdamped & $T > 0$ & \checkmark & $\cross$ \\
    \hline
   Inertial & $T=0$ & \checkmark & \checkmark \\
   \hline
    Inertial & $T > 0$ & $\cross$ & \checkmark \\
    \hline
\end{tabular}
    \caption{Under what conditions we can integrate power and net momentum transfer (linear and angular) for the MNP system. Comparing to direct energy and momenta computations is useful for validation of models and benchmarking of implementations.}\label{tab:integration_methods}
\end{table}

\begin{figure*}[t]
    \centering
    \begin{minipage}[l]{0.45\textwidth}
        \includegraphics[width=\textwidth]{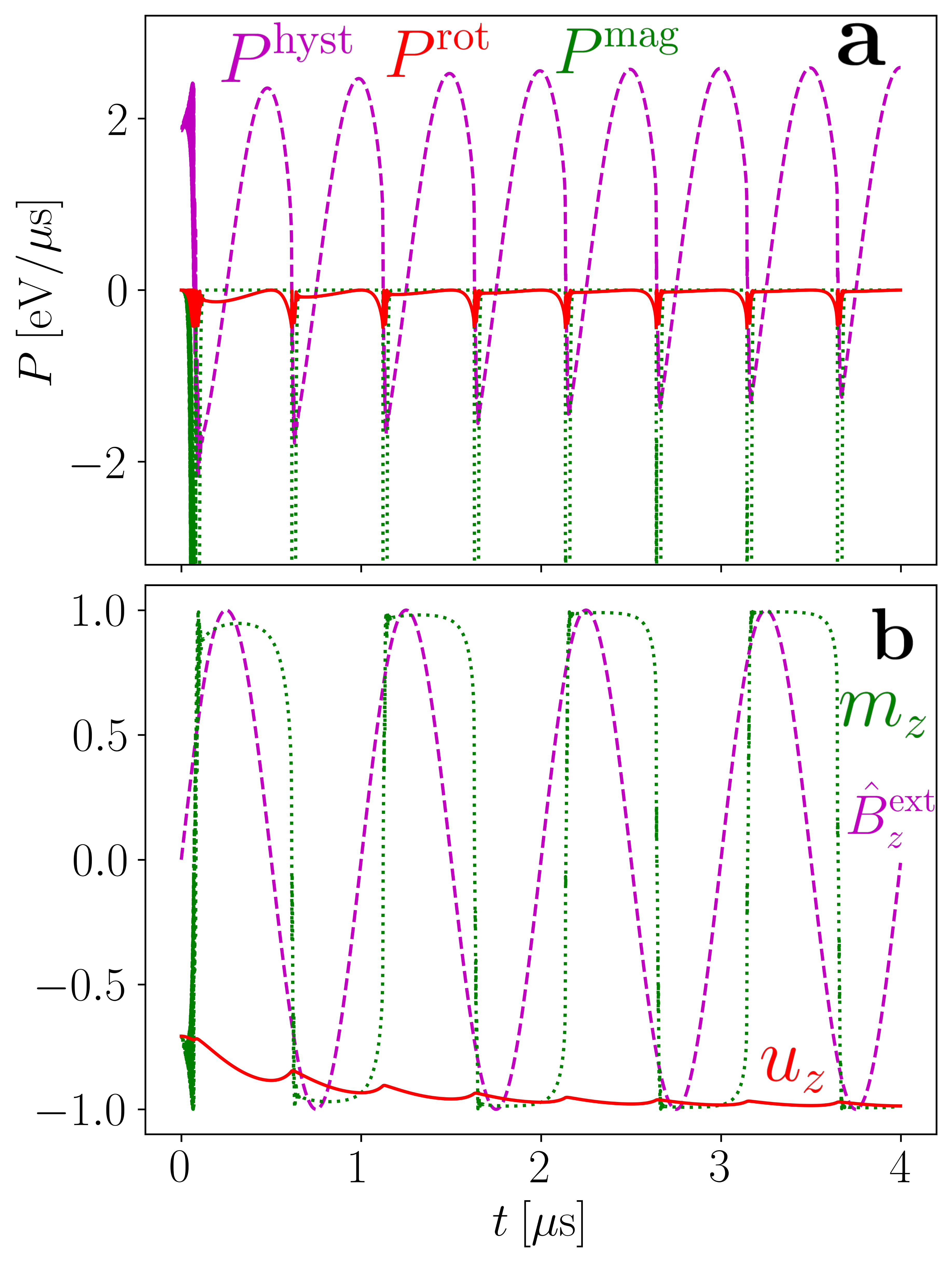}
    \end{minipage}
    \begin{minipage}[r]{0.45\textwidth}
        \includegraphics[width=\textwidth]{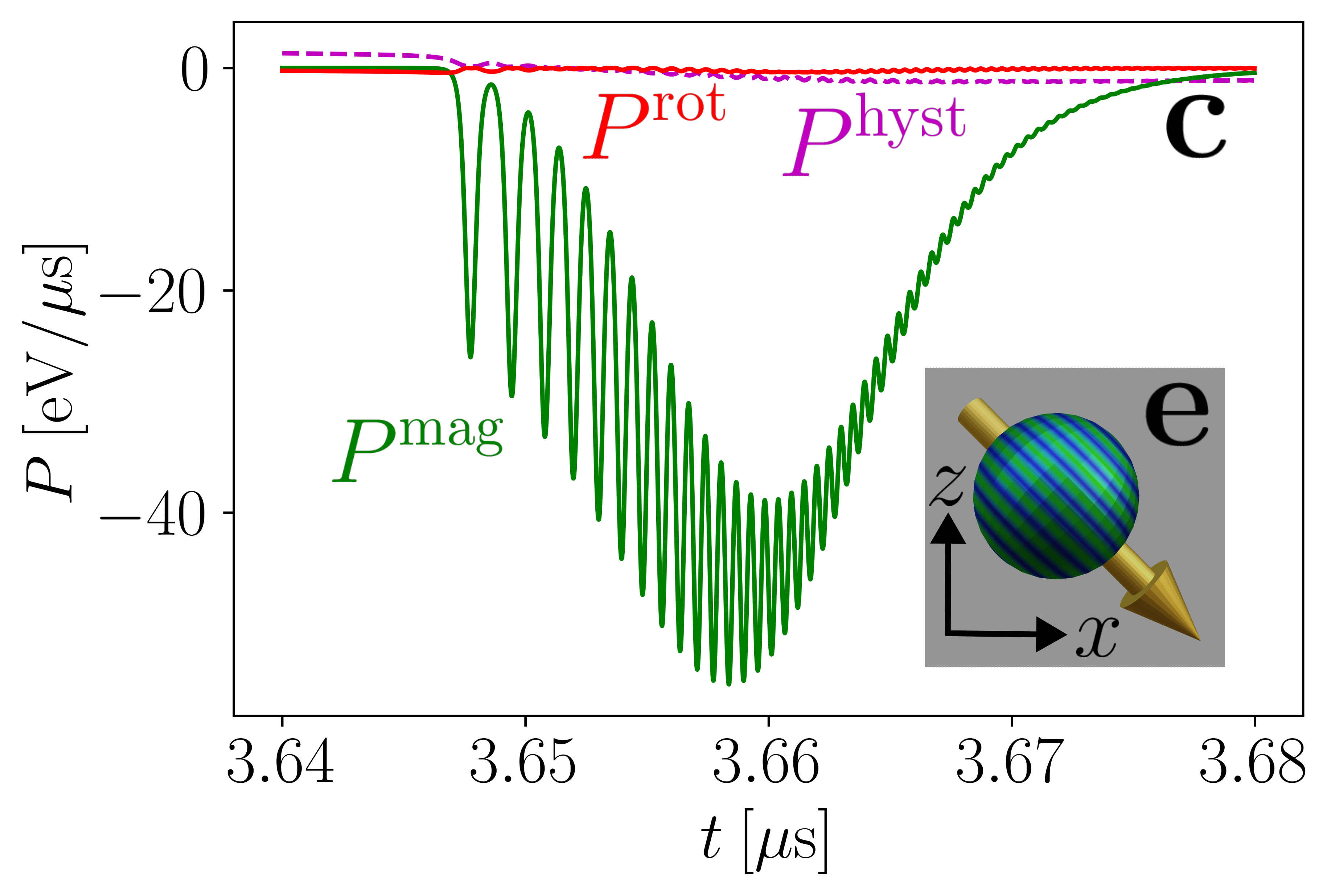}
        \includegraphics[width=\textwidth]{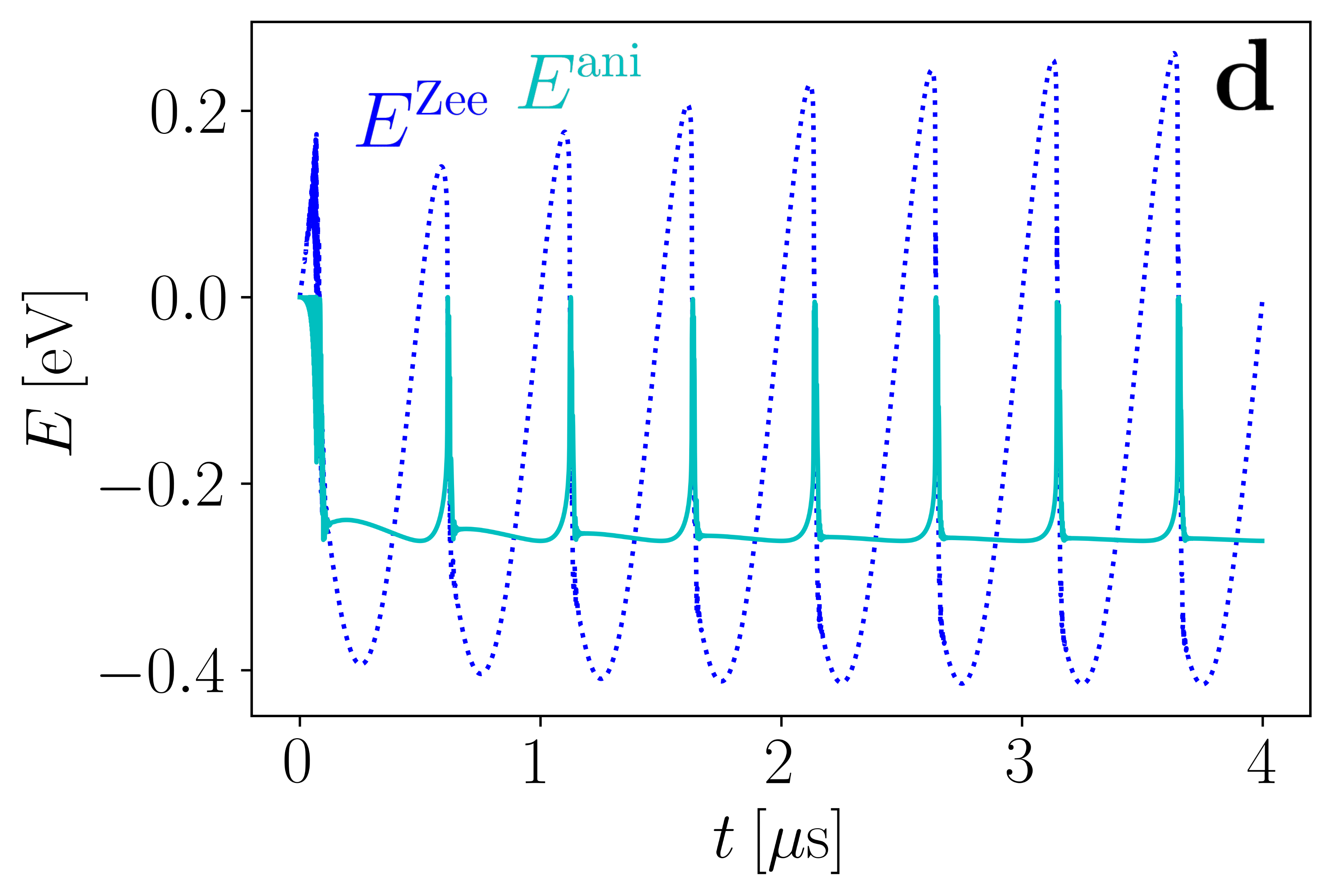}
    \end{minipage}
    \caption{Single MNP in liquid. Same parameters as in \cref{fig:conservation_law_tests} except $T=0$ and $B^\text{ext}=40\:\mathrm{mT}$. The initial state is shown in \textbf{e)} and given by $\vb{m}=\frac{1}{\sqrt{2}}(1, 0, -1), \vb{u} = \frac{1}{\sqrt{2}}(-1, 0, -1)$. The plots are identical for inertial- and overdamped simulations. \textbf{a,c)} Energy transfer vs. time, decomposed into Gilbert damping, rotational viscous damping and input power from the driving field. \textbf{a)} shows the full 4 $\mathrm{\mu s}$ but a subset of the power axis for clarity, while \textbf{c)} shows a 40 ns zoom onto the last moment reversal. \textbf{b)} Same simulation in terms of the z components of the normalised moment $\vb{m}$, orientation vector $\vb{u}$ and driving field $\vb{\hat{B}}^\text{ext}$ \textbf{d)} Energy decomposed into Zeeman- and anisotropy energy.}
    \label{fig:1MNP_hysteresis}
\end{figure*}

\subsection{Conservation law tests \label{subsec:Conservationa_law_tests}}

In \cref{fig:conservation_law_tests} we compare the two presented methods for computing conserved quantities at different timesteps. Note that the test simulations include moment relaxation, an alternating external field, dipole interactions, MNP collisions and thermal fluctuations in all degrees of freedom, all with experimentally relevant parameter values.

In \cref{fig:conservation_law_tests}a we see excellent agreement in the energy integration for short times. At longer times, variations in the red and blue curve are still nearly identical over short intervals, but the red is shifted down. This indicates a persistent, erroneous drift in method 2.
In \cref{fig:conservation_law_tests}c we see that regardless of timestep, the error increases linearly in time at long time, like thermal drift. But unlike thermal drift the error decreases with decreasing timestep, so as $\Delta t \xrightarrow{} 0$ the methods converge. Furthermore, using our modified Heuns' method for the moment rotations, the drift is reduced by several orders of magnitude. We conclude that the 2 integration methods are formally identical and the drift is a purely numerical error. This self-consistency check in turn validates our analytical energy transfer analysis (cf. \cref{eq:E_dot}).

In \cref{fig:conservation_law_tests}b we see the time-variation in $S_z$ which is proportional to the $z$-component of the systems net moment. Interestingly, while the external driving field is sinusoidal, the moment switches in a step-like manner. In some cases the switching happens in several smaller steps, as thermal fluctuations and dipole-interactions make some of the 10 MNPs flip before others.

We also see excellent agreement between method 1 and 2 throughout, i.e.\ midpoint integrating $\vb{\dot{S}}$ has eliminated thermal drift. Thus we can meaningfully consider the time-averaged deviation of the two methods for momenta. In \cref{fig:conservation_law_tests}d it is seen that the methods converge with decreasing timestep as expected and at $\Delta t = 0.1 \: \mathrm{ps}$ the relative deviation is less than $1\:\%$. The same qualitative behaviour is found for the $x$ and $y$ components. This validates the analysis in \cref{sec:conservation_laws}, so we conclude that all momenta and momentum transfer in the model is accounted for.

\subsubsection{The computational bottleneck}

From \cref{fig:conservation_law_tests}d we see that the moment dynamics are more timestep sensitive than the mechanical motion. Similarly at $T=0$ noticeable error in method 2 occurs only during the brief periods of transient moment dynamics (cf. \cref{fig:1MNP_hysteresis}), but at $T>0$ the moments are never quite in equilibrium hence the continuous error. It is unclear why the result is drift rather than random noise (see \cref{fig:conservation_law_tests}c).
With our modified Heuns' method just on moment dynamics, the drift remains, but is reduced by approximately a factor of 100.

The reason the moment dynamics are critical is that moment precession at the Larmor frequency $\gamma B^\text{eff}$ is by far the fastest dynamical behaviour in the system (cf. \cref{subsec:characteristic_frequencies}). This leads to rapid variation of $P^\text{mag}$, making the integration of power losses particularly timestep sensitive (see \cref{fig:1MNP_hysteresis}c). The issue is exacerbated when the magnetic field $\vb{B}$ and anisotropy axis are non-collinear, as both the anisotropy field magnitude $B^\text{ani} \sim \vb{m} \vdot \vb{u}$ and the moment direction then oscillate at the Larmor frequency, but not in a self-consistent manner.

For zero temperature integration or when studying the instantaneous power, one can beneficially use a reduced timestep during transient moment dynamics. At finite temperature, one can compute the total energy change directly from \cref{eq:E_multi_MNP}, integrate hysteresis power and mechanical losses, then use energy conservation to estimate the integrated magnetic losses.
This works at the system level or for non-interacting particles. 

To integrate magnetic losses directly at finite temperature and reasonable computational expense, an advanced timestep method is necessary for the moment rotation. We recommend comparing energy estimates from method 1 and 2 like in \cref{fig:conservation_law_tests}a,c for method validation.

\subsection{Single-particle hysteresis \label{subsec:Single-particle_hysteresis}}

\Cref{fig:1MNP_hysteresis} shows a single MNP under a sinusoidal driving field. In the first 100 ns we have the transient dynamics of the moment relaxing towards the anisotropy axis, as $\vb{m}, \vb{u}$ start with a $90^\circ$ relative angle. This results in major magnetic losses (green curve) rapid variation in the energy transfer from the external field (purple curve) and a smaller quantity of viscous losses. In general magnetic losses are dominant in this parameter regime. Rapid variation is also seen in Zeeman and anisotropy energy (blue and lightblue curves respectively).

The green spikes in \cref{fig:1MNP_hysteresis} signify moment reversals. Before each reversal, viscous losses increase as the MNP mechanically rotates towards $\vb{B}^\text{ext}$ (tails in red curve), but as the field reaches some critical value, the moment flips to the other anisotropy minimum instead. This flip is followed by about 5 ns of precessional motion and associated Gilbert damping, as seen in \cref{fig:1MNP_hysteresis}b. Interestingly the precession frequency (oscillations of green curve) increases over time as $\vb{B}^\text{eff}$ changes. The maximal frequency is about $1.3\:\mathrm{GHz}$ which is consistent with the Larmor frequency $\gamma B^\text{ext}/(2\pi) = 1.4\:\mathrm{GHz}$. The reversal events are also marked by spikes in $E^\text{ani}$, sudden drops in $E^\text{Zee}$ and seen directly on the $m_z$ curve. 

The integrated area of $P^\text{hyst}=-\mu m_z \dot{B}^\text{ext}_z$ is positive in every half-cycle, so there is a net transfer of energy from driving field to MNP. This asymmetry is only possible because the MNP moves in response to the field. In particular when $m_z$ is positive, $B^\text{ext}$ is mostly decreasing and for $m_z$ negative, $B^\text{ext}$ is mostly increasing. Without changes in $\vb{u}$ or $\vb{m}$, the $P^\text{hyst}$ and $E^\text{Zee}$ curves would both follow the $B^\text{ext}$ curve and average to 0. Indeed $P^\text{hyst}, E^\text{Zee}$ vary sinusoidally between reversal events like $B^\text{ext}$.

Over time, the purple curve is shifted upwards, which means the net energy transfer increases, both input and subsequent dissipation. This is accompanied by an increase in the scale of $E^\text{Zee}$. The reason is the gradual relaxation of the anisotropy axis towards vertical ($u_z \xrightarrow{} -1$). The corresponding mechanical rotation also explains the viscous losses and $E^\text{ani}$-changes between reversal events, which are visible in the first $1.5 \: \mathrm{\mu s}$. Finally, we note that reversal events happen twice every field cycle, so all the power and energy contributions have double the frequency of the driving field.

To conclude, while the system energy builds gradually over each half-cycle of the driving field, it is dissipated in the span of a few ns by Gilbert damping. This means the internal temperature of the MNP will increase in sudden jumps, after which the energy is gradually conducted to the surrounding fluid.

This analysis is specific to the chosen parameters, in fact it has been shown in previous numerical studies that viscous, rotational losses dominate in some regimes \cite{usov_dynamics_2012,helbig_self-consistent_2023}. Also, the number of moment oscillations during magnetic relaxation, and hence the timescale of a reversal event, is proportional to $\alpha$, which varies greatly between materials.

\subsection{Zero-field aggregation \label{subsec:Zero-field_aggregation}}

\begin{figure}[t]
    \centering
    \includegraphics[width=0.45\textwidth]{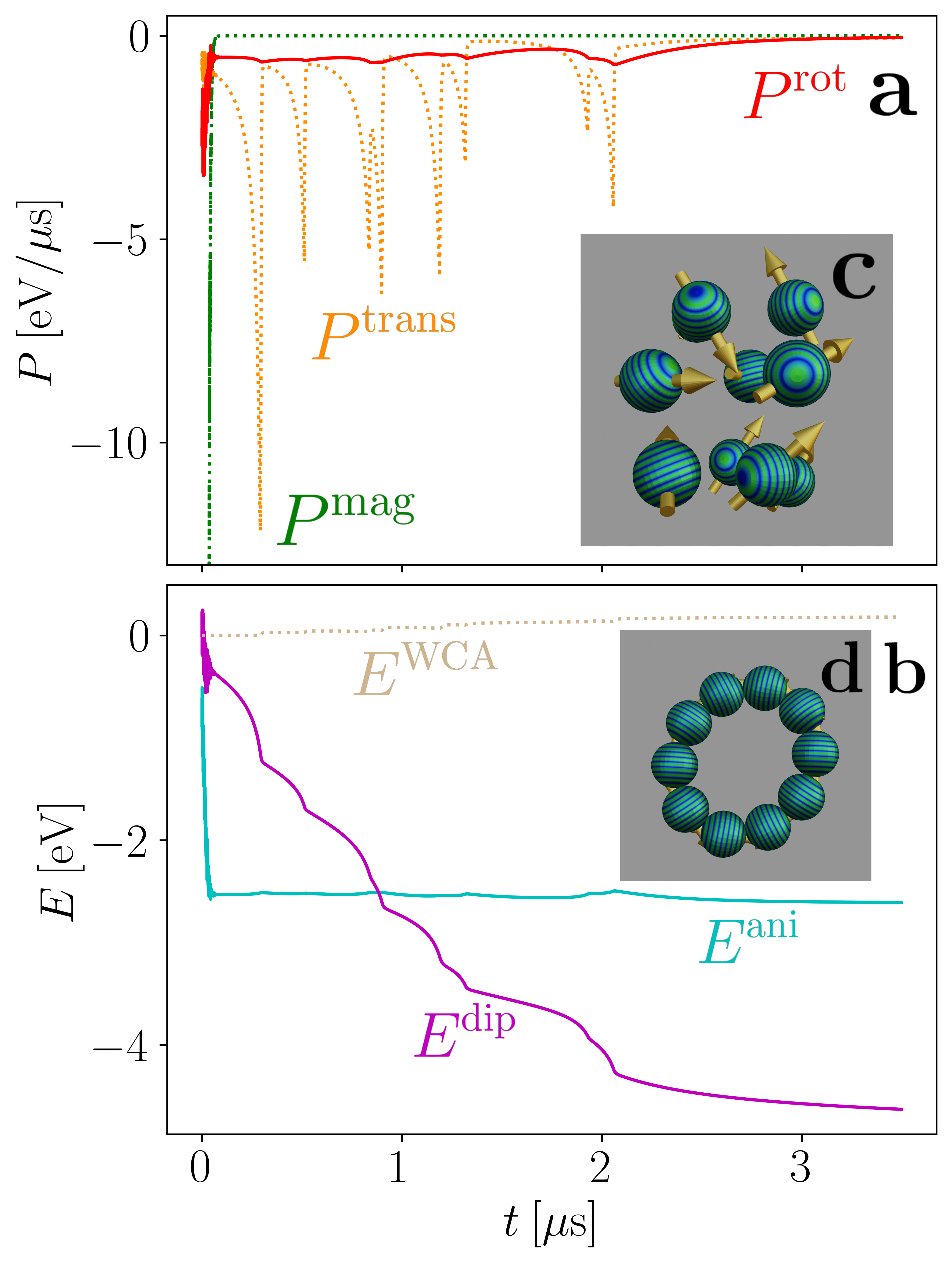}
    \caption{10 MNP collision. Same parameters as in \cref{fig:conservation_law_tests}, except $T,B^\text{ext}=0$. \textbf{a)} dissipated power decomposed into Gilbert damping, rotational- and translational viscous losses. \textbf{b)} System energy decomposed into anisotropy-, WCA-, and dipole interaction energy.
    \textbf{c)} Initial state. Randomly generated at a volume fraction of 0.1. \textbf{d)} Final configuration after $15\:\mathrm{\mu s}$. Only the first $3.5\:\mathrm{\mu s}$ are shown in \textbf{a,b)}, because aside from a collision around $t=10\:\mathrm{\mu s}$ all the curves are nearly flat after $t=3.0\mathrm{\mu s}$. The plots are identical for inertial- and overdamped simulations.}
    \label{fig:10MNP_collision}
\end{figure}

In \cref{fig:10MNP_collision} we consider 10 randomly initiated MNPs at zero-temperature and no external field. Consequently no energy is added to the system, so this is a pure relaxation. The particles are initiated randomly as seen in \cref{fig:10MNP_collision}c, form into a single chain within the first 3 $\mathrm{\mu s}$, then after about $10\:\mathrm{\mu s}$ the two ends snap together, forming the ring seen in \cref{fig:10MNP_collision}d.

The initial spike in $P^\text{mag}$ and rapid decrease in $E^\text{ani}$ correspond to the initial relaxation of all moments towards their respective anisotropy axes. After this relaxation there are virtually no Gilbert losses and $E^\text{ani}$ is nearly constant, i.e.\ even though the LLG equation is used throughout, the moment rotation remains locked to the mechanical rotation. This would have justified using the rigid dipole approximation. Especially since the random, initial moment configuration is implausible in an experiment.  

For the first 3 $\mathrm{\mu s}$ there are steady rotational losses, while the translational losses are marked by spikes. The reason is that whenever two or more MNPs collide, the translational loss rate increases as the MNPs accelerate, then drops sharply while they settle in the combined minimum of the dipole- and WCA potentials. Indeed $E^\text{WCA}$ is seen to increase as more and more MNPs reach surface contact. The dissipated energy largely comes from the dipolar interactions, as seen by the consistent decrease of $E^\text{dip}$. The curves all remain nearly flat after $3\:\mathrm{\mu s}$, except around $t = 10\:\mathrm{\mu s}$ when the MNP-chain snaps into a ring causing one more spike in $P^\text{trans}$.
This illustrates how the relaxation of MNP systems occurs across many timescales.

This example simulation is in stark contrast to \cref{fig:1MNP_hysteresis}, as the losses from a few ns and onwards are exclusively viscous. Thus for a more realistic initial state with relaxed moments, instead of heating the MNPs directly, the energy goes to exciting the surrounding fluid. In reality thermal fluctuations will transfer kinetic energy from the fluid back into the particles, so the whole system reaches a uniform temperature eventually, but at no point will the MNP cores be hotter than the surrounding fluid.

\subsection{Hysteresis during collision \label{subsec:Hysteresis_during_collision}}

\begin{figure*}[t]
    \centering
    \begin{minipage}{0.47\textwidth}
        \includegraphics[width=\textwidth]{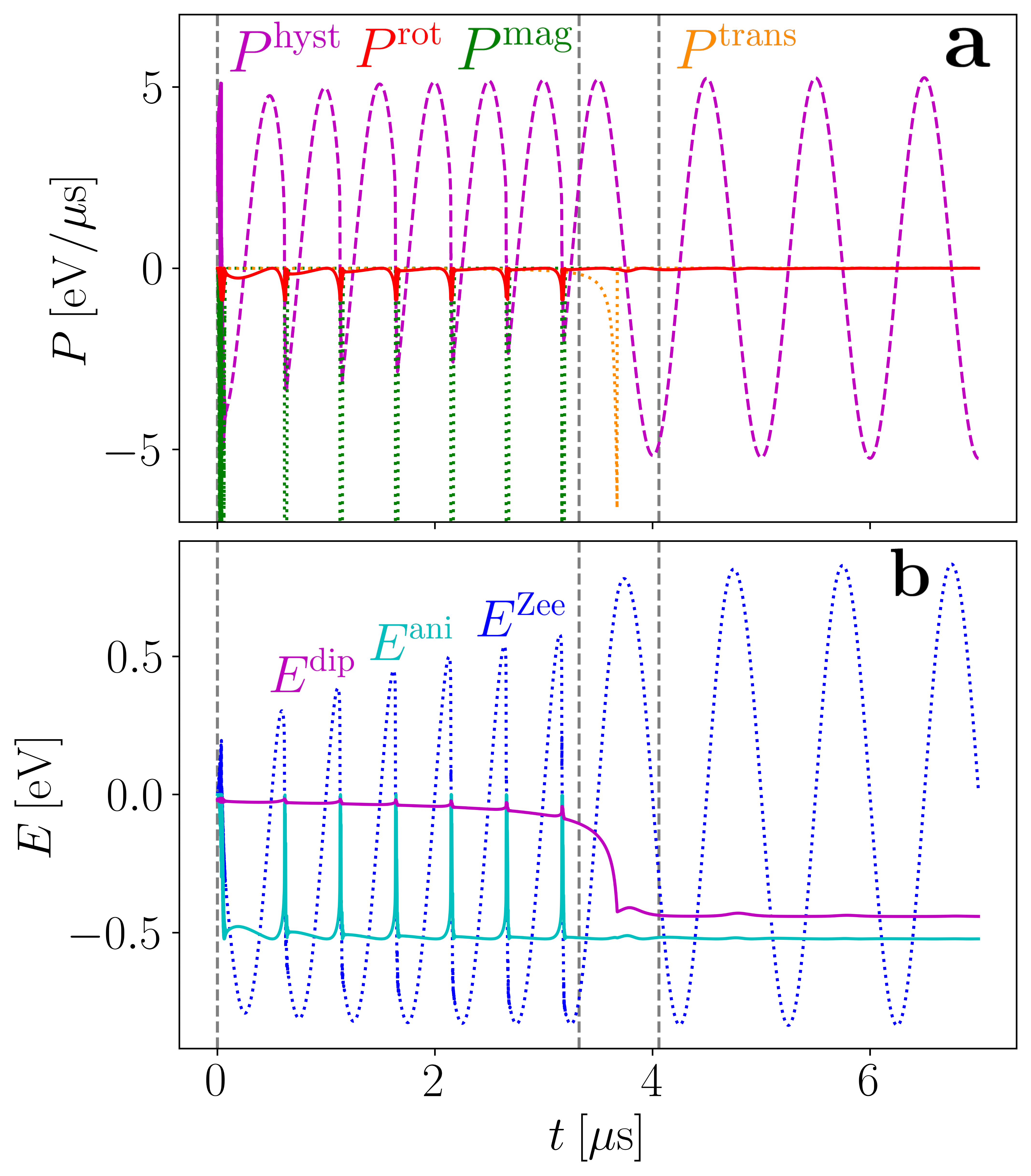}
    \end{minipage}
    \begin{minipage}{0.45\textwidth}
        \includegraphics[width=0.8\textwidth]{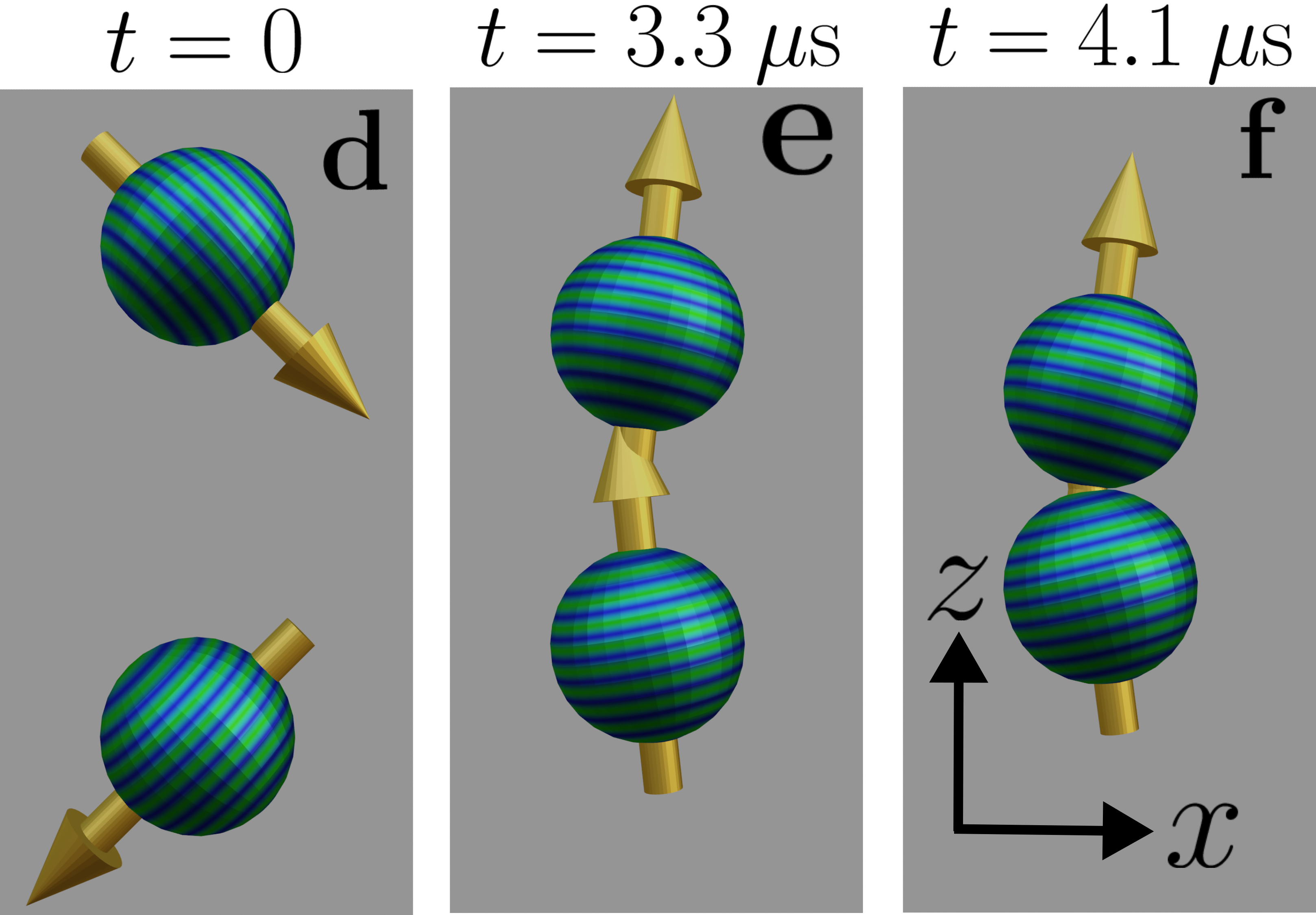}
        \includegraphics[width=0.95\textwidth]{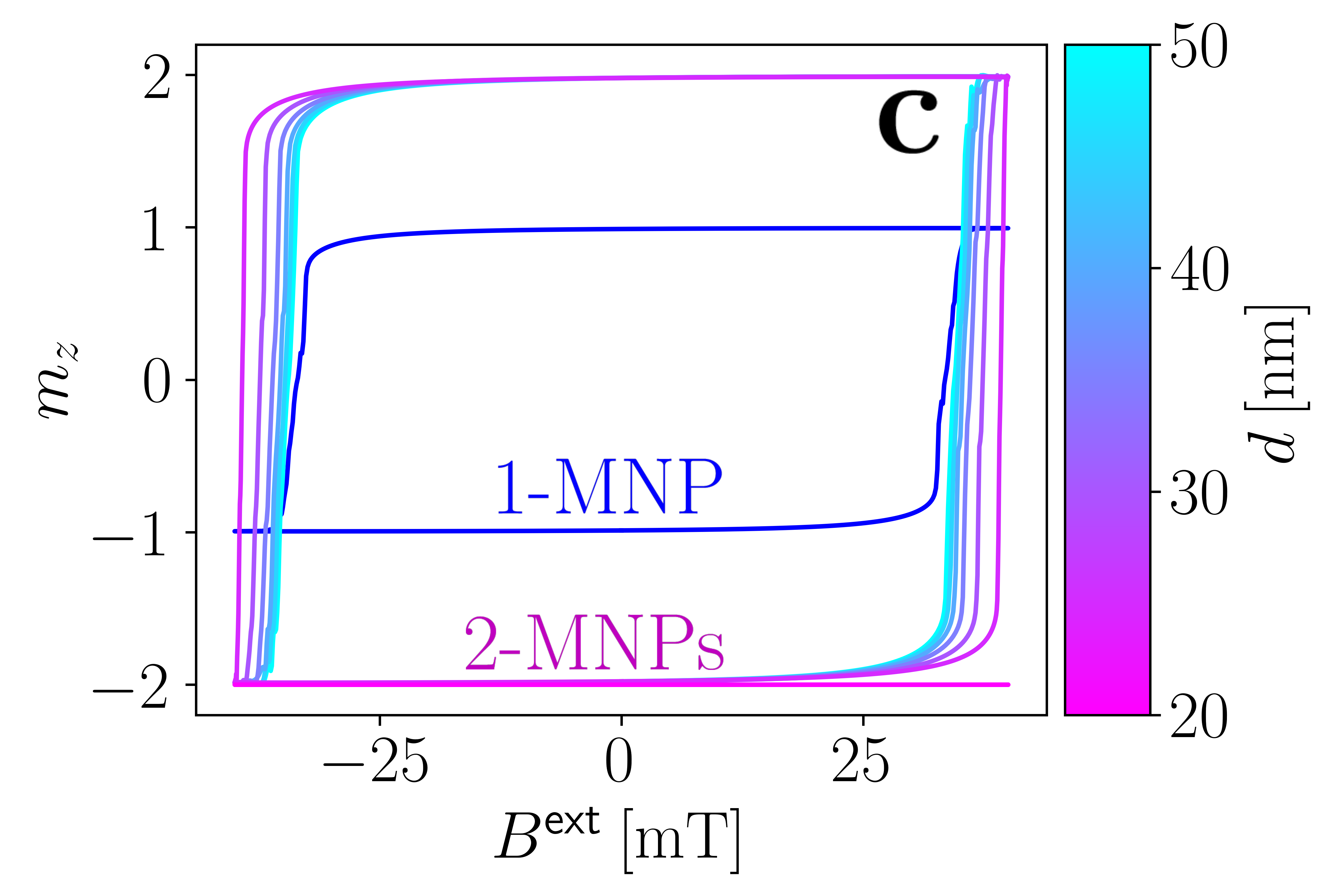}
    \end{minipage}
    \caption{2MNPs colliding in an alternating field. Same parameters as in \cref{fig:1MNP_hysteresis}. The initial state is seen in \textbf{d)} with coordinate axes in \textbf{f)}. The plots are identical for inertial- and overdamped simulations. \textbf{a-b)} evolution of power and energy contributions in time. Vertical, gray lines correspond to the configuration pictures in \textbf{d-f)}. The anisotropy axes are perpendicular to the blue circles, the arrows represent magnetic moments and the sizes and distances are to scale. \textbf{c)} Hysteresis curves simulated with the MNPs free to rotate, but fixed at a center-to-center distance of $d$. $m_z$ is the z-component of system moment relative to the moment per particle. A single-particle hysteresis curve is shown for comparison.}
\label{fig:2MNP_hysteresis_collision}
\end{figure*}

As a final use case, in \cref{fig:2MNP_hysteresis_collision} we consider 2 MNPs colliding under a 1 MHz, sinusoidal driving field. We have $B^\text{ani} = 50\:\mathrm{mT}$ and  found by repeated simulations that for $B^\text{ext}$ between $25$ and $50$ mT, a sudden change in behaviour occurs as the MNPs get nearer. In \cref{fig:2MNP_hysteresis_collision} we chose $B^\text{ext} = 40\:\mathrm{mT}$ to illustrate this.

The first part of the curve is similar to \cref{fig:1MNP_hysteresis} even though there are now 2 interacting MNPs. This is because at the initial distance of $d=50\:\mathrm{nm}$ the field from one MNP on the other is at most $2\mu_0 \mu/(4\pi d^3)=2.7\:\mathrm{mT}$ (cf. \cref{eq:B_dip}), i.e.\ their interaction is much weaker than the driving field, so they respond nearly as if isolated. From \cref{fig:2MNP_hysteresis_collision}b we note spikes in $E^\text{dip}$ during each moment reversal, which are small compared to the $E^\text{ani}$ spikes. The reason is that the MNPs flip simultaneously, so their moments remain nearly aligned, thus limiting the variation in $E^\text{dip}$. The $E^\text{dip}$ spikes grow with decreasing distance, indicating increased losses from the dipolar coupling.

In \cref{fig:2MNP_hysteresis_collision}c, we confirm that the hysteresis losses grow as the MNPs approach each other, by computing hysteresis curves with positions fixed. The height is locked by the total moment, but the width increases with decreasing $d$. That is the dipole interaction increases the coercive field, making the 2-MNP system a harder magnet. Indeed when the coercive field exceeds $B^\text{ext}$, the loop closes entirely and the MNP behaviour qualitatively changes.

In the second part of \cref{fig:2MNP_hysteresis_collision}a,b moment reversal ceases as the MNPs collide.
The collision is signified by a drop in $E^\text{dip}$, due to the rapidly diminishing distance (cf. \cref{fig:2MNP_hysteresis_collision}e-f), and a spike in translational dissipation, due to the velocity changes around impact. This same decrease in distance means stronger dipole coupling, which is why moment reversal ceases. 
Because both velocity and dipole field strength scale non-linearly with distance, most of the coercivity change occurs in the last fraction of the collision. Therefore the moment locking nearly coincides with the impact for a broad parameter range.
After some final rotational relaxation, the MNPs cease all motion. $P^\text{hyst}$ and $E^\text{Zee}$ still oscillate in tune with the driving field, but they are symmetric around $0$, so there is no net energy transfer, nor any dissipation.

In summary, the dipole coupling increases the coercive field as the MNPs draw nearer, resulting in a sudden dynamical phase transition which ends all magnetic losses. The MNPs then collide and stop moving entirely. We note that while some of this was elucidated by simulating multiple hysteresis curves, the entire story can be inferred from \cref{fig:2MNP_hysteresis_collision}a alone, which is a single simulation that also shows the relative importance of magnetic and mechanical losses, and the point of impact. Also, as demonstrated in \cref{fig:10MNP_collision}, the presented plotting techniques scale well to many-particle systems.

\section{Model generalisations
\label{sec:Model_generalisations}}

In the rest of the present paper, we restricted all analysis to the model we implemented and tested numerically. However both the formal procedure for deriving our model and the formal results on transfer rates for conserved quantities readily generalise to a broader setting.

\subsection{Spherical particles}

For single-domain MNPs of multiple sizes and materials, the only difference is a subscript on $\alpha_i, \gamma'_i, I_i, \mathfrak{m}_i, \zeta^\text{t(r)}_i$ and $\mu_i$ to signify that these parameters vary between particles. A coating layer can be included by using the hydrodynamic radius for the damping coefficients in \cref{eq:zeta_sphere}, and for surface forces like \cref{eq:F_WCA}, but core radius elsewhere as in Ref. \cite{durhuus_simulated_2021}.

For nanomagnets with non-uniform magnetisation, the notion of a single particle moment is at best an approximation, so a rigorous treatment would require considerable further analysis.

Depending on the crystal symmetries of the magnetic core, a large number of higher order magnetocrystalline anisotropy terms are possible, enabling any number of easy- or hard axes\cite{chikazumi1997physics,cullity2011introduction,bedanta_supermagnetism_2015}. For uniformly magnetised objects, shape anisotropy only modifies the $\mathrm{K}$ tensor from \cref{subsec:Uniaxial_anisotropy}, yielding uni- or triaxial anisotropy\cite{keshtgar_magnetomechanical_2017}. These additional anisotropies only change the functional form of $\vb{B}^\text{ani}$, not how it enters the remaining equations.

For a non-uniform external field a magnetic gradient force enters in \cref{eq:r_ddot} and the subsequent momentum analyses. We invoke that any quasistatic magnetic field averaged over a sphere is equal to the field at the spheres center\cite[problem 5.59]{griffiths_EM}. It follows that the Zeeman energy takes the form $E^\text{Zee}_i = -\bs{\mu}_i \vdot \vb{B}^\text{ext}_i$ for uniformly magnetised spheres where $\vb{B}^\text{ext}_i$ is the external field evaluated at the center of the $i$'th MNP. The gradient force does not change \cref{appsec:E_dot}, so the power expressions \cref{eq:p_dot,eq:J_dot,eq:E_dot} apply for non-uniform external fields.

\subsection{Non-spherical particles}
For non-spherical particles in liquid suspension, viscous drag becomes a tensor quantity. In general, at low Reynolds number in a stationary fluid, the viscous force and torque on an isolated particle are of the form \cite{kraft_brownian_2013,wittkowski_self-propelled_2012} 
\begin{align}
\mqty(\vb{F}^\text{visc}_i \\ \bs{\tau}^\text{visc}_i) = -\mathrm{Z} \mqty(\vb{v}_i \\ \bs{\omega}_i) \qq{,} \mathrm{Z} = \mqty(\mathrm{Z}^\text{tt} & \mathrm{Z}^\text{tr} \\ \mathrm{Z}^\text{rt} & \mathrm{Z}^\text{rr}),    \label{eq:viscous_drag_non_spherical}
\end{align}
where $\mathrm{Z}$ is a symmetric 6-by-6 matrix and $\mathrm{Z}^\text{t(r),t(r)}$ are 3-by-3 symmetric matrices. One can then straightforwardly generalise the derivation in \cref{appsec:E_dot} (see Supplementary Material). The result is
\begin{align}
\dot{E} = \sum_i (P^\text{hyst}_i + P^\text{mag}_i + P^\text{visc}_i) \label{eq:E_dot_generalised}
\end{align}
where $P^\text{hyst}_i,P^\text{mag}_i$ are still given by \cref{eq:P_in,eq:P_mag} while, using $\vb{f}_i^\text{th} = (\vb{F}_i^\text{th}, \bs{\tau}_i^\text{th})$ and $\vb{V}_i = (\vb{v}_i, \bs{\omega}_i)$,
\begin{align*}
    P_i^\text{visc} = -\vb{V}_i \vdot \mathrm{Z} \vb{V}_i + \vb{f}_i^\text{th} \vdot \vb{V}_i.
\end{align*}
$P_i^\text{visc}$ is consistent with Refs. \cite{hill_extremum_1956} and \cite[sec.\ 2.2]{kim_microhydrodynamics_1991}. Similarly, the transfer of linear and angular momentum generalise to (see Supplementary Material)
\begin{align}
    \vb{\dot{p}} &= \sum_i \left(\vb{F}^\text{visc}_i + \vb{F}^\text{th}_i\right)
    \label{eq:p_dot_generalised}\\
    \vb{\dot{J}} &= \sum_i \bigg(\bs{\mu}_i \cross\vb{B}^\text{ext} + \bs{\tau}^\text{visc}_i + \bs{\tau}^\text{th}_i 
    \notag\\
    &\hspace{1cm} + \vb{r}_i \cross [\vb{F}_i^\text{visc} + \vb{F}^\text{th}_i]\bigg) \label{eq:J_dot_generalised}
\end{align}

Note that because of the coupling between the drag on different components of $\vb{V}$, there are also statistical correlations between components of the thermal fluctuations\cite{ermak_brownian_1978}. Therefore, instead of \cref{eq:F_th,eq:tau_th} we have
\begin{align}
    \expval{f_\alpha^\text{th}(t)f_\beta^\text{th}(t')} = 2k_BT Z_{\alpha \beta}\delta(t - t')   \label{eq:f_th}
\end{align}

Another effect of non-spherical particles is that magnetic fields from the MNPs and the interaction energy $E_{ij}^\text{int}$ are close to dipolar only at long range\cite{beleggia_magnetostatic_2004}. However \cref{appsec:E_dot,appsec:conservation_laws_in_magnetostatics} never use the dipole approximation, so the derivations are still valid. That is, \cref{appeq:J_dot} becomes an integral equation for non-spherical magnets\cite{beleggia_magnetostatic_2004}, but the transfer equations \cref{eq:J_dot_generalised,eq:p_dot_generalised,eq:E_dot_generalised} are unchanged.

Finally, we consider the greatest simplifying assumption in the present work, namely the neglect of hydrodynamic interactions between MNPs. 
With multiple particles in liquid suspension, the velocity distribution of the fluid depends on the position and motion of every suspended particle. This leads to a long-range, many-body interaction \cite{cichocki_friction_1994,kim_microhydrodynamics_1991}, which completely changes the equations of motion. One can make the problem computationally tractable by the methods and simplifications of Stokesian dynamics\cite{durlofsky_dynamic_1987,banchio_accelerated_2003,brady_stokesian_1988}, which have been applied to MNP suspensions both with\cite{satoh_brownian_1999} and without \cite{satoh_stokesian_1998,sand_stokesian_2016} thermal fluctuations. Alternatively, one can use various approximations to the 2-sphere problem to derive pair-interactions. We refer to Ref. \cite{goddard_singular_2020} for a historical overview, Ref. \cite{russel_colloidal_1989} for lubrication theory (lowest order, short-range model) and Refs. \cite{jeffrey_calculation_1984,jeffrey_calculation_1992,townsend_generating_2019} for a more complete analysis. Regardless of the level of approximation, the mathematical form of viscous drag is altered, so \cref{eq:E_dot_generalised,eq:p_dot_generalised,eq:J_dot_generalised} are not directly applicable.

In summary; the transfer equations \cref{eq:J_dot_generalised,eq:p_dot_generalised,eq:E_dot_generalised} hold for a collection of arbitrarily shaped but uniformly magnetised nanoparticles, when neglecting electrodynamic- and hydrodynamic interactions and assuming that the driving field $\vb{B}^\text{ext}$ is uniform across a particle. 

\subsection{Additional interactions}

Besides magnetic interactions, MNPs in liquid suspension are subject to various surface-to-surface interactions, depending on the details of surface coating and fluid medium\cite{min2010role,bishop_nanoscale_2009,israelachvili_intermolecular_2011,morup_magnetic_2010}. When modelling MNP aggregation, a bare minimum is a model of steric repulsion to prevent MNP overlap, but simulations have also included ligand bonding\cite{rozhkov_self-assembly_2018,novikau_influence_2020}, as well as electrostatic-\cite{chuan_lim_agglomeration_2012} and van der Waals forces\cite{durhuus_simulated_2021} among others. All 3 can be modelled by interaction potentials, which depend only on the relative positions and orientations of MNP pairs. 
Such contributions are guaranteed to conserve the energy and mechanical momentum of the whole MNP collection.

In general, interaction potentials can be appended to the system energy \cref{eq:E_multi_MNP}. Then one may repeat the procedure in \cref{sec:Model} to derive forces and torques, and the analysis in \cref{sec:conservation_laws} for non-conservative interactions to modify the equations for transfer of energy and momenta.

\section{Conclusion}

We derived a general model of the energy, linear- and angular momentum for a collection of interacting magnetic nanoparticles in an external magnetic field. Starting from these conserved quantities, we used a formal, easily generalisable procedure for deriving the equations of motion in liquid suspension, at finite temperature. The result is a set of Langevin equations which couple the LLG equation for moment rotation with the mechanical translation and rotation of the particles.

Within this model we derived expressions for the transfer of energy and momenta between system and environment, which we validated numerically by timestep integration. This both demonstrates self-consistency of the model, and that we can numerically analyse the transfer rates at finite temperature, despite the technical difficulties of integrating stochastic processes.

We note that a number of established models and analyses can be derived as special cases or approximations to our results. For example the RDA, the hysteresis curve area as a measure of energy dissipation or the recent analyses by Leliaert et. al. \cite{leliaert_individual_2021} and Helbig et. al.\cite{helbig_self-consistent_2023} on energy transfer. 

In addition to model development and verification, tracking the transfer of conserved quantities gives new insight on the physical system. Using our formulas, one can calculate the instantaneous power at each MNP and decompose into different transfer channels. In particular input power from the driving field, magnetic losses from Gilbert damping and viscous losses to the fluid; both translational and rotational. Whether the losses are magnetic or viscous determines whether heating occurs inside the particles or in the surrounding fluid. By simulating zero-temperature hysteresis and aggregation in 1-, 2- and 10-particle systems we demonstrate how tracking power and energy contributions gives new ways to visualise and interpret MNP dynamics, which are applicable to many-particle simulations.

Our results give a novel perspective on known steady-state phenomena, useful tools for analysing MNPs as mechanical actuators and hyperthermia agents, and may facilitate new studies on the transient and driven dynamics of nanomagnets.

\appendix

\section{Rewriting the LLG equation \label{appsec:rewriting_LLG_equation}}

\Cref{eq:mu_dot_mid} defines $\vb{\dot{m}}$ recursively. Taking the recursion one step further, it may be rewritten
\begin{align*}
    \vb{\dot{m}}_i = -\gamma \vb{m}_i \cross \vb{B}^\text{eff}_i + \alpha_i \vb{m}_i \cross [-\gamma \vb{m}_i \cross \vb{B}^\text{eff}_i + \alpha_i \vb{m}_i \cross \vb{\dot{m}}_i]
\end{align*}
We recall that $\vb{m}$ has constant magnitude; $m^2 = 1$. Differentiating yields $\vb{m} \vdot \vb{\dot{m}} = 0$, i.e.\ $\vb{\dot{m}}$ is perpendicular to $\vb{m}$. It follows that $\vb{m} \cross [\vb{m} \cross \vb{\dot{m}}] = - m^2\vb{\dot{m}} = -\vb{\dot{m}}$. Thus we can isolate $\vb{\dot{m}}$, which yields \cref{eq:mu_dot}.

\section{Mechanical equations of motion \label{appsec:mechanical_equations_of_motion}}

For translation, simply add the damping and thermal force from \cref{sec:dissipation_and_fluctuations} to the rhs.\ of \cref{eq:r_ddot_no_damping} to get \cref{eq:r_ddot}. For rotation it follows from \cref{eq:Gilbert_torque_1} that
\begin{align*}
    \vb{\dot{L}}_i = \eval{\vb{\dot{L}}_i}_{\alpha=0} + \eval{\vb{\dot{S}}_i}_{\alpha = 0} - \vb{\dot{S}}_i
\end{align*}
Writing out the viscous and damping torque, then using the angular momentum definitions, \cref{eq:J_i},
\begin{align*}
    I \bs{\dot{\omega}}_i = \eval{I\bs{\dot{\omega}}_i}_{\zeta^\text{r}=\alpha=0} + \frac{\mu}{\gamma} \left(\vb{\dot{m}}_i - \eval{\vb{\dot{m}}_i}_{\alpha = 0}\right) - \zeta^\text{r} \bs{\omega}_i + \bs{\tau}^\text{th}_i
\end{align*}
Inserting the equations without damping, \cref{eq:omega_dot_no_damping,eq:mu_dot_no_damping}, yields \cref{eq:omega_dot}.

\section{Conservation laws in magnetostatics \label{appsec:conservation_laws_in_magnetostatics}}

In the framework of electrodynamics, linear- and angular momentum conservation can be shown formally for any finite distribution of charges and currents\cite{jackson_classical_1999}.
Here, we use a simpler version of the same argument for a finite current distribution in magnetostatics.

Maxwells laws take the form
\begin{align*}
    \div{\vb{B}} = 0 \qq{,} \curl{\vb{B}} = \mu_0 \vb{J}_e \qq{,} \vb{E} = 0
\end{align*}
where $\vb{J}_e$ is electrical current density.
From the Lorentz force law, the force density is
\begin{align*}
    \vb{f} = \vb{J}_e \cross \vb{B} = \frac{1}{\mu_0}(\curl{\vb{B}}) \cross \vb{B} = \frac{1}{\mu_0}\left[(\vb{B} \vdot \grad)\vb{B} - \frac{1}{2}\grad B^2\right]
\end{align*}
In component form, with $\partial_{\alpha} = \pdv{r_{\alpha}}$, this may be written
\begin{align}
    f_{\alpha} = \sum_\beta \partial_\beta T_{\beta\alpha}, \quad T_{\beta\alpha}
    = \frac{1}{\mu_0}\left[B_\beta B_\alpha - \frac{1}{2}\delta_{\beta\alpha} B^2 \right] \label{eq:force_density_component}
\end{align}
The tensor $\mathrm{T}$ is Maxwells stress-tensor in the absence of $\vb{E}$-fields. The momentum density in the EM-fields is $\epsilon_0 \vb{E} \cross \vb{B} = 0$, hence the rate of change of momentum is given by $\dv{t}p_\alpha = \int f_i\alpha \dd\vb{r}$. Integrating \cref{eq:force_density_component} over all space and using Gauss theorem thus proves conservation of linear momentum.

Similarly, the components of the torque density are
\begin{align*}
    \tau_\alpha = (\vb{r} \cross \vb{f})_\alpha = \sum_\beta \partial_\beta M_{\beta\alpha} \qq{,} M_{\beta\alpha} = \sum_{\gamma\delta} \epsilon_{\gamma\delta\alpha} x_\gamma T_{\beta\delta},
\end{align*}
where $\epsilon_{\alpha\beta\gamma}$ is the Levi-Civita symbol. The change in angular momentum is given by $\dv{t}J_\alpha = \int \tau_\alpha \dd\vb{r}$, which also gives zero by Gauss theorem. 

In conclusion; in magnetostatics, linear- and angular momentum are conserved without any field momentum. 

Since a magnetisation distribution is equivalent to a collection of bound currents\cite{griffiths_EM}, the proof also applies to MNPs and other permanent magnets. In particular, for a collection of dipole magnets at zero temperature in vacuum it follows that
\begin{align}
    \dv{t} \vb{J}^\text{vacuum}_{T=0} = \sum_i \left(\vb{r}_i \cross \vb{F}_i^\text{dip} + \vb{m}_i \cross \vb{B}^\text{dip}_i\right) = 0 \label{appeq:J_dot}
\end{align}
This can also be shown by inserting \cref{eq:B_dip,eq:F_dip} for $\vb{B}^\text{dip},\vb{F}^\text{dip}$, however the present argument can be used to derive analogous identities for arbitrarily shaped magnets with higher order multipole interactions.

\section{Time derivative of system energy \label{appsec:E_dot}}

The system energy, $E$, is a function of the independent variables $\vb{m}_i, \vb{u}_i, \bs{\omega}_i, \vb{r}_i, \vb{v}_i, t$, so the time derivative is
\begin{align}
    \dv{t}E = \sum_i\bigg[\pdv{E}{\vb{m}_i} \vdot \vb{\dot{m}}_i &+ \pdv{E}{\vb{u}_i} \vdot \vb{\dot{u}}_i + \pdv{E}{\bs{\omega}_i} \vdot \bs{\dot{\omega}}_i 
    \notag\\
    +\pdv{E}{\vb{r}_i} \vdot \vb{v}_i &+ \pdv{E}{\vb{v}_i} \vdot \vb{\dot{v}}_i\bigg] + \pdv{E}{t} \label{appeq:dE_dt}
\end{align}
Here we assume the Stratonovich interpretation, which means the ordinary rules of calculus apply despite the stochastic terms\cite{gardiner_handbook_2002,sarkka_applied_2019}. In \cref{subsec:Conservationa_law_tests}, we verify numerically that the Stratonovich interpretation is correct in the overdamped case.

We consider a single term and suppress the $i$-subscript to avoid notational clutter.
Let $\vb{\dot{m}}_0,\bs{\dot{\omega}}_0, \vb{\dot{p}}_0 = \mathfrak{m} \vb{\dot{v}}_0$ denote the derivatives in absence of damping (\cref{eq:mu_dot_no_damping,eq:omega_dot_no_damping,eq:r_ddot_no_damping}). The derivatives with damping are given in \cref{eq:mu_dot_mid,eq:omega_dot,eq:r_ddot} and $\vb{\dot{u}} = \bs{\omega} \cross \vb{u}$ applies in both cases. For reference we recall the general vector identity 
\begin{align*}
\vb{a} \vdot (\vb{b} \cross \vb{c}) = \vb{b} \vdot (\vb{c} \cross \vb{a}) = \vb{c} \vdot (\vb{a} \cross \vb{b})
\end{align*}

The only explicit time-dependence is in the external field, so
\begin{align}
    \pdv{E}{t} = -\mu \vb{m} \vdot \vb{\dot{B}}^\text{ext} \label{appeq:pdv_E_t}
\end{align}
By definition $\partial_{\vb{r}} E = -\vb{\dot{p}}_0$, hence
\begin{align*}
    \pdv{E}{\vb{r}} \vdot \vb{v} &= -\vb{\dot{p}}_0 \vdot \vb{v} = -(\mathfrak{m}\vb{\dot{v}} + \zeta^\text{t}\vb{v} - \vb{F}^\text{th}) \vdot \vb{v} 
    \\
    &= - \vb{p} \vdot \vb{\dot{v}} -\zeta^\text{t}v^2 + \vb{F}^\text{th} \vdot \vb{v}
\end{align*}
Using \cref{eq:E_multi_MNP} $\partial_{\vb{v}}E = \vb{p}$, thus
\begin{align}
    \pdv{E}{\vb{r}} \vdot \vb{v} + \pdv{E}{\vb{v}} \vdot \vb{\dot{v}} = -\zeta^\text{t}v^2 + \vb{F}^\text{th} \vdot \vb{v}  \label{appeq:diss_trans}
\end{align}
\Cref{appeq:diss_trans} is the final result for the translational degrees of freedom.

For mechanical rotation we find, using eqs. \eqref{eq:L_dot_calculation}, \eqref{eq:u_dot} and $\vb{L}_0 = I\bs{\omega}_0$, that
\begin{align*}
    \pdv{E}{\vb{u}} \vdot \vb{\dot{u}} &= \pdv{E}{\vb{u}} \vdot (\bs{\omega} \cross \vb{u}) = -\bs{\omega} \vdot \left(\pdv{E}{\vb{u}} \cross \vb{u} \right) = -I\bs{\omega} \vdot \dot{\boldsymbol{\omega}}_0
\end{align*}
while 
\begin{align*}
    \pdv{E}{\boldsymbol{\omega}} \vdot \bs{\dot{\omega}} = I \bs{\omega} \vdot \bs{\dot{\omega}}
\end{align*}
Inserting \cref{eq:omega_dot_no_damping,eq:omega_dot} yields
\begin{align}
    \pdv{E}{\vb{u}} \vdot \vb{\dot{u}} + \pdv{E}{\boldsymbol{\omega}} \vdot \bs{\dot{\omega}} &= I\bs{\omega} \vdot (\dot{\boldsymbol{\omega}} - \dot{\boldsymbol{\omega}}_0) 
    \label{appeq:mech_rot_pdv}\\
    &= \frac{\mu}{\gamma}\bs{\omega} \vdot (\vb{\dot{m}} - \vb{\dot{m}}_0) - \zeta^\text{r}\omega^2 + \bs{\omega} \vdot \bs{\tau}^\text{th} \notag
\end{align}

For magnetic rotation, we find analogously that
\begin{align*}
    \pdv{E}{\vb{m}} \vdot \vb{\dot{m}} &= \pdv{E}{\vb{m}} \vdot (\bs{\Omega} \cross \vb{m}) = -\bs{\Omega} \vdot \vb{\dot{S}}_0 = \frac{\mu}{\gamma} \bs{\Omega} \vdot \vb{\dot{m}}_0
\end{align*}
where $\bs{\Omega}$ is defined in \cref{eq:Omega} and $\vb{\dot{S}}_0 = \partial_{\vb{m}} E \cross \vb{m}_0$. We note that $\bs{\Omega} \vdot \vb{\dot{m}} = \bs{\Omega} \vdot (\bs{\Omega} \cross \vb{m}) = 0$, so
\begin{align}
    \pdv{E}{\vb{m}} \vdot \vb{\dot{m}} = \frac{\mu}{\gamma} \bs{\Omega} \vdot (\vb{\dot{m}}_0 - \vb{\dot{m}})   \label{appeq:mag_rot_pdv}
\end{align}

Now, adding \cref{appeq:mech_rot_pdv,appeq:mag_rot_pdv} we get 
\begin{align}
    \pdv{E}{\vb{u}} \vdot \vb{\dot{u}} + &\pdv{E}{\bs{\omega}} \vdot \bs{\dot{\omega}} + \pdv{E}{\vb{m}} \vdot \vb{\dot{m}} 
    \label{appeq:rot_pdv}\\
    &= \frac{\mu}{\gamma}(\bs{\Omega} - \bs{\omega}) \vdot (\vb{\dot{m}}_0 - \vb{\dot{m}}) - \zeta^\text{r} \omega^2 + \bs{\omega} \vdot \bs{\tau}^\text{th} \notag
\end{align}
It follows from \cref{eq:mu_dot_no_damping,eq:mu_dot_mid,eq:B_eff} that
\begin{align*}
    \vb{\dot{m}}_0 - \vb{\dot{m}} = \gamma \vb{m} \cross \vb{B}^\text{th} - \alpha \vb{m} \cross (\bs{\Omega} \cross \vb{m} - \gamma\vb{B}^\text{bar}).
\end{align*}
Inserting the definition of the Barnett field, \cref{eq:B_bar}, this can be written
\begin{align*}
    \vb{\dot{m}}_0 - \vb{\dot{m}} = \gamma \vb{m} \cross \left[\vb{B}^\text{th} - \alpha (\bs{\Omega} - \bs{\omega}) \cross \vb{m}\right].
\end{align*}
Hence
\begin{align}
\frac{\mu}{\gamma}(\bs{\Omega} &- \bs{\omega}) \vdot (\vb{\dot{m}}_0 - \vb{\dot{m}}) \label{appeq:mag_mech_rot_identity}\\
&= \mu \left[(\bs{\Omega} - \bs{\omega}) \cross \vb{m} \right] \vdot \vb{B}^\text{th} - \alpha \mu \gamma^{-1} ([\bs{\Omega} - \bs{\omega}] \cross \vb{m})^2 \notag
\end{align}
Inserting \cref{appeq:mag_mech_rot_identity} in \cref{appeq:rot_pdv} then adding \cref{appeq:pdv_E_t,appeq:diss_trans} gives the energy transfer expressed in \cref{eq:E_dot,eq:P_in,eq:P_mag,eq:P_rot,eq:P_trans}.

We emphasize that no approximations were used, so within the model of \cref{sec:Full_Langevin_dynamics} the derived energy transfer is exact under the Stratonovich interpretation.
The result is unchanged in the overdamped limit, i.e.\ when setting $\mathfrak{m}=I=0$, and we never explicitly used the form of the dipole interactions, so the proof applies to any magnetostatic interactions.

We note that the same derivation applies to a single, isolated MNP, which shows that the energy transfer described by the $i$'th term of \cref{eq:E_dot} occurs locally at and around the $i$'th MNP.

\section{Numerical implementation \label{appsec:implementation}}

For simulations in the overdamped limit, we use the time-stepping procedure:
\begin{itemize}
    \itemsep-0.15em
    \item[] \textbf{Overdamped limit}
    \item Update $\vb{F}^\text{th}, \bs{\tau}^\text{th}, \vb{B}^\text{th}$
    \item Compute $\vb{v}_i^n, \bs{\omega}_i^n, \bs{\Omega}_i^n$ \quad [\cref{eq:r_dot_overdamped,eq:omega_overdamped,eq:Omega}]
    \item $(\vb{m}_i^n, \vb{u}_i^n, \vb{r}_i^n) \xrightarrow{} (\vb{m}_i^{n+1/2}, \vb{u}_i^{n+1/2}, \vb{r}_i^{n+1/2})$
    \item Compute $\vb{v}_i^{n+1/2}, \bs{\omega}_i^{n+1/2}, \bs{\Omega}_i^{n+1/2}$
    \item Compute $\dot{E}^{n+1/2}$ \quad [\cref{eq:E_dot}]
    \item $E^{n+1} = E^n + \Delta t \dot{E}^{n+1/2}$
    \item $(\vb{m}_i^{n+1/2}, \vb{u}_i^{n+1/2}, \vb{r}_i^{n+1/2}) \xrightarrow{} (\vb{m}_i^{n+1}, \vb{u}_i^{n+1}, \vb{r}_i^{n+1})$
    \item Repeat from top
\end{itemize}
Here $n$ is the timestep index, i.e.\ $t_n = n \Delta t$ so for example $E^n = E(t = n \Delta t)$.
We use the Euler method for translations and the Euler-Rodriguez formula for rotations\cite{cheng_historical_1989}, i.e.
\begin{align*}
    \vb{m}^{n+1/2}_i &= \mathrm{Rot}(\vb{m}_i^n, \Delta t \bs{\Omega}_i^n/2)
    \\
    \vb{u}^{n+1/2}_i &= \mathrm{Rot}(\vb{u}_i^n, \Delta t \bs{\omega}_i^n/2)
    \\
    \vb{r}^{n+1/2}_i &= \vb{r}^n_i + \frac{1}{2}\Delta t \vb{v}_i^n
\end{align*}
where
\begin{align*}
    \mathrm{Rot}(\vb{u}, \bs{\theta}) = \vb{u} \cos \theta + (\bs{\hat{\theta}} \cross \vb{u}) \sin \theta + (\bs{\hat{\theta}} \vdot \vb{u})(1 - \cos \theta)\bs{\hat{\theta}}.
\end{align*}
For the stochastic vectors, each component is drawn independently from a Gaussian distribution with zero mean and variances given in \cref{eq:B_th,eq:F_th,eq:tau_th}. Since the algorithm uses discrete time $\delta(t-t') \xrightarrow{} \frac{1}{\Delta t} \delta_{nn'}$, so for example $F_x^\text{th},F_y^\text{th}$ and $F_z^\text{th}$ all have a variance of $2k_B T \zeta^\text{t}/\Delta t$. The meaning of $\delta_{nn'}$ is that $\vb{F}^\text{th}, \bs{\tau}^\text{th}, \vb{B}^\text{th}$ should be redrawn at random every timestep.

Note that when computing $\vb{v}, \bs{\omega}, \bs{\Omega}$ we compute all deterministic contributions to $\vb{F}$, $\bs{\tau}$ and $\vb{B}^\text{eff}$ both at $t_n$ and $t_{n+1/2}$, but the stochastic contributions are only changed at $t_n$. Since power is evaluated at $t_{n+1/2}$, this amounts to a midpoint integration of energy, also known as a Stratonovich integral. If $\dot{E}$ was instead evaluated at the same point in time where we update $\vb{F}^\text{th}, \bs{\tau}^\text{th}, \vb{B}^\text{th}$ it would be an Itô integral, which entails artificial thermal drift in the energy, even at vanishingly small timestep.

We also tested a combination of the Euler-Rodriguez formula and Heuns' method:
\begin{align*}
    \vb{\Tilde{m}}^{n+1/2}_i &= \text{Rot}(\vb{m}_i^n, \Delta t \bs{\Omega}_i^n/2) \\
    \overline{\bs{\Omega}_i^n} &= \frac{1}{2}\left(\boldsymbol{\Omega}^n_i + \bs{\Omega}[\vb{\Tilde{m}}^{n+1/2}_i, \vb{u}^n_i,\vb{r}_n^i] \right)
    \\
    \vb{m}^{n+1/2}_i &= \text{Rot}(\vb{m}_i^n, \Delta t \overline{\bs{\Omega}_i^n}/2)
\end{align*}
That is, instead of using the angular velocity at $t_n$, we use the predicted moment $\vb{\Tilde{m}}_i^{n+1/2}$ to estimate the average angular velocity $\overline{\bs{\Omega}_i^n}$ between $t_n$ and $t_{n+1/2}$.

For calculations with inertia, we use the procedure:

\begin{itemize}
\setlength{\itemindent}{-0.75em}
    \itemsep-0.15em
    \item[] \textbf{With inertia}
    \item Update $\vb{F}^\text{th}, \bs{\tau}^\text{th}, \vb{B}^\text{th}$
    \item $(\vb{m}_i^n, \vb{u}_i^n, \vb{r}_i^n) \xrightarrow{} (\vb{m}_i^{n+1/2}, \vb{u}_i^{n+1}, \vb{r}_i^{n+1})$
    \item Compute $\vb{\dot{v}}_i^{n+1}, \bs{\dot{\omega}}_i^{n}, \bs{\Omega}_i^{n+1/2}$ [\text{\cref{eq:r_ddot,eq:omega_dot,eq:Omega}}]
    \item $(\vb{v}_i^n, \bs{\omega}_i^n) \xrightarrow{} (\vb{v}_i^{n+1}, \bs{\omega}_i^{n+1})$
    \item Compute $\vb{\dot{J}}^{n+1/2}$ \quad [\text{\cref{eq:J_dot}}]
    \item $\vb{m}_i^{n+1/2} \xrightarrow{} \vb{m}_i^{n+1}$
    \item Repeat from top
\end{itemize}
We use the Euler method for updating $\vb{v}, \bs{\omega}, \vb{r}$ and the Euler-Rodriguez formula for rotating $\vb{m}, \vb{u}$. 

Our momentum integration is an endpoint integration in terms of the mechanical variables and midpoint in terms of magnetic moment. This works, because we only observed thermal drift in the magnetic angular momentum $\vb{S}$, not in the mechanical $\vb{L}$. We have not found a method for eliminating thermal drift in the energy integration, when both temperature and inertia is included.

\newpage

\twocolumngrid

\bibliography{references.bib}

\begin{thebibliography}{120}%
\makeatletter
\providecommand \@ifxundefined [1]{%
 \@ifx{#1\undefined}
}%
\providecommand \@ifnum [1]{%
 \ifnum #1\expandafter \@firstoftwo
 \else \expandafter \@secondoftwo
 \fi
}%
\providecommand \@ifx [1]{%
 \ifx #1\expandafter \@firstoftwo
 \else \expandafter \@secondoftwo
 \fi
}%
\providecommand \natexlab [1]{#1}%
\providecommand \enquote  [1]{``#1''}%
\providecommand \bibnamefont  [1]{#1}%
\providecommand \bibfnamefont [1]{#1}%
\providecommand \citenamefont [1]{#1}%
\providecommand \href@noop [0]{\@secondoftwo}%
\providecommand \href [0]{\begingroup \@sanitize@url \@href}%
\providecommand \@href[1]{\@@startlink{#1}\@@href}%
\providecommand \@@href[1]{\endgroup#1\@@endlink}%
\providecommand \@sanitize@url [0]{\catcode `\\12\catcode `\$12\catcode
  `\&12\catcode `\#12\catcode `\^12\catcode `\_12\catcode `\%12\relax}%
\providecommand \@@startlink[1]{}%
\providecommand \@@endlink[0]{}%
\providecommand \url  [0]{\begingroup\@sanitize@url \@url }%
\providecommand \@url [1]{\endgroup\@href {#1}{\urlprefix }}%
\providecommand \urlprefix  [0]{URL }%
\providecommand \Eprint [0]{\href }%
\providecommand \doibase [0]{https://doi.org/}%
\providecommand \selectlanguage [0]{\@gobble}%
\providecommand \bibinfo  [0]{\@secondoftwo}%
\providecommand \bibfield  [0]{\@secondoftwo}%
\providecommand \translation [1]{[#1]}%
\providecommand \BibitemOpen [0]{}%
\providecommand \bibitemStop [0]{}%
\providecommand \bibitemNoStop [0]{.\EOS\space}%
\providecommand \EOS [0]{\spacefactor3000\relax}%
\providecommand \BibitemShut  [1]{\csname bibitem#1\endcsname}%
\let\auto@bib@innerbib\@empty
\bibitem [{\citenamefont {Sanusi}\ \emph {et~al.}(2023)\citenamefont {Sanusi},
  \citenamefont {Zambach}, \citenamefont {Frandsen}, \citenamefont {Beleggia},
  \citenamefont {Michael~J{\o}rgensen},\ and\ \citenamefont
  {Ouyang}}]{sanusi_investigation_2023}%
  \BibitemOpen
  \bibfield  {author} {\bibinfo {author} {\bibfnamefont {B.~N.}\ \bibnamefont
  {Sanusi}}, \bibinfo {author} {\bibfnamefont {M.}~\bibnamefont {Zambach}},
  \bibinfo {author} {\bibfnamefont {C.}~\bibnamefont {Frandsen}}, \bibinfo
  {author} {\bibfnamefont {M.}~\bibnamefont {Beleggia}}, \bibinfo {author}
  {\bibfnamefont {A.}~\bibnamefont {Michael~J{\o}rgensen}},\ and\ \bibinfo
  {author} {\bibfnamefont {Z.}~\bibnamefont {Ouyang}},\ }\href
  {https://doi.org/10.1109/TPEL.2023.3249106} {\bibfield  {journal} {\bibinfo
  {journal} {IEEE Transactions on Power Electronics}\ }\textbf {\bibinfo
  {volume} {38}},\ \bibinfo {pages} {7444} (\bibinfo {year}
  {2023})}\BibitemShut {NoStop}%
\bibitem [{\citenamefont {Almind}\ \emph {et~al.}(2021)\citenamefont {Almind},
  \citenamefont {Vinum}, \citenamefont {Wismann}, \citenamefont {Hansen},
  \citenamefont {Vendelbo}, \citenamefont {Engb{\ae}k}, \citenamefont
  {Mortensen}, \citenamefont {Chorkendorff},\ and\ \citenamefont
  {Frandsen}}]{almind_optimized_2021}%
  \BibitemOpen
  \bibfield  {author} {\bibinfo {author} {\bibfnamefont {M.~R.}\ \bibnamefont
  {Almind}}, \bibinfo {author} {\bibfnamefont {M.~G.}\ \bibnamefont {Vinum}},
  \bibinfo {author} {\bibfnamefont {S.~T.}\ \bibnamefont {Wismann}}, \bibinfo
  {author} {\bibfnamefont {M.~F.}\ \bibnamefont {Hansen}}, \bibinfo {author}
  {\bibfnamefont {S.~B.}\ \bibnamefont {Vendelbo}}, \bibinfo {author}
  {\bibfnamefont {J.~S.}\ \bibnamefont {Engb{\ae}k}}, \bibinfo {author}
  {\bibfnamefont {P.~M.}\ \bibnamefont {Mortensen}}, \bibinfo {author}
  {\bibfnamefont {I.}~\bibnamefont {Chorkendorff}},\ and\ \bibinfo {author}
  {\bibfnamefont {C.}~\bibnamefont {Frandsen}},\ }\href
  {https://doi.org/10.1021/acsanm.1c01941} {\bibfield  {journal} {\bibinfo
  {journal} {ACS Applied Nano Materials}\ }\textbf {\bibinfo {volume} {4}},\
  \bibinfo {pages} {11537} (\bibinfo {year} {2021})}\BibitemShut {NoStop}%
\bibitem [{\citenamefont {Yassine}\ \emph {et~al.}(2020)\citenamefont
  {Yassine}, \citenamefont {Fatfat}, \citenamefont {Darwish},\ and\
  \citenamefont {Karam}}]{yassine_localized_2020}%
  \BibitemOpen
  \bibfield  {author} {\bibinfo {author} {\bibfnamefont {S.~R.}\ \bibnamefont
  {Yassine}}, \bibinfo {author} {\bibfnamefont {Z.}~\bibnamefont {Fatfat}},
  \bibinfo {author} {\bibfnamefont {G.~H.}\ \bibnamefont {Darwish}},\ and\
  \bibinfo {author} {\bibfnamefont {P.}~\bibnamefont {Karam}},\ }\href
  {https://doi.org/10.1039/D0CY00439A} {\bibfield  {journal} {\bibinfo
  {journal} {Catalysis Science \& Technology}\ }\textbf {\bibinfo {volume}
  {10}},\ \bibinfo {pages} {3890} (\bibinfo {year} {2020})}\BibitemShut
  {NoStop}%
\bibitem [{\citenamefont {Tietze}\ \emph {et~al.}(2015)\citenamefont {Tietze},
  \citenamefont {Zaloga}, \citenamefont {Unterweger}, \citenamefont {Lyer},
  \citenamefont {Friedrich}, \citenamefont {Janko}, \citenamefont
  {P{\"o}ttler}, \citenamefont {D{\"u}rr},\ and\ \citenamefont
  {Alexiou}}]{tietze_magnetic_2015}%
  \BibitemOpen
  \bibfield  {author} {\bibinfo {author} {\bibfnamefont {R.}~\bibnamefont
  {Tietze}}, \bibinfo {author} {\bibfnamefont {J.}~\bibnamefont {Zaloga}},
  \bibinfo {author} {\bibfnamefont {H.}~\bibnamefont {Unterweger}}, \bibinfo
  {author} {\bibfnamefont {S.}~\bibnamefont {Lyer}}, \bibinfo {author}
  {\bibfnamefont {R.~P.}\ \bibnamefont {Friedrich}}, \bibinfo {author}
  {\bibfnamefont {C.}~\bibnamefont {Janko}}, \bibinfo {author} {\bibfnamefont
  {M.}~\bibnamefont {P{\"o}ttler}}, \bibinfo {author} {\bibfnamefont
  {S.}~\bibnamefont {D{\"u}rr}},\ and\ \bibinfo {author} {\bibfnamefont
  {C.}~\bibnamefont {Alexiou}},\ }\href
  {https://doi.org/10.1016/j.bbrc.2015.08.022} {\bibfield  {journal} {\bibinfo
  {journal} {Biochemical and Biophysical Research Communications}\ }\textbf
  {\bibinfo {volume} {468}},\ \bibinfo {pages} {463} (\bibinfo {year}
  {2015})}\BibitemShut {NoStop}%
\bibitem [{\citenamefont {Pankhurst}\ \emph {et~al.}(2003)\citenamefont
  {Pankhurst}, \citenamefont {Connolly}, \citenamefont {Jones},\ and\
  \citenamefont {Dobson}}]{pankhurst_applications_2003}%
  \BibitemOpen
  \bibfield  {author} {\bibinfo {author} {\bibfnamefont {Q.~A.}\ \bibnamefont
  {Pankhurst}}, \bibinfo {author} {\bibfnamefont {J.}~\bibnamefont {Connolly}},
  \bibinfo {author} {\bibfnamefont {S.~K.}\ \bibnamefont {Jones}},\ and\
  \bibinfo {author} {\bibfnamefont {J.}~\bibnamefont {Dobson}},\ }\href
  {https://doi.org/10.1088/0022-3727/36/13/201} {\bibfield  {journal} {\bibinfo
   {journal} {Journal of Physics D: Applied Physics}\ }\textbf {\bibinfo
  {volume} {36}},\ \bibinfo {pages} {R167} (\bibinfo {year} {2003})},\ \bibinfo
  {note} {publisher: IOP Publishing}\BibitemShut {NoStop}%
\bibitem [{\citenamefont {Pankhurst}\ \emph {et~al.}(2009)\citenamefont
  {Pankhurst}, \citenamefont {Thanh}, \citenamefont {Jones},\ and\
  \citenamefont {Dobson}}]{pankhurst_progress_2009}%
  \BibitemOpen
  \bibfield  {author} {\bibinfo {author} {\bibfnamefont {Q.~A.}\ \bibnamefont
  {Pankhurst}}, \bibinfo {author} {\bibfnamefont {N.~T.~K.}\ \bibnamefont
  {Thanh}}, \bibinfo {author} {\bibfnamefont {S.~K.}\ \bibnamefont {Jones}},\
  and\ \bibinfo {author} {\bibfnamefont {J.}~\bibnamefont {Dobson}},\ }\href
  {https://doi.org/10.1088/0022-3727/42/22/224001} {\bibfield  {journal}
  {\bibinfo  {journal} {Journal of Physics D: Applied Physics}\ }\textbf
  {\bibinfo {volume} {42}},\ \bibinfo {pages} {224001} (\bibinfo {year}
  {2009})}\BibitemShut {NoStop}%
\bibitem [{\citenamefont {Périgo}\ \emph {et~al.}(2015)\citenamefont
  {Périgo}, \citenamefont {Hemery}, \citenamefont {Sandre}, \citenamefont
  {Ortega}, \citenamefont {Garaio}, \citenamefont {Plazaola},\ and\
  \citenamefont {Teran}}]{perigo_fundamentals_2015}%
  \BibitemOpen
  \bibfield  {author} {\bibinfo {author} {\bibfnamefont {E.~A.}\ \bibnamefont
  {Périgo}}, \bibinfo {author} {\bibfnamefont {G.}~\bibnamefont {Hemery}},
  \bibinfo {author} {\bibfnamefont {O.}~\bibnamefont {Sandre}}, \bibinfo
  {author} {\bibfnamefont {D.}~\bibnamefont {Ortega}}, \bibinfo {author}
  {\bibfnamefont {E.}~\bibnamefont {Garaio}}, \bibinfo {author} {\bibfnamefont
  {F.}~\bibnamefont {Plazaola}},\ and\ \bibinfo {author} {\bibfnamefont
  {F.~J.}\ \bibnamefont {Teran}},\ }\href {https://doi.org/10.1063/1.4935688}
  {\bibfield  {journal} {\bibinfo  {journal} {Applied Physics Reviews}\
  }\textbf {\bibinfo {volume} {2}},\ \bibinfo {pages} {041302} (\bibinfo {year}
  {2015})},\ \Eprint
  {https://arxiv.org/abs/https://pubs.aip.org/aip/apr/article-pdf/doi/10.1063/1.4935688/13246585/041302\_1\_online.pdf}
  {https://pubs.aip.org/aip/apr/article-pdf/doi/10.1063/1.4935688/13246585/041302\_1\_online.pdf}
  \BibitemShut {NoStop}%
\bibitem [{\citenamefont {Wu}\ \emph {et~al.}(2019)\citenamefont {Wu},
  \citenamefont {Su}, \citenamefont {Liu}, \citenamefont {Saha},\ and\
  \citenamefont {Wang}}]{wu_magnetic_2019}%
  \BibitemOpen
  \bibfield  {author} {\bibinfo {author} {\bibfnamefont {K.}~\bibnamefont
  {Wu}}, \bibinfo {author} {\bibfnamefont {D.}~\bibnamefont {Su}}, \bibinfo
  {author} {\bibfnamefont {J.}~\bibnamefont {Liu}}, \bibinfo {author}
  {\bibfnamefont {R.}~\bibnamefont {Saha}},\ and\ \bibinfo {author}
  {\bibfnamefont {J.-P.}\ \bibnamefont {Wang}},\ }\href
  {https://doi.org/10.1088/1361-6528/ab4241} {\bibfield  {journal} {\bibinfo
  {journal} {Nanotechnology}\ }\textbf {\bibinfo {volume} {30}},\ \bibinfo
  {pages} {502003} (\bibinfo {year} {2019})}\BibitemShut {NoStop}%
\bibitem [{\citenamefont {Panagiotopoulos}\ \emph {et~al.}(2015)\citenamefont
  {Panagiotopoulos}, \citenamefont {Duschka}, \citenamefont {Ahlborg},
  \citenamefont {Bringout}, \citenamefont {Debbeler}, \citenamefont {Graeser},
  \citenamefont {Kaethner}, \citenamefont {{L{\"u}dtke-Buzug}}, \citenamefont
  {Medimagh}, \citenamefont {Stelzner}, \citenamefont {Buzug}, \citenamefont
  {Barkhausen}, \citenamefont {Vogt},\ and\ \citenamefont
  {Haegele}}]{panagiotopoulos_magnetic_2015}%
  \BibitemOpen
  \bibfield  {author} {\bibinfo {author} {\bibfnamefont {N.}~\bibnamefont
  {Panagiotopoulos}}, \bibinfo {author} {\bibfnamefont {R.~L.}\ \bibnamefont
  {Duschka}}, \bibinfo {author} {\bibfnamefont {M.}~\bibnamefont {Ahlborg}},
  \bibinfo {author} {\bibfnamefont {G.}~\bibnamefont {Bringout}}, \bibinfo
  {author} {\bibfnamefont {C.}~\bibnamefont {Debbeler}}, \bibinfo {author}
  {\bibfnamefont {M.}~\bibnamefont {Graeser}}, \bibinfo {author} {\bibfnamefont
  {C.}~\bibnamefont {Kaethner}}, \bibinfo {author} {\bibfnamefont
  {K.}~\bibnamefont {{L{\"u}dtke-Buzug}}}, \bibinfo {author} {\bibfnamefont
  {H.}~\bibnamefont {Medimagh}}, \bibinfo {author} {\bibfnamefont
  {J.}~\bibnamefont {Stelzner}}, \bibinfo {author} {\bibfnamefont {T.~M.}\
  \bibnamefont {Buzug}}, \bibinfo {author} {\bibfnamefont {J.}~\bibnamefont
  {Barkhausen}}, \bibinfo {author} {\bibfnamefont {F.~M.}\ \bibnamefont
  {Vogt}},\ and\ \bibinfo {author} {\bibfnamefont {J.}~\bibnamefont
  {Haegele}},\ }\href {https://doi.org/10.2147/IJN.S70488} {\bibfield
  {journal} {\bibinfo  {journal} {International Journal of Nanomedicine}\
  }\textbf {\bibinfo {volume} {10}},\ \bibinfo {pages} {3097} (\bibinfo {year}
  {2015})}\BibitemShut {NoStop}%
\bibitem [{\citenamefont {Rosensweig}(2002)}]{rosensweig_heating_2002}%
  \BibitemOpen
  \bibfield  {author} {\bibinfo {author} {\bibfnamefont {R.~E.}\ \bibnamefont
  {Rosensweig}},\ }\href {https://doi.org/10.1016/S0304-8853(02)00706-0}
  {\bibfield  {journal} {\bibinfo  {journal} {Journal of Magnetism and Magnetic
  Materials}\ }\bibinfo {series} {Proceedings of the 9th {International}
  {Conference} on {Magnetic} {Fluids}},\ \textbf {\bibinfo {volume} {252}},\
  \bibinfo {pages} {370} (\bibinfo {year} {2002})}\BibitemShut {NoStop}%
\bibitem [{\citenamefont {Carrey}\ \emph {et~al.}(2011)\citenamefont {Carrey},
  \citenamefont {Mehdaoui},\ and\ \citenamefont
  {Respaud}}]{carrey_simple_2011}%
  \BibitemOpen
  \bibfield  {author} {\bibinfo {author} {\bibfnamefont {J.}~\bibnamefont
  {Carrey}}, \bibinfo {author} {\bibfnamefont {B.}~\bibnamefont {Mehdaoui}},\
  and\ \bibinfo {author} {\bibfnamefont {M.}~\bibnamefont {Respaud}},\ }\href
  {https://doi.org/10.1063/1.3551582} {\bibfield  {journal} {\bibinfo
  {journal} {Journal of Applied Physics}\ }\textbf {\bibinfo {volume} {109}},\
  \bibinfo {pages} {083921} (\bibinfo {year} {2011})}\BibitemShut {NoStop}%
\bibitem [{\citenamefont {{Cazares-Cortes}}\ \emph {et~al.}(2019)\citenamefont
  {{Cazares-Cortes}}, \citenamefont {Cabana}, \citenamefont {Boitard},
  \citenamefont {Nehlig}, \citenamefont {Griffete}, \citenamefont {Fresnais},
  \citenamefont {Wilhelm}, \citenamefont {{Abou-Hassan}},\ and\ \citenamefont
  {M{\'e}nager}}]{cazares-cortes_recent_2019}%
  \BibitemOpen
  \bibfield  {author} {\bibinfo {author} {\bibfnamefont {E.}~\bibnamefont
  {{Cazares-Cortes}}}, \bibinfo {author} {\bibfnamefont {S.}~\bibnamefont
  {Cabana}}, \bibinfo {author} {\bibfnamefont {C.}~\bibnamefont {Boitard}},
  \bibinfo {author} {\bibfnamefont {E.}~\bibnamefont {Nehlig}}, \bibinfo
  {author} {\bibfnamefont {N.}~\bibnamefont {Griffete}}, \bibinfo {author}
  {\bibfnamefont {J.}~\bibnamefont {Fresnais}}, \bibinfo {author}
  {\bibfnamefont {C.}~\bibnamefont {Wilhelm}}, \bibinfo {author} {\bibfnamefont
  {A.}~\bibnamefont {{Abou-Hassan}}},\ and\ \bibinfo {author} {\bibfnamefont
  {C.}~\bibnamefont {M{\'e}nager}},\ }\href
  {https://doi.org/10.1016/j.addr.2018.10.016} {\bibfield  {journal} {\bibinfo
  {journal} {Advanced Drug Delivery Reviews}\ }\textbf {\bibinfo {volume}
  {138}},\ \bibinfo {pages} {233} (\bibinfo {year} {2019})}\BibitemShut
  {NoStop}%
\bibitem [{\citenamefont {Haase}\ and\ \citenamefont
  {Nowak}(2012)}]{haase_role_2012}%
  \BibitemOpen
  \bibfield  {author} {\bibinfo {author} {\bibfnamefont {C.}~\bibnamefont
  {Haase}}\ and\ \bibinfo {author} {\bibfnamefont {U.}~\bibnamefont {Nowak}},\
  }\href {https://doi.org/10.1103/PhysRevB.85.045435} {\bibfield  {journal}
  {\bibinfo  {journal} {Physical Review B}\ }\textbf {\bibinfo {volume} {85}},\
  \bibinfo {pages} {045435} (\bibinfo {year} {2012})}\BibitemShut {NoStop}%
\bibitem [{\citenamefont {Muñoz-Menendez}\ \emph {et~al.}(2020)\citenamefont
  {Muñoz-Menendez}, \citenamefont {Serantes}, \citenamefont
  {Chubykalo-Fesenko}, \citenamefont {Ruta}, \citenamefont {Hovorka},
  \citenamefont {Nieves}, \citenamefont {Livesey}, \citenamefont {Baldomir},\
  and\ \citenamefont {Chantrell}}]{munoz-menendez_disentangling_2020}%
  \BibitemOpen
  \bibfield  {author} {\bibinfo {author} {\bibfnamefont {C.}~\bibnamefont
  {Muñoz-Menendez}}, \bibinfo {author} {\bibfnamefont {D.}~\bibnamefont
  {Serantes}}, \bibinfo {author} {\bibfnamefont {O.}~\bibnamefont
  {Chubykalo-Fesenko}}, \bibinfo {author} {\bibfnamefont {S.}~\bibnamefont
  {Ruta}}, \bibinfo {author} {\bibfnamefont {O.}~\bibnamefont {Hovorka}},
  \bibinfo {author} {\bibfnamefont {P.}~\bibnamefont {Nieves}}, \bibinfo
  {author} {\bibfnamefont {K.~L.}\ \bibnamefont {Livesey}}, \bibinfo {author}
  {\bibfnamefont {D.}~\bibnamefont {Baldomir}},\ and\ \bibinfo {author}
  {\bibfnamefont {R.}~\bibnamefont {Chantrell}},\ }\href
  {https://doi.org/10.1103/PhysRevB.102.214412} {\bibfield  {journal} {\bibinfo
   {journal} {Physical Review B}\ }\textbf {\bibinfo {volume} {102}},\ \bibinfo
  {pages} {214412} (\bibinfo {year} {2020})}\BibitemShut {NoStop}%
\bibitem [{\citenamefont {Torche}\ \emph {et~al.}(2020)\citenamefont {Torche},
  \citenamefont {{Munoz-Menendez}}, \citenamefont {Serantes}, \citenamefont
  {Baldomir}, \citenamefont {Livesey}, \citenamefont {{Chubykalo-Fesenko}},
  \citenamefont {Ruta}, \citenamefont {Chantrell},\ and\ \citenamefont
  {Hovorka}}]{torche_thermodynamics_2020}%
  \BibitemOpen
  \bibfield  {author} {\bibinfo {author} {\bibfnamefont {P.}~\bibnamefont
  {Torche}}, \bibinfo {author} {\bibfnamefont {C.}~\bibnamefont
  {{Munoz-Menendez}}}, \bibinfo {author} {\bibfnamefont {D.}~\bibnamefont
  {Serantes}}, \bibinfo {author} {\bibfnamefont {D.}~\bibnamefont {Baldomir}},
  \bibinfo {author} {\bibfnamefont {K.~L.}\ \bibnamefont {Livesey}}, \bibinfo
  {author} {\bibfnamefont {O.}~\bibnamefont {{Chubykalo-Fesenko}}}, \bibinfo
  {author} {\bibfnamefont {S.}~\bibnamefont {Ruta}}, \bibinfo {author}
  {\bibfnamefont {R.}~\bibnamefont {Chantrell}},\ and\ \bibinfo {author}
  {\bibfnamefont {O.}~\bibnamefont {Hovorka}},\ }\href
  {https://doi.org/10.1103/PhysRevB.101.224429} {\bibfield  {journal} {\bibinfo
   {journal} {Physical Review B}\ }\textbf {\bibinfo {volume} {101}},\ \bibinfo
  {pages} {224429} (\bibinfo {year} {2020})}\BibitemShut {NoStop}%
\bibitem [{\citenamefont {{Ortega-Julia}}\ \emph {et~al.}(2023)\citenamefont
  {{Ortega-Julia}}, \citenamefont {Ortega},\ and\ \citenamefont
  {Leliaert}}]{ortega-julia_estimating_2023}%
  \BibitemOpen
  \bibfield  {author} {\bibinfo {author} {\bibfnamefont {J.}~\bibnamefont
  {{Ortega-Julia}}}, \bibinfo {author} {\bibfnamefont {D.}~\bibnamefont
  {Ortega}},\ and\ \bibinfo {author} {\bibfnamefont {J.}~\bibnamefont
  {Leliaert}},\ }\href {https://doi.org/10.1039/D3NR01269G} {\bibfield
  {journal} {\bibinfo  {journal} {Nanoscale}\ }\textbf {\bibinfo {volume}
  {15}},\ \bibinfo {pages} {10342} (\bibinfo {year} {2023})}\BibitemShut
  {NoStop}%
\bibitem [{\citenamefont {Coral}\ \emph {et~al.}(2016)\citenamefont {Coral},
  \citenamefont {Mendoza~Zélis}, \citenamefont {Marciello}, \citenamefont
  {Morales}, \citenamefont {Craievich}, \citenamefont {Sánchez},\ and\
  \citenamefont {Fernández~van Raap}}]{coral_effect_2016}%
  \BibitemOpen
  \bibfield  {author} {\bibinfo {author} {\bibfnamefont {D.~F.}\ \bibnamefont
  {Coral}}, \bibinfo {author} {\bibfnamefont {P.}~\bibnamefont
  {Mendoza~Zélis}}, \bibinfo {author} {\bibfnamefont {M.}~\bibnamefont
  {Marciello}}, \bibinfo {author} {\bibfnamefont {M.~d.~P.}\ \bibnamefont
  {Morales}}, \bibinfo {author} {\bibfnamefont {A.}~\bibnamefont {Craievich}},
  \bibinfo {author} {\bibfnamefont {F.~H.}\ \bibnamefont {Sánchez}},\ and\
  \bibinfo {author} {\bibfnamefont {M.~B.}\ \bibnamefont {Fernández~van
  Raap}},\ }\href {https://doi.org/10.1021/acs.langmuir.5b03559} {\bibfield
  {journal} {\bibinfo  {journal} {Langmuir}\ }\textbf {\bibinfo {volume}
  {32}},\ \bibinfo {pages} {1201} (\bibinfo {year} {2016})}\BibitemShut
  {NoStop}%
\bibitem [{\citenamefont {Guibert}\ \emph {et~al.}(2015)\citenamefont
  {Guibert}, \citenamefont {Dupuis}, \citenamefont {Peyre},\ and\ \citenamefont
  {Fresnais}}]{guibert_hyperthermia_2015}%
  \BibitemOpen
  \bibfield  {author} {\bibinfo {author} {\bibfnamefont {C.}~\bibnamefont
  {Guibert}}, \bibinfo {author} {\bibfnamefont {V.}~\bibnamefont {Dupuis}},
  \bibinfo {author} {\bibfnamefont {V.}~\bibnamefont {Peyre}},\ and\ \bibinfo
  {author} {\bibfnamefont {J.}~\bibnamefont {Fresnais}},\ }\href
  {https://doi.org/10.1021/acs.jpcc.5b07796} {\bibfield  {journal} {\bibinfo
  {journal} {The Journal of Physical Chemistry C}\ }\textbf {\bibinfo {volume}
  {119}},\ \bibinfo {pages} {28148} (\bibinfo {year} {2015})}\BibitemShut
  {NoStop}%
\bibitem [{\citenamefont {Naud}\ \emph {et~al.}(2020)\citenamefont {Naud},
  \citenamefont {Th{\'e}bault}, \citenamefont {Carri{\`e}re}, \citenamefont
  {Hou}, \citenamefont {Morel}, \citenamefont {Berger}, \citenamefont
  {Di{\'e}ny},\ and\ \citenamefont {Joisten}}]{naud_cancer_2020}%
  \BibitemOpen
  \bibfield  {author} {\bibinfo {author} {\bibfnamefont {C.}~\bibnamefont
  {Naud}}, \bibinfo {author} {\bibfnamefont {C.}~\bibnamefont {Th{\'e}bault}},
  \bibinfo {author} {\bibfnamefont {M.}~\bibnamefont {Carri{\`e}re}}, \bibinfo
  {author} {\bibfnamefont {Y.}~\bibnamefont {Hou}}, \bibinfo {author}
  {\bibfnamefont {R.}~\bibnamefont {Morel}}, \bibinfo {author} {\bibfnamefont
  {F.}~\bibnamefont {Berger}}, \bibinfo {author} {\bibfnamefont
  {B.}~\bibnamefont {Di{\'e}ny}},\ and\ \bibinfo {author} {\bibfnamefont
  {H.}~\bibnamefont {Joisten}},\ }\href {https://doi.org/10.1039/D0NA00187B}
  {\bibfield  {journal} {\bibinfo  {journal} {Nanoscale Advances}\ }\textbf
  {\bibinfo {volume} {2}},\ \bibinfo {pages} {3632} (\bibinfo {year}
  {2020})}\BibitemShut {NoStop}%
\bibitem [{\citenamefont {Golovin}\ \emph {et~al.}(2015)\citenamefont
  {Golovin}, \citenamefont {Gribanovsky}, \citenamefont {Golovin},
  \citenamefont {Klyachko}, \citenamefont {Majouga}, \citenamefont {Master},
  \citenamefont {Sokolsky},\ and\ \citenamefont
  {Kabanov}}]{golovin_towards_2015}%
  \BibitemOpen
  \bibfield  {author} {\bibinfo {author} {\bibfnamefont {Y.~I.}\ \bibnamefont
  {Golovin}}, \bibinfo {author} {\bibfnamefont {S.~L.}\ \bibnamefont
  {Gribanovsky}}, \bibinfo {author} {\bibfnamefont {D.~Y.}\ \bibnamefont
  {Golovin}}, \bibinfo {author} {\bibfnamefont {N.~L.}\ \bibnamefont
  {Klyachko}}, \bibinfo {author} {\bibfnamefont {A.~G.}\ \bibnamefont
  {Majouga}}, \bibinfo {author} {\bibfnamefont {A.~M.}\ \bibnamefont {Master}},
  \bibinfo {author} {\bibfnamefont {M.}~\bibnamefont {Sokolsky}},\ and\
  \bibinfo {author} {\bibfnamefont {A.~V.}\ \bibnamefont {Kabanov}},\ }\href
  {https://doi.org/10.1016/j.jconrel.2015.09.038} {\bibfield  {journal}
  {\bibinfo  {journal} {Journal of Controlled Release}\ }\textbf {\bibinfo
  {volume} {219}},\ \bibinfo {pages} {43} (\bibinfo {year} {2015})}\BibitemShut
  {NoStop}%
\bibitem [{\citenamefont {Coffey}\ and\ \citenamefont
  {Kalmykov}(2017)}]{coffey_langevin_2017}%
  \BibitemOpen
  \bibfield  {author} {\bibinfo {author} {\bibfnamefont {W.}~\bibnamefont
  {Coffey}}\ and\ \bibinfo {author} {\bibfnamefont {Y.~P.}\ \bibnamefont
  {Kalmykov}},\ }\href@noop {} {\emph {\bibinfo {title} {The {{Langevin}}
  Equation: With Applications to Stochastic Problems in Physics, Chemistry, and
  Electrical Engineering}}},\ \bibinfo {edition} {fourth edition}\ ed.,\
  \bibinfo {series} {World {{Scientific}} Series in Contemporary Chemical
  Physics}\ No.\ \bibinfo {number} {vol. 28}\ (\bibinfo  {publisher} {{World
  Scientific}},\ \bibinfo {address} {{New Jersey}},\ \bibinfo {year}
  {2017})\BibitemShut {NoStop}%
\bibitem [{\citenamefont {Lyutyy}\ \emph {et~al.}(2017)\citenamefont {Lyutyy},
  \citenamefont {Denisova},\ and\ \citenamefont
  {Kvasnina}}]{lyutyy_forced_2017}%
  \BibitemOpen
  \bibfield  {author} {\bibinfo {author} {\bibfnamefont {T.~V.}\ \bibnamefont
  {Lyutyy}}, \bibinfo {author} {\bibfnamefont {E.~S.}\ \bibnamefont
  {Denisova}},\ and\ \bibinfo {author} {\bibfnamefont {A.~V.}\ \bibnamefont
  {Kvasnina}},\ }in\ \href {https://doi.org/10.1109/NAP.2017.8190416} {\emph
  {\bibinfo {booktitle} {2017 {IEEE} 7th {International} {Conference}
  {Nanomaterials}: {Application} {Properties} ({NAP})}}}\ (\bibinfo {year}
  {2017})\ pp.\ \bibinfo {pages} {02MFPM09--1--02MFPM09--4}\BibitemShut
  {NoStop}%
\bibitem [{\citenamefont {Lyutyy}\ \emph {et~al.}(2018)\citenamefont {Lyutyy},
  \citenamefont {Hryshko},\ and\ \citenamefont {Kovner}}]{lyutyy_power_2018}%
  \BibitemOpen
  \bibfield  {author} {\bibinfo {author} {\bibfnamefont {T.~V.}\ \bibnamefont
  {Lyutyy}}, \bibinfo {author} {\bibfnamefont {O.~M.}\ \bibnamefont
  {Hryshko}},\ and\ \bibinfo {author} {\bibfnamefont {A.~A.}\ \bibnamefont
  {Kovner}},\ }\href {https://doi.org/10.1016/j.jmmm.2017.09.021} {\bibfield
  {journal} {\bibinfo  {journal} {Journal of Magnetism and Magnetic Materials}\
  }\textbf {\bibinfo {volume} {446}},\ \bibinfo {pages} {87} (\bibinfo {year}
  {2018})},\ \bibinfo {note} {arXiv: 1706.00777}\BibitemShut {NoStop}%
\bibitem [{\citenamefont {Keshtgar}\ \emph {et~al.}(2017)\citenamefont
  {Keshtgar}, \citenamefont {Streib}, \citenamefont {Kamra}, \citenamefont
  {Blanter},\ and\ \citenamefont {Bauer}}]{keshtgar_magnetomechanical_2017}%
  \BibitemOpen
  \bibfield  {author} {\bibinfo {author} {\bibfnamefont {H.}~\bibnamefont
  {Keshtgar}}, \bibinfo {author} {\bibfnamefont {S.}~\bibnamefont {Streib}},
  \bibinfo {author} {\bibfnamefont {A.}~\bibnamefont {Kamra}}, \bibinfo
  {author} {\bibfnamefont {Y.~M.}\ \bibnamefont {Blanter}},\ and\ \bibinfo
  {author} {\bibfnamefont {G.~E.~W.}\ \bibnamefont {Bauer}},\ }\href
  {https://doi.org/10.1103/PhysRevB.95.134447} {\bibfield  {journal} {\bibinfo
  {journal} {Physical Review B}\ }\textbf {\bibinfo {volume} {95}},\ \bibinfo
  {pages} {134447} (\bibinfo {year} {2017})}\BibitemShut {NoStop}%
\bibitem [{\citenamefont {Denisov}\ \emph {et~al.}(2020)\citenamefont
  {Denisov}, \citenamefont {Lyutyy},\ and\ \citenamefont
  {Liutyi}}]{denisov_dynamics_2020-1}%
  \BibitemOpen
  \bibfield  {author} {\bibinfo {author} {\bibfnamefont {S.~I.}\ \bibnamefont
  {Denisov}}, \bibinfo {author} {\bibfnamefont {T.~V.}\ \bibnamefont
  {Lyutyy}},\ and\ \bibinfo {author} {\bibfnamefont {A.~T.}\ \bibnamefont
  {Liutyi}},\ }\href {https://doi.org/10.21272/jnep.12(6).06028} {\bibfield
  {journal} {\bibinfo  {journal} {Journal of Nano- and Electronic Physics}\
  }\textbf {\bibinfo {volume} {12}},\ \bibinfo {pages} {06028} (\bibinfo {year}
  {2020})},\ \Eprint {https://arxiv.org/abs/2012.10491} {arxiv:2012.10491
  [cond-mat]} \BibitemShut {NoStop}%
\bibitem [{\citenamefont {de~Châtel}\ \emph {et~al.}(2009)\citenamefont
  {de~Châtel}, \citenamefont {Nándori}, \citenamefont {Hakl}, \citenamefont
  {Mészáros},\ and\ \citenamefont {Vad}}]{de_chatel_magnetic_2009}%
  \BibitemOpen
  \bibfield  {author} {\bibinfo {author} {\bibfnamefont {P.~F.}\ \bibnamefont
  {de~Châtel}}, \bibinfo {author} {\bibfnamefont {I.}~\bibnamefont
  {Nándori}}, \bibinfo {author} {\bibfnamefont {J.}~\bibnamefont {Hakl}},
  \bibinfo {author} {\bibfnamefont {S.}~\bibnamefont {Mészáros}},\ and\
  \bibinfo {author} {\bibfnamefont {K.}~\bibnamefont {Vad}},\ }\href
  {https://doi.org/10.1088/0953-8984/21/12/124202} {\bibfield  {journal}
  {\bibinfo  {journal} {Journal of Physics: Condensed Matter}\ }\textbf
  {\bibinfo {volume} {21}},\ \bibinfo {pages} {124202} (\bibinfo {year}
  {2009})}\BibitemShut {NoStop}%
\bibitem [{\citenamefont {Shasha}\ and\ \citenamefont
  {Krishnan}(2021)}]{shasha_nonequilibrium_2021}%
  \BibitemOpen
  \bibfield  {author} {\bibinfo {author} {\bibfnamefont {C.}~\bibnamefont
  {Shasha}}\ and\ \bibinfo {author} {\bibfnamefont {K.~M.}\ \bibnamefont
  {Krishnan}},\ }\href {https://doi.org/10.1002/adma.201904131} {\bibfield
  {journal} {\bibinfo  {journal} {Advanced Materials}\ }\textbf {\bibinfo
  {volume} {33}},\ \bibinfo {pages} {1904131} (\bibinfo {year}
  {2021})}\BibitemShut {NoStop}%
\bibitem [{\citenamefont {Leliaert}\ \emph {et~al.}(2017)\citenamefont
  {Leliaert}, \citenamefont {Mulkers}, \citenamefont {De~Clercq}, \citenamefont
  {Coene}, \citenamefont {Dvornik},\ and\ \citenamefont
  {Van~Waeyenberge}}]{leliaert_adaptively_2017}%
  \BibitemOpen
  \bibfield  {author} {\bibinfo {author} {\bibfnamefont {J.}~\bibnamefont
  {Leliaert}}, \bibinfo {author} {\bibfnamefont {J.}~\bibnamefont {Mulkers}},
  \bibinfo {author} {\bibfnamefont {J.}~\bibnamefont {De~Clercq}}, \bibinfo
  {author} {\bibfnamefont {A.}~\bibnamefont {Coene}}, \bibinfo {author}
  {\bibfnamefont {M.}~\bibnamefont {Dvornik}},\ and\ \bibinfo {author}
  {\bibfnamefont {B.}~\bibnamefont {Van~Waeyenberge}},\ }\href
  {https://doi.org/10.1063/1.5003957} {\bibfield  {journal} {\bibinfo
  {journal} {AIP Advances}\ }\textbf {\bibinfo {volume} {7}},\ \bibinfo {pages}
  {125010} (\bibinfo {year} {2017})}\BibitemShut {NoStop}%
\bibitem [{\citenamefont {Ruta}\ \emph {et~al.}(2015)\citenamefont {Ruta},
  \citenamefont {Chantrell},\ and\ \citenamefont
  {Hovorka}}]{ruta_unified_2015}%
  \BibitemOpen
  \bibfield  {author} {\bibinfo {author} {\bibfnamefont {S.}~\bibnamefont
  {Ruta}}, \bibinfo {author} {\bibfnamefont {R.}~\bibnamefont {Chantrell}},\
  and\ \bibinfo {author} {\bibfnamefont {O.}~\bibnamefont {Hovorka}},\ }\href
  {https://doi.org/10.1038/srep09090} {\bibfield  {journal} {\bibinfo
  {journal} {Scientific Reports}\ }\textbf {\bibinfo {volume} {5}},\ \bibinfo
  {pages} {9090} (\bibinfo {year} {2015})}\BibitemShut {NoStop}%
\bibitem [{\citenamefont {Anand}(2021)}]{anand_hysteresis_2021}%
  \BibitemOpen
  \bibfield  {author} {\bibinfo {author} {\bibfnamefont {M.}~\bibnamefont
  {Anand}},\ }\href {https://doi.org/10.1016/j.jmmm.2021.168461} {\bibfield
  {journal} {\bibinfo  {journal} {Journal of Magnetism and Magnetic Materials}\
  }\textbf {\bibinfo {volume} {540}},\ \bibinfo {pages} {168461} (\bibinfo
  {year} {2021})}\BibitemShut {NoStop}%
\bibitem [{\citenamefont {Durhuus}\ \emph {et~al.}(2021)\citenamefont
  {Durhuus}, \citenamefont {Wandall}, \citenamefont {Boisen}, \citenamefont
  {Kure}, \citenamefont {Beleggia},\ and\ \citenamefont
  {Frandsen}}]{durhuus_simulated_2021}%
  \BibitemOpen
  \bibfield  {author} {\bibinfo {author} {\bibfnamefont {F.~L.}\ \bibnamefont
  {Durhuus}}, \bibinfo {author} {\bibfnamefont {L.~H.}\ \bibnamefont
  {Wandall}}, \bibinfo {author} {\bibfnamefont {M.~H.}\ \bibnamefont {Boisen}},
  \bibinfo {author} {\bibfnamefont {M.}~\bibnamefont {Kure}}, \bibinfo {author}
  {\bibfnamefont {M.}~\bibnamefont {Beleggia}},\ and\ \bibinfo {author}
  {\bibfnamefont {C.}~\bibnamefont {Frandsen}},\ }\href
  {https://doi.org/10.1039/D0NR08561H} {\bibfield  {journal} {\bibinfo
  {journal} {Nanoscale}\ }\textbf {\bibinfo {volume} {13}},\ \bibinfo {pages}
  {1970} (\bibinfo {year} {2021})}\BibitemShut {NoStop}%
\bibitem [{\citenamefont {Rozhkov}\ \emph {et~al.}(2018)\citenamefont
  {Rozhkov}, \citenamefont {Pyanzina}, \citenamefont {Novak}, \citenamefont
  {Cerdà}, \citenamefont {Sintes}, \citenamefont {Ronti}, \citenamefont
  {Sánchez},\ and\ \citenamefont {Kantorovich}}]{rozhkov_self-assembly_2018}%
  \BibitemOpen
  \bibfield  {author} {\bibinfo {author} {\bibfnamefont {D.~A.}\ \bibnamefont
  {Rozhkov}}, \bibinfo {author} {\bibfnamefont {E.~S.}\ \bibnamefont
  {Pyanzina}}, \bibinfo {author} {\bibfnamefont {E.~V.}\ \bibnamefont {Novak}},
  \bibinfo {author} {\bibfnamefont {J.~J.}\ \bibnamefont {Cerdà}}, \bibinfo
  {author} {\bibfnamefont {T.}~\bibnamefont {Sintes}}, \bibinfo {author}
  {\bibfnamefont {M.}~\bibnamefont {Ronti}}, \bibinfo {author} {\bibfnamefont
  {P.~A.}\ \bibnamefont {Sánchez}},\ and\ \bibinfo {author} {\bibfnamefont
  {S.~S.}\ \bibnamefont {Kantorovich}},\ }\href
  {https://doi.org/10.1080/08927022.2017.1378815} {\bibfield  {journal}
  {\bibinfo  {journal} {Molecular Simulation}\ }\textbf {\bibinfo {volume}
  {44}},\ \bibinfo {pages} {507} (\bibinfo {year} {2018})}\BibitemShut
  {NoStop}%
\bibitem [{\citenamefont {Anderson}\ \emph {et~al.}(2021)\citenamefont
  {Anderson}, \citenamefont {Davidson}, \citenamefont {Louie}, \citenamefont
  {Serantes},\ and\ \citenamefont {Livesey}}]{anderson_simulating_2021}%
  \BibitemOpen
  \bibfield  {author} {\bibinfo {author} {\bibfnamefont {N.~R.}\ \bibnamefont
  {Anderson}}, \bibinfo {author} {\bibfnamefont {J.}~\bibnamefont {Davidson}},
  \bibinfo {author} {\bibfnamefont {D.~R.}\ \bibnamefont {Louie}}, \bibinfo
  {author} {\bibfnamefont {D.}~\bibnamefont {Serantes}},\ and\ \bibinfo
  {author} {\bibfnamefont {K.~L.}\ \bibnamefont {Livesey}},\ }\href
  {https://doi.org/10.3390/nano11112870} {\bibfield  {journal} {\bibinfo
  {journal} {Nanomaterials}\ }\textbf {\bibinfo {volume} {11}},\ \bibinfo
  {pages} {2870} (\bibinfo {year} {2021})},\ \bibinfo {note} {number: 11
  Publisher: Multidisciplinary Digital Publishing Institute}\BibitemShut
  {NoStop}%
\bibitem [{\citenamefont {Satoh}\ \emph {et~al.}(1999)\citenamefont {Satoh},
  \citenamefont {Chantrell},\ and\ \citenamefont
  {Coverdale}}]{satoh_brownian_1999}%
  \BibitemOpen
  \bibfield  {author} {\bibinfo {author} {\bibfnamefont {A.}~\bibnamefont
  {Satoh}}, \bibinfo {author} {\bibfnamefont {R.~W.}\ \bibnamefont
  {Chantrell}},\ and\ \bibinfo {author} {\bibfnamefont {G.~N.}\ \bibnamefont
  {Coverdale}},\ }\href {https://doi.org/10.1006/jcis.1998.5826} {\bibfield
  {journal} {\bibinfo  {journal} {Journal of Colloid and Interface Science}\
  }\textbf {\bibinfo {volume} {209}},\ \bibinfo {pages} {44} (\bibinfo {year}
  {1999})}\BibitemShut {NoStop}%
\bibitem [{\citenamefont {S.~Andreu}\ \emph {et~al.}(2011)\citenamefont
  {S.~Andreu}, \citenamefont {Camacho},\ and\ \citenamefont
  {Faraudo}}]{sandreu_aggregation_2011}%
  \BibitemOpen
  \bibfield  {author} {\bibinfo {author} {\bibfnamefont {J.}~\bibnamefont
  {S.~Andreu}}, \bibinfo {author} {\bibfnamefont {J.}~\bibnamefont {Camacho}},\
  and\ \bibinfo {author} {\bibfnamefont {J.}~\bibnamefont {Faraudo}},\ }\href
  {https://doi.org/10.1039/C0SM01424A} {\bibfield  {journal} {\bibinfo
  {journal} {Soft Matter}\ }\textbf {\bibinfo {volume} {7}},\ \bibinfo {pages}
  {2336} (\bibinfo {year} {2011})}\BibitemShut {NoStop}%
\bibitem [{\citenamefont {Novikau}\ \emph {et~al.}(2020)\citenamefont
  {Novikau}, \citenamefont {S{\'a}nchez},\ and\ \citenamefont
  {Kantorovich}}]{novikau_influence_2020}%
  \BibitemOpen
  \bibfield  {author} {\bibinfo {author} {\bibfnamefont {I.~S.}\ \bibnamefont
  {Novikau}}, \bibinfo {author} {\bibfnamefont {P.~A.}\ \bibnamefont
  {S{\'a}nchez}},\ and\ \bibinfo {author} {\bibfnamefont {S.~S.}\ \bibnamefont
  {Kantorovich}},\ }\href {https://doi.org/10.1016/j.molliq.2020.112902}
  {\bibfield  {journal} {\bibinfo  {journal} {Journal of Molecular Liquids}\
  }\textbf {\bibinfo {volume} {307}},\ \bibinfo {pages} {112902} (\bibinfo
  {year} {2020})}\BibitemShut {NoStop}%
\bibitem [{\citenamefont {Xue}\ \emph {et~al.}(2015)\citenamefont {Xue},
  \citenamefont {Wang},\ and\ \citenamefont
  {Furlani}}]{xue_self-assembly_2015}%
  \BibitemOpen
  \bibfield  {author} {\bibinfo {author} {\bibfnamefont {X.}~\bibnamefont
  {Xue}}, \bibinfo {author} {\bibfnamefont {J.}~\bibnamefont {Wang}},\ and\
  \bibinfo {author} {\bibfnamefont {E.~P.}\ \bibnamefont {Furlani}},\ }\href
  {https://doi.org/10.1021/acsami.5b08310} {\bibfield  {journal} {\bibinfo
  {journal} {ACS Applied Materials \& Interfaces}\ }\textbf {\bibinfo {volume}
  {7}},\ \bibinfo {pages} {22515} (\bibinfo {year} {2015})}\BibitemShut
  {NoStop}%
\bibitem [{\citenamefont {Usov}\ and\ \citenamefont
  {Liubimov}(2012)}]{usov_dynamics_2012}%
  \BibitemOpen
  \bibfield  {author} {\bibinfo {author} {\bibfnamefont {N.~A.}\ \bibnamefont
  {Usov}}\ and\ \bibinfo {author} {\bibfnamefont {B.~Y.}\ \bibnamefont
  {Liubimov}},\ }\href {https://doi.org/10.1063/1.4737126} {\bibfield
  {journal} {\bibinfo  {journal} {Journal of Applied Physics}\ }\textbf
  {\bibinfo {volume} {112}},\ \bibinfo {pages} {023901} (\bibinfo {year}
  {2012})}\BibitemShut {NoStop}%
\bibitem [{\citenamefont {Usadel}\ and\ \citenamefont
  {Usadel}(2015)}]{usadel_dynamics_2015}%
  \BibitemOpen
  \bibfield  {author} {\bibinfo {author} {\bibfnamefont {K.~D.}\ \bibnamefont
  {Usadel}}\ and\ \bibinfo {author} {\bibfnamefont {C.}~\bibnamefont
  {Usadel}},\ }\href {https://doi.org/10.1063/1.4937919} {\bibfield  {journal}
  {\bibinfo  {journal} {Journal of Applied Physics}\ }\textbf {\bibinfo
  {volume} {118}},\ \bibinfo {pages} {234303} (\bibinfo {year} {2015})},\
  \bibinfo {note} {arXiv: 1509.05233}\BibitemShut {NoStop}%
\bibitem [{\citenamefont {Usadel}(2017)}]{usadel_dynamics_2017}%
  \BibitemOpen
  \bibfield  {author} {\bibinfo {author} {\bibfnamefont {K.~D.}\ \bibnamefont
  {Usadel}},\ }\href {https://doi.org/10.1103/PhysRevB.95.104430} {\bibfield
  {journal} {\bibinfo  {journal} {Physical Review B}\ }\textbf {\bibinfo
  {volume} {95}},\ \bibinfo {pages} {104430} (\bibinfo {year}
  {2017})}\BibitemShut {NoStop}%
\bibitem [{\citenamefont {Helbig}\ \emph {et~al.}(2023)\citenamefont {Helbig},
  \citenamefont {Abert}, \citenamefont {S\'anchez}, \citenamefont
  {Kantorovich},\ and\ \citenamefont {Suess}}]{helbig_self-consistent_2023}%
  \BibitemOpen
  \bibfield  {author} {\bibinfo {author} {\bibfnamefont {S.}~\bibnamefont
  {Helbig}}, \bibinfo {author} {\bibfnamefont {C.}~\bibnamefont {Abert}},
  \bibinfo {author} {\bibfnamefont {P.~A.}\ \bibnamefont {S\'anchez}}, \bibinfo
  {author} {\bibfnamefont {S.~S.}\ \bibnamefont {Kantorovich}},\ and\ \bibinfo
  {author} {\bibfnamefont {D.}~\bibnamefont {Suess}},\ }\href
  {https://doi.org/10.1103/PhysRevB.107.054416} {\bibfield  {journal} {\bibinfo
   {journal} {Phys. Rev. B}\ }\textbf {\bibinfo {volume} {107}},\ \bibinfo
  {pages} {054416} (\bibinfo {year} {2023})}\BibitemShut {NoStop}%
\bibitem [{\citenamefont {Lyutyy}\ and\ \citenamefont
  {Reva}(2018)}]{lyutyy_energy_2018}%
  \BibitemOpen
  \bibfield  {author} {\bibinfo {author} {\bibfnamefont {T.~V.}\ \bibnamefont
  {Lyutyy}}\ and\ \bibinfo {author} {\bibfnamefont {V.~V.}\ \bibnamefont
  {Reva}},\ }\href {https://doi.org/10.1103/PhysRevE.97.052611} {\bibfield
  {journal} {\bibinfo  {journal} {Physical Review E}\ }\textbf {\bibinfo
  {volume} {97}},\ \bibinfo {pages} {052611} (\bibinfo {year}
  {2018})}\BibitemShut {NoStop}%
\bibitem [{\citenamefont {Cabrera}\ \emph {et~al.}(2017)\citenamefont
  {Cabrera}, \citenamefont {Lak}, \citenamefont {Yoshida}, \citenamefont
  {Materia}, \citenamefont {Ortega}, \citenamefont {Ludwig}, \citenamefont
  {Guardia}, \citenamefont {Sathya}, \citenamefont {Pellegrino},\ and\
  \citenamefont {Teran}}]{cabrera_unraveling_2017}%
  \BibitemOpen
  \bibfield  {author} {\bibinfo {author} {\bibfnamefont {D.}~\bibnamefont
  {Cabrera}}, \bibinfo {author} {\bibfnamefont {A.}~\bibnamefont {Lak}},
  \bibinfo {author} {\bibfnamefont {T.}~\bibnamefont {Yoshida}}, \bibinfo
  {author} {\bibfnamefont {M.~E.}\ \bibnamefont {Materia}}, \bibinfo {author}
  {\bibfnamefont {D.}~\bibnamefont {Ortega}}, \bibinfo {author} {\bibfnamefont
  {F.}~\bibnamefont {Ludwig}}, \bibinfo {author} {\bibfnamefont
  {P.}~\bibnamefont {Guardia}}, \bibinfo {author} {\bibfnamefont
  {A.}~\bibnamefont {Sathya}}, \bibinfo {author} {\bibfnamefont
  {T.}~\bibnamefont {Pellegrino}},\ and\ \bibinfo {author} {\bibfnamefont
  {F.~J.}\ \bibnamefont {Teran}},\ }\href {https://doi.org/10.1039/C7NR00810D}
  {\bibfield  {journal} {\bibinfo  {journal} {Nanoscale}\ }\textbf {\bibinfo
  {volume} {9}},\ \bibinfo {pages} {5094} (\bibinfo {year} {2017})}\BibitemShut
  {NoStop}%
\bibitem [{\citenamefont {Kim}\ \emph {et~al.}(2018)\citenamefont {Kim},
  \citenamefont {Sim}, \citenamefont {Lee}, \citenamefont {Kim},\ and\
  \citenamefont {Kim}}]{kim_dynamical_2018}%
  \BibitemOpen
  \bibfield  {author} {\bibinfo {author} {\bibfnamefont {M.-K.}\ \bibnamefont
  {Kim}}, \bibinfo {author} {\bibfnamefont {J.}~\bibnamefont {Sim}}, \bibinfo
  {author} {\bibfnamefont {J.-H.}\ \bibnamefont {Lee}}, \bibinfo {author}
  {\bibfnamefont {M.}~\bibnamefont {Kim}},\ and\ \bibinfo {author}
  {\bibfnamefont {S.-K.}\ \bibnamefont {Kim}},\ }\href
  {https://doi.org/10.1103/PhysRevApplied.9.054037} {\bibfield  {journal}
  {\bibinfo  {journal} {Physical Review Applied}\ }\textbf {\bibinfo {volume}
  {9}},\ \bibinfo {pages} {054037} (\bibinfo {year} {2018})}\BibitemShut
  {NoStop}%
\bibitem [{\citenamefont {Leliaert}\ \emph {et~al.}(2021)\citenamefont
  {Leliaert}, \citenamefont {Ortega-Julia},\ and\ \citenamefont
  {Ortega}}]{leliaert_individual_2021}%
  \BibitemOpen
  \bibfield  {author} {\bibinfo {author} {\bibfnamefont {J.}~\bibnamefont
  {Leliaert}}, \bibinfo {author} {\bibfnamefont {J.}~\bibnamefont
  {Ortega-Julia}},\ and\ \bibinfo {author} {\bibfnamefont {D.}~\bibnamefont
  {Ortega}},\ }\href {https://doi.org/10.1039/D1NR05311F} {\bibfield  {journal}
  {\bibinfo  {journal} {Nanoscale}\ }\textbf {\bibinfo {volume} {13}},\
  \bibinfo {pages} {14734} (\bibinfo {year} {2021})}\BibitemShut {NoStop}%
\bibitem [{\citenamefont {Gilbert}(2004)}]{gilbert_LLG_2004}%
  \BibitemOpen
  \bibfield  {author} {\bibinfo {author} {\bibfnamefont {T.}~\bibnamefont
  {Gilbert}},\ }\href {https://doi.org/10.1109/TMAG.2004.836740} {\bibfield
  {journal} {\bibinfo  {journal} {IEEE Transactions on Magnetics}\ }\textbf
  {\bibinfo {volume} {40}},\ \bibinfo {pages} {3443} (\bibinfo {year}
  {2004})}\BibitemShut {NoStop}%
\bibitem [{\citenamefont {Kustura}\ \emph {et~al.}(2022)\citenamefont
  {Kustura}, \citenamefont {Wachter}, \citenamefont {Rubio~L{\'o}pez},\ and\
  \citenamefont {Rusconi}}]{kustura_stability_2022}%
  \BibitemOpen
  \bibfield  {author} {\bibinfo {author} {\bibfnamefont {K.}~\bibnamefont
  {Kustura}}, \bibinfo {author} {\bibfnamefont {V.}~\bibnamefont {Wachter}},
  \bibinfo {author} {\bibfnamefont {A.~E.}\ \bibnamefont {Rubio~L{\'o}pez}},\
  and\ \bibinfo {author} {\bibfnamefont {C.~C.}\ \bibnamefont {Rusconi}},\
  }\href {https://doi.org/10.1103/PhysRevB.105.174439} {\bibfield  {journal}
  {\bibinfo  {journal} {Physical Review B}\ }\textbf {\bibinfo {volume}
  {105}},\ \bibinfo {pages} {174439} (\bibinfo {year} {2022})}\BibitemShut
  {NoStop}%
\bibitem [{\citenamefont {Usov}\ and\ \citenamefont
  {Liubimov}(2015)}]{usov_magnetic_2015}%
  \BibitemOpen
  \bibfield  {author} {\bibinfo {author} {\bibfnamefont {N.}~\bibnamefont
  {Usov}}\ and\ \bibinfo {author} {\bibfnamefont {B.~Y.}\ \bibnamefont
  {Liubimov}},\ }\href
  {https://doi.org/https://doi.org/10.1016/j.jmmm.2015.03.035} {\bibfield
  {journal} {\bibinfo  {journal} {Journal of Magnetism and Magnetic Materials}\
  }\textbf {\bibinfo {volume} {385}},\ \bibinfo {pages} {339} (\bibinfo {year}
  {2015})}\BibitemShut {NoStop}%
\bibitem [{\citenamefont {Berkov}\ \emph {et~al.}(2006)\citenamefont {Berkov},
  \citenamefont {Gorn}, \citenamefont {Schmitz},\ and\ \citenamefont
  {Stock}}]{berkov_langevin_2006}%
  \BibitemOpen
  \bibfield  {author} {\bibinfo {author} {\bibfnamefont {D.~V.}\ \bibnamefont
  {Berkov}}, \bibinfo {author} {\bibfnamefont {N.~L.}\ \bibnamefont {Gorn}},
  \bibinfo {author} {\bibfnamefont {R.}~\bibnamefont {Schmitz}},\ and\ \bibinfo
  {author} {\bibfnamefont {D.}~\bibnamefont {Stock}},\ }\href
  {https://doi.org/10.1088/0953-8984/18/38/S05} {\bibfield  {journal} {\bibinfo
   {journal} {Journal of Physics: Condensed Matter}\ }\textbf {\bibinfo
  {volume} {18}},\ \bibinfo {pages} {S2595} (\bibinfo {year}
  {2006})}\BibitemShut {NoStop}%
\bibitem [{\citenamefont {Umeda}\ \emph {et~al.}(2023)\citenamefont {Umeda},
  \citenamefont {Chudo}, \citenamefont {Imai}, \citenamefont {Sato},\ and\
  \citenamefont {Saitoh}}]{umeda_temperature-variable_2023}%
  \BibitemOpen
  \bibfield  {author} {\bibinfo {author} {\bibfnamefont {M.}~\bibnamefont
  {Umeda}}, \bibinfo {author} {\bibfnamefont {H.}~\bibnamefont {Chudo}},
  \bibinfo {author} {\bibfnamefont {M.}~\bibnamefont {Imai}}, \bibinfo {author}
  {\bibfnamefont {N.}~\bibnamefont {Sato}},\ and\ \bibinfo {author}
  {\bibfnamefont {E.}~\bibnamefont {Saitoh}},\ }\href
  {https://doi.org/10.1063/5.0142318} {\bibfield  {journal} {\bibinfo
  {journal} {Review of Scientific Instruments}\ }\textbf {\bibinfo {volume}
  {94}},\ \bibinfo {pages} {063906} (\bibinfo {year} {2023})}\BibitemShut
  {NoStop}%
\bibitem [{\citenamefont {Rusconi}\ \emph {et~al.}(2017)\citenamefont
  {Rusconi}, \citenamefont {P\"ochhacker}, \citenamefont {Kustura},
  \citenamefont {Cirac},\ and\ \citenamefont
  {Romero-Isart}}]{rusconi_quantum_2017}%
  \BibitemOpen
  \bibfield  {author} {\bibinfo {author} {\bibfnamefont {C.~C.}\ \bibnamefont
  {Rusconi}}, \bibinfo {author} {\bibfnamefont {V.}~\bibnamefont
  {P\"ochhacker}}, \bibinfo {author} {\bibfnamefont {K.}~\bibnamefont
  {Kustura}}, \bibinfo {author} {\bibfnamefont {J.~I.}\ \bibnamefont {Cirac}},\
  and\ \bibinfo {author} {\bibfnamefont {O.}~\bibnamefont {Romero-Isart}},\
  }\href {https://doi.org/10.1103/PhysRevLett.119.167202} {\bibfield  {journal}
  {\bibinfo  {journal} {Phys. Rev. Lett.}\ }\textbf {\bibinfo {volume} {119}},\
  \bibinfo {pages} {167202} (\bibinfo {year} {2017})}\BibitemShut {NoStop}%
\bibitem [{\citenamefont {Kubo}(1966)}]{kubo_fluctuation-dissipation_1966}%
  \BibitemOpen
  \bibfield  {author} {\bibinfo {author} {\bibfnamefont {R.}~\bibnamefont
  {Kubo}},\ }\href@noop {} {\bibfield  {journal} {\bibinfo  {journal} {Reports
  on Progress in Physics}\ }\textbf {\bibinfo {volume} {255}},\ \bibinfo
  {pages} {31} (\bibinfo {year} {1966})}\BibitemShut {NoStop}%
\bibitem [{\citenamefont {Purcell}(1977)}]{purcell_life_1977}%
  \BibitemOpen
  \bibfield  {author} {\bibinfo {author} {\bibfnamefont {E.~M.}\ \bibnamefont
  {Purcell}},\ }\href {https://doi.org/10.1119/1.10903} {\bibfield  {journal}
  {\bibinfo  {journal} {American Journal of Physics}\ }\textbf {\bibinfo
  {volume} {45}},\ \bibinfo {pages} {3} (\bibinfo {year} {1977})}\BibitemShut
  {NoStop}%
\bibitem [{\citenamefont {Community}(2018)}]{Blender2018}%
  \BibitemOpen
  \bibfield  {author} {\bibinfo {author} {\bibfnamefont {B.~O.}\ \bibnamefont
  {Community}},\ }\href {http://www.blender.org} {\emph {\bibinfo {title}
  {Blender - a 3D modelling and rendering package}}},\ \bibinfo {organization}
  {Blender Foundation},\ \bibinfo {address} {Stichting Blender Foundation,
  Amsterdam} (\bibinfo {year} {2018})\BibitemShut {NoStop}%
\bibitem [{\citenamefont {Edwards}\ \emph {et~al.}(2017)\citenamefont
  {Edwards}, \citenamefont {Riffe}, \citenamefont {Ji},\ and\ \citenamefont
  {Booth}}]{edwards_interactions_2017}%
  \BibitemOpen
  \bibfield  {author} {\bibinfo {author} {\bibfnamefont {B.~F.}\ \bibnamefont
  {Edwards}}, \bibinfo {author} {\bibfnamefont {D.~M.}\ \bibnamefont {Riffe}},
  \bibinfo {author} {\bibfnamefont {J.-Y.}\ \bibnamefont {Ji}},\ and\ \bibinfo
  {author} {\bibfnamefont {W.~A.}\ \bibnamefont {Booth}},\ }\href
  {https://doi.org/10.1119/1.4973409} {\bibfield  {journal} {\bibinfo
  {journal} {American Journal of Physics}\ }\textbf {\bibinfo {volume} {85}},\
  \bibinfo {pages} {130} (\bibinfo {year} {2017})}\BibitemShut {NoStop}%
\bibitem [{\citenamefont {Griffiths}(2013)}]{griffiths_EM}%
  \BibitemOpen
  \bibfield  {author} {\bibinfo {author} {\bibfnamefont {D.~J.}\ \bibnamefont
  {Griffiths}},\ }\href@noop {} {\emph {\bibinfo {title} {Introduction to
  electrodynamics}}},\ \bibinfo {edition} {fourth edition}\ ed.\ (\bibinfo
  {publisher} {Pearson},\ \bibinfo {address} {Boston},\ \bibinfo {year}
  {2013})\BibitemShut {NoStop}%
\bibitem [{\citenamefont {Griffiths}(1992)}]{griffiths_dipoles_1992}%
  \BibitemOpen
  \bibfield  {author} {\bibinfo {author} {\bibfnamefont {D.~J.}\ \bibnamefont
  {Griffiths}},\ }\href {https://doi.org/10.1119/1.17001} {\bibfield  {journal}
  {\bibinfo  {journal} {American Journal of Physics}\ }\textbf {\bibinfo
  {volume} {60}},\ \bibinfo {pages} {979} (\bibinfo {year} {1992})}\BibitemShut
  {NoStop}%
\bibitem [{\citenamefont {Bedanta}\ \emph {et~al.}(2015)\citenamefont
  {Bedanta}, \citenamefont {Petracic},\ and\ \citenamefont
  {Kleemann}}]{bedanta_supermagnetism_2015}%
  \BibitemOpen
  \bibfield  {author} {\bibinfo {author} {\bibfnamefont {S.}~\bibnamefont
  {Bedanta}}, \bibinfo {author} {\bibfnamefont {O.}~\bibnamefont {Petracic}},\
  and\ \bibinfo {author} {\bibfnamefont {W.}~\bibnamefont {Kleemann}},\ }in\
  \href {https://doi.org/10.1016/B978-0-444-63528-0.00001-2} {\emph {\bibinfo
  {booktitle} {Handbook of {Magnetic} {Materials}}}},\ Vol.~\bibinfo {volume}
  {23}\ (\bibinfo  {publisher} {Elsevier},\ \bibinfo {year} {2015})\ pp.\
  \bibinfo {pages} {1--83}\BibitemShut {NoStop}%
\bibitem [{\citenamefont {Landau}\ and\ \citenamefont
  {Lifshitz}(1960)}]{landau_vol8_1960}%
  \BibitemOpen
  \bibfield  {author} {\bibinfo {author} {\bibfnamefont {L.~D.}\ \bibnamefont
  {Landau}}\ and\ \bibinfo {author} {\bibfnamefont {E.~M.}\ \bibnamefont
  {Lifshitz}},\ }\href@noop {} {\emph {\bibinfo {title} {Vol.8 -
  {Electrodynamics} {Of} {Continuous} {Media}}}}\ (\bibinfo {year}
  {1960})\BibitemShut {NoStop}%
\bibitem [{\citenamefont {Chikazumi}\ \emph {et~al.}(1997)\citenamefont
  {Chikazumi}, \citenamefont {Chikazumi},\ and\ \citenamefont
  {Graham}}]{chikazumi1997physics}%
  \BibitemOpen
  \bibfield  {author} {\bibinfo {author} {\bibfnamefont {S.}~\bibnamefont
  {Chikazumi}}, \bibinfo {author} {\bibfnamefont {S.}~\bibnamefont
  {Chikazumi}},\ and\ \bibinfo {author} {\bibfnamefont {C.~D.}\ \bibnamefont
  {Graham}},\ }\href@noop {} {\emph {\bibinfo {title} {Physics of
  ferromagnetism}}},\ \bibinfo {number} {94}\ (\bibinfo  {publisher} {Oxford
  University Press},\ \bibinfo {year} {1997})\BibitemShut {NoStop}%
\bibitem [{\citenamefont {Cullity}\ and\ \citenamefont
  {Graham}(2011)}]{cullity2011introduction}%
  \BibitemOpen
  \bibfield  {author} {\bibinfo {author} {\bibfnamefont {B.~D.}\ \bibnamefont
  {Cullity}}\ and\ \bibinfo {author} {\bibfnamefont {C.~D.}\ \bibnamefont
  {Graham}},\ }\href@noop {} {\emph {\bibinfo {title} {Introduction to magnetic
  materials}}}\ (\bibinfo  {publisher} {John Wiley \& Sons},\ \bibinfo {year}
  {2011})\BibitemShut {NoStop}%
\bibitem [{\citenamefont {{Grifftihs, D. J.}}(2005)}]{grifftihs_QM}%
  \BibitemOpen
  \bibfield  {author} {\bibinfo {author} {\bibnamefont {{Grifftihs, D. J.}}},\
  }\href@noop {} {\emph {\bibinfo {title} {Introduction to quantum
  mechanics}}},\ \bibinfo {edition} {second edition}\ ed.\ (\bibinfo
  {publisher} {Pearson},\ \bibinfo {year} {2005})\BibitemShut {NoStop}%
\bibitem [{\citenamefont {Foot}(2004)}]{foot2004atomic}%
  \BibitemOpen
  \bibfield  {author} {\bibinfo {author} {\bibfnamefont {C.~J.}\ \bibnamefont
  {Foot}},\ }\href@noop {} {\emph {\bibinfo {title} {Atomic physics}}},\
  Vol.~\bibinfo {volume} {7}\ (\bibinfo  {publisher} {OUP Oxford},\ \bibinfo
  {year} {2004})\BibitemShut {NoStop}%
\bibitem [{\citenamefont {Landecker}\ \emph {et~al.}(1970)\citenamefont
  {Landecker}, \citenamefont {Villani},\ and\ \citenamefont
  {Yung}}]{landecker1970analytic}%
  \BibitemOpen
  \bibfield  {author} {\bibinfo {author} {\bibfnamefont {P.~B.}\ \bibnamefont
  {Landecker}}, \bibinfo {author} {\bibfnamefont {D.~D.}\ \bibnamefont
  {Villani}},\ and\ \bibinfo {author} {\bibfnamefont {K.~W.}\ \bibnamefont
  {Yung}},\ }\href@noop {} {\bibfield  {journal} {\bibinfo  {journal} {Magnetic
  and Electrical Separation}\ }\textbf {\bibinfo {volume} {10}} (\bibinfo
  {year} {1970})}\BibitemShut {NoStop}%
\bibitem [{\citenamefont {Barnett}(1915)}]{barnett_magnetization_1915}%
  \BibitemOpen
  \bibfield  {author} {\bibinfo {author} {\bibfnamefont {S.~J.}\ \bibnamefont
  {Barnett}},\ }\href {http://www.jstor.org/stable/1641123} {\bibfield
  {journal} {\bibinfo  {journal} {Science}\ }\textbf {\bibinfo {volume} {42}},\
  \bibinfo {pages} {163} (\bibinfo {year} {1915})}\BibitemShut {NoStop}%
\bibitem [{\citenamefont {Rubinow}\ and\ \citenamefont
  {Keller}(1961)}]{rubinow_transverse_1961}%
  \BibitemOpen
  \bibfield  {author} {\bibinfo {author} {\bibfnamefont {S.~I.}\ \bibnamefont
  {Rubinow}}\ and\ \bibinfo {author} {\bibfnamefont {J.~B.}\ \bibnamefont
  {Keller}},\ }\href {https://doi.org/10.1017/S0022112061000640} {\bibfield
  {journal} {\bibinfo  {journal} {Journal of Fluid Mechanics}\ }\textbf
  {\bibinfo {volume} {11}},\ \bibinfo {pages} {447} (\bibinfo {year} {1961})},\
  \bibinfo {note} {publisher: Cambridge University Press}\BibitemShut {NoStop}%
\bibitem [{\citenamefont {Brown}(1963)}]{brown_thermal_1963}%
  \BibitemOpen
  \bibfield  {author} {\bibinfo {author} {\bibfnamefont {W.~F.}\ \bibnamefont
  {Brown}},\ }\href {https://doi.org/10.1103/PhysRev.130.1677} {\bibfield
  {journal} {\bibinfo  {journal} {Physical Review}\ }\textbf {\bibinfo {volume}
  {130}},\ \bibinfo {pages} {1677} (\bibinfo {year} {1963})}\BibitemShut
  {NoStop}%
\bibitem [{\citenamefont {Blundell}\ and\ \citenamefont
  {Blundell}(2010)}]{blundell2010concepts}%
  \BibitemOpen
  \bibfield  {author} {\bibinfo {author} {\bibfnamefont {S.~J.}\ \bibnamefont
  {Blundell}}\ and\ \bibinfo {author} {\bibfnamefont {K.~M.}\ \bibnamefont
  {Blundell}},\ }\href@noop {} {\emph {\bibinfo {title} {Concepts in thermal
  physics}}}\ (\bibinfo  {publisher} {Oxford University Press on Demand},\
  \bibinfo {year} {2010})\BibitemShut {NoStop}%
\bibitem [{\citenamefont {Kamenev}(2011)}]{kamenev_field_2011}%
  \BibitemOpen
  \bibfield  {author} {\bibinfo {author} {\bibfnamefont {A.}~\bibnamefont
  {Kamenev}},\ }\href {https://doi.org/10.1017/CBO9781139003667} {\emph
  {\bibinfo {title} {Field {Theory} of {Non}-{Equilibrium} {Systems}}}}\
  (\bibinfo  {publisher} {Cambridge University Press},\ \bibinfo {address}
  {Cambridge},\ \bibinfo {year} {2011})\BibitemShut {NoStop}%
\bibitem [{\citenamefont {Han}\ \emph {et~al.}(2006)\citenamefont {Han},
  \citenamefont {Alsayed}, \citenamefont {Nobili}, \citenamefont {Zhang},
  \citenamefont {Lubensky},\ and\ \citenamefont {Yodh}}]{han_brownian_2006}%
  \BibitemOpen
  \bibfield  {author} {\bibinfo {author} {\bibfnamefont {Y.}~\bibnamefont
  {Han}}, \bibinfo {author} {\bibfnamefont {A.~M.}\ \bibnamefont {Alsayed}},
  \bibinfo {author} {\bibfnamefont {M.}~\bibnamefont {Nobili}}, \bibinfo
  {author} {\bibfnamefont {J.}~\bibnamefont {Zhang}}, \bibinfo {author}
  {\bibfnamefont {T.~C.}\ \bibnamefont {Lubensky}},\ and\ \bibinfo {author}
  {\bibfnamefont {A.~G.}\ \bibnamefont {Yodh}},\ }\href@noop {} {\bibfield
  {journal} {\bibinfo  {journal} {Science}\ }\textbf {\bibinfo {volume}
  {314}},\ \bibinfo {pages} {6} (\bibinfo {year} {2006})}\BibitemShut {NoStop}%
\bibitem [{\citenamefont {Happel}\ and\ \citenamefont
  {Brenner}(1981)}]{happel_low_1981}%
  \BibitemOpen
  \bibfield  {author} {\bibinfo {author} {\bibfnamefont {J.}~\bibnamefont
  {Happel}}\ and\ \bibinfo {author} {\bibfnamefont {H.}~\bibnamefont
  {Brenner}},\ }\href {https://doi.org/10.1007/978-94-009-8352-6} {\emph
  {\bibinfo {title} {Low {{Reynolds}} Number Hydrodynamics}}},\ edited by\
  \bibinfo {editor} {\bibfnamefont {R.~J.}\ \bibnamefont {Moreau}},\ \bibinfo
  {series} {Mechanics of Fluids and Transport Processes}, Vol.~\bibinfo
  {volume} {1}\ (\bibinfo  {publisher} {{Springer Netherlands}},\ \bibinfo
  {address} {{Dordrecht}},\ \bibinfo {year} {1981})\BibitemShut {NoStop}%
\bibitem [{\citenamefont {Babson}\ \emph {et~al.}(2009)\citenamefont {Babson},
  \citenamefont {Reynolds}, \citenamefont {Bjorkquist},\ and\ \citenamefont
  {Griffiths}}]{babson_hidden_2009}%
  \BibitemOpen
  \bibfield  {author} {\bibinfo {author} {\bibfnamefont {D.}~\bibnamefont
  {Babson}}, \bibinfo {author} {\bibfnamefont {S.~P.}\ \bibnamefont
  {Reynolds}}, \bibinfo {author} {\bibfnamefont {R.}~\bibnamefont
  {Bjorkquist}},\ and\ \bibinfo {author} {\bibfnamefont {D.~J.}\ \bibnamefont
  {Griffiths}},\ }\href {https://doi.org/10.1119/1.3152712} {\bibfield
  {journal} {\bibinfo  {journal} {American Journal of Physics}\ }\textbf
  {\bibinfo {volume} {77}},\ \bibinfo {pages} {826} (\bibinfo {year}
  {2009})}\BibitemShut {NoStop}%
\bibitem [{\citenamefont {Romer}(1966)}]{romer_angular_1966}%
  \BibitemOpen
  \bibfield  {author} {\bibinfo {author} {\bibfnamefont {R.~H.}\ \bibnamefont
  {Romer}},\ }\href {https://doi.org/10.1119/1.1973478} {\bibfield  {journal}
  {\bibinfo  {journal} {American Journal of Physics}\ }\textbf {\bibinfo
  {volume} {34}},\ \bibinfo {pages} {772} (\bibinfo {year} {1966})}\BibitemShut
  {NoStop}%
\bibitem [{\citenamefont {Hnizdo}(1992)}]{hnizdo_conservation_1992}%
  \BibitemOpen
  \bibfield  {author} {\bibinfo {author} {\bibfnamefont {V.}~\bibnamefont
  {Hnizdo}},\ }\href {https://doi.org/10.1119/1.16902} {\bibfield  {journal}
  {\bibinfo  {journal} {American Journal of Physics}\ }\textbf {\bibinfo
  {volume} {60}},\ \bibinfo {pages} {242} (\bibinfo {year} {1992})}\BibitemShut
  {NoStop}%
\bibitem [{\citenamefont {Christiansen}\ \emph {et~al.}(2022)\citenamefont
  {Christiansen}, \citenamefont {Mirkhani}, \citenamefont {Hornslien},\ and\
  \citenamefont {Schuerle}}]{christiansen_theoretical_2022}%
  \BibitemOpen
  \bibfield  {author} {\bibinfo {author} {\bibfnamefont {M.~G.}\ \bibnamefont
  {Christiansen}}, \bibinfo {author} {\bibfnamefont {N.}~\bibnamefont
  {Mirkhani}}, \bibinfo {author} {\bibfnamefont {W.}~\bibnamefont
  {Hornslien}},\ and\ \bibinfo {author} {\bibfnamefont {S.}~\bibnamefont
  {Schuerle}},\ }\href {https://doi.org/10.1063/5.0102153} {\bibfield
  {journal} {\bibinfo  {journal} {Journal of Applied Physics}\ }\textbf
  {\bibinfo {volume} {132}},\ \bibinfo {pages} {174304} (\bibinfo {year}
  {2022})}\BibitemShut {NoStop}%
\bibitem [{\citenamefont {Jackson}(1999)}]{jackson_classical_1999}%
  \BibitemOpen
  \bibfield  {author} {\bibinfo {author} {\bibfnamefont {J.~D.}\ \bibnamefont
  {Jackson}},\ }\href@noop {} {\emph {\bibinfo {title} {Classical
  {Electrodynamics}.pdf}}},\ \bibinfo {edition} {third edition}\ ed.\ (\bibinfo
   {publisher} {American Association of Physics Teachers},\ \bibinfo {year}
  {1999})\BibitemShut {NoStop}%
\bibitem [{\citenamefont {Sharma}(1988)}]{sharma_field_1988}%
  \BibitemOpen
  \bibfield  {author} {\bibinfo {author} {\bibfnamefont {N.~L.}\ \bibnamefont
  {Sharma}},\ }\href {https://doi.org/10.1119/1.15592} {\bibfield  {journal}
  {\bibinfo  {journal} {American Journal of Physics}\ }\textbf {\bibinfo
  {volume} {56}},\ \bibinfo {pages} {420} (\bibinfo {year} {1988})}\BibitemShut
  {NoStop}%
\bibitem [{\citenamefont {Streib}\ \emph {et~al.}(2019)\citenamefont {Streib},
  \citenamefont {Vidal-Silva}, \citenamefont {Shen},\ and\ \citenamefont
  {Bauer}}]{streib_magnon-phonon_2019}%
  \BibitemOpen
  \bibfield  {author} {\bibinfo {author} {\bibfnamefont {S.}~\bibnamefont
  {Streib}}, \bibinfo {author} {\bibfnamefont {N.}~\bibnamefont {Vidal-Silva}},
  \bibinfo {author} {\bibfnamefont {K.}~\bibnamefont {Shen}},\ and\ \bibinfo
  {author} {\bibfnamefont {G.~E.~W.}\ \bibnamefont {Bauer}},\ }\href
  {https://doi.org/10.1103/PhysRevB.99.184442} {\bibfield  {journal} {\bibinfo
  {journal} {Physical Review B}\ }\textbf {\bibinfo {volume} {99}},\ \bibinfo
  {pages} {184442} (\bibinfo {year} {2019})}\BibitemShut {NoStop}%
\bibitem [{\citenamefont {Simensen}\ \emph {et~al.}(2020)\citenamefont
  {Simensen}, \citenamefont {Kamra}, \citenamefont {Troncoso},\ and\
  \citenamefont {Brataas}}]{simensen_magnon_2020}%
  \BibitemOpen
  \bibfield  {author} {\bibinfo {author} {\bibfnamefont {H.~T.}\ \bibnamefont
  {Simensen}}, \bibinfo {author} {\bibfnamefont {A.}~\bibnamefont {Kamra}},
  \bibinfo {author} {\bibfnamefont {R.~E.}\ \bibnamefont {Troncoso}},\ and\
  \bibinfo {author} {\bibfnamefont {A.}~\bibnamefont {Brataas}},\ }\href
  {https://doi.org/10.1103/PhysRevB.101.020403} {\bibfield  {journal} {\bibinfo
   {journal} {Physical Review B}\ }\textbf {\bibinfo {volume} {101}},\ \bibinfo
  {pages} {020403} (\bibinfo {year} {2020})}\BibitemShut {NoStop}%
\bibitem [{\citenamefont {HILL}\ and\ \citenamefont
  {POWER}(1956)}]{hill_extremum_1956}%
  \BibitemOpen
  \bibfield  {author} {\bibinfo {author} {\bibfnamefont {R.}~\bibnamefont
  {HILL}}\ and\ \bibinfo {author} {\bibfnamefont {G.}~\bibnamefont {POWER}},\
  }\href {https://doi.org/10.1093/qjmam/9.3.313} {\bibfield  {journal}
  {\bibinfo  {journal} {The Quarterly Journal of Mechanics and Applied
  Mathematics}\ }\textbf {\bibinfo {volume} {9}},\ \bibinfo {pages} {313}
  (\bibinfo {year} {1956})}\BibitemShut {NoStop}%
\bibitem [{\citenamefont {Kim}\ and\ \citenamefont
  {Karrila}(1991)}]{kim_microhydrodynamics_1991}%
  \BibitemOpen
  \bibfield  {author} {\bibinfo {author} {\bibfnamefont {S.}~\bibnamefont
  {Kim}}\ and\ \bibinfo {author} {\bibfnamefont {S.~J.}\ \bibnamefont
  {Karrila}},\ }\href@noop {} {\emph {\bibinfo {title} {Microhydrodynamics:
  Principles and Selected Applications}}},\ Butterworth-{{Heinemann}} Series in
  Chemical Engineering\ (\bibinfo  {publisher} {{Butterworth-Heinemann}},\
  \bibinfo {address} {{Boston}},\ \bibinfo {year} {1991})\BibitemShut {NoStop}%
\bibitem [{\citenamefont {Singamaneni}\ \emph {et~al.}(2011)\citenamefont
  {Singamaneni}, \citenamefont {Bliznyuk}, \citenamefont {Binek},\ and\
  \citenamefont {Tsymbal}}]{singamaneni_magnetic_2011}%
  \BibitemOpen
  \bibfield  {author} {\bibinfo {author} {\bibfnamefont {S.}~\bibnamefont
  {Singamaneni}}, \bibinfo {author} {\bibfnamefont {V.~N.}\ \bibnamefont
  {Bliznyuk}}, \bibinfo {author} {\bibfnamefont {C.}~\bibnamefont {Binek}},\
  and\ \bibinfo {author} {\bibfnamefont {E.~Y.}\ \bibnamefont {Tsymbal}},\
  }\href {https://doi.org/10.1039/c1jm11845e} {\bibfield  {journal} {\bibinfo
  {journal} {Journal of Materials Chemistry}\ }\textbf {\bibinfo {volume}
  {21}},\ \bibinfo {pages} {16819} (\bibinfo {year} {2011})}\BibitemShut
  {NoStop}%
\bibitem [{\citenamefont {Michels}(2021)}]{michels_magnetic_2021}%
  \BibitemOpen
  \bibfield  {author} {\bibinfo {author} {\bibfnamefont {A.}~\bibnamefont
  {Michels}},\ }\href@noop {} {\emph {\bibinfo {title} {Magnetic Small-Angle
  Neutron Scattering: A Probe for Mesoscale Magnetism Analysis}}},\
  Vol.~\bibinfo {volume} {16}\ (\bibinfo  {publisher} {Oxford University
  Press},\ \bibinfo {year} {2021})\BibitemShut {NoStop}%
\bibitem [{\citenamefont {Disch}\ \emph {et~al.}(2012)\citenamefont {Disch},
  \citenamefont {Wetterskog}, \citenamefont {Hermann}, \citenamefont
  {Wiedenmann}, \citenamefont {Vainio}, \citenamefont {{Salazar-Alvarez}},
  \citenamefont {Bergstr{\"o}m},\ and\ \citenamefont
  {Br{\"u}ckel}}]{disch_quantitative_2012}%
  \BibitemOpen
  \bibfield  {author} {\bibinfo {author} {\bibfnamefont {S.}~\bibnamefont
  {Disch}}, \bibinfo {author} {\bibfnamefont {E.}~\bibnamefont {Wetterskog}},
  \bibinfo {author} {\bibfnamefont {R.~P.}\ \bibnamefont {Hermann}}, \bibinfo
  {author} {\bibfnamefont {A.}~\bibnamefont {Wiedenmann}}, \bibinfo {author}
  {\bibfnamefont {U.}~\bibnamefont {Vainio}}, \bibinfo {author} {\bibfnamefont
  {G.}~\bibnamefont {{Salazar-Alvarez}}}, \bibinfo {author} {\bibfnamefont
  {L.}~\bibnamefont {Bergstr{\"o}m}},\ and\ \bibinfo {author} {\bibfnamefont
  {T.}~\bibnamefont {Br{\"u}ckel}},\ }\href
  {https://doi.org/10.1088/1367-2630/14/1/013025} {\bibfield  {journal}
  {\bibinfo  {journal} {New Journal of Physics}\ }\textbf {\bibinfo {volume}
  {14}},\ \bibinfo {pages} {013025} (\bibinfo {year} {2012})}\BibitemShut
  {NoStop}%
\bibitem [{\citenamefont {B{\o}dker}\ \emph {et~al.}(1994)\citenamefont
  {B{\o}dker}, \citenamefont {M{\o}rup},\ and\ \citenamefont
  {Linderoth}}]{bodker_surface_1994}%
  \BibitemOpen
  \bibfield  {author} {\bibinfo {author} {\bibfnamefont {F.}~\bibnamefont
  {B{\o}dker}}, \bibinfo {author} {\bibfnamefont {S.}~\bibnamefont
  {M{\o}rup}},\ and\ \bibinfo {author} {\bibfnamefont {S.}~\bibnamefont
  {Linderoth}},\ }\href {https://doi.org/10.1103/PhysRevLett.72.282} {\bibfield
   {journal} {\bibinfo  {journal} {Physical Review Letters}\ }\textbf {\bibinfo
  {volume} {72}},\ \bibinfo {pages} {282} (\bibinfo {year} {1994})}\BibitemShut
  {NoStop}%
\bibitem [{\citenamefont {Pisane}\ \emph {et~al.}(2017)\citenamefont {Pisane},
  \citenamefont {Singh},\ and\ \citenamefont {Seehra}}]{pisane_unusual_2017}%
  \BibitemOpen
  \bibfield  {author} {\bibinfo {author} {\bibfnamefont {K.~L.}\ \bibnamefont
  {Pisane}}, \bibinfo {author} {\bibfnamefont {S.}~\bibnamefont {Singh}},\ and\
  \bibinfo {author} {\bibfnamefont {M.~S.}\ \bibnamefont {Seehra}},\ }\href
  {https://doi.org/10.1063/1.4984903} {\bibfield  {journal} {\bibinfo
  {journal} {Applied Physics Letters}\ }\textbf {\bibinfo {volume} {110}},\
  \bibinfo {pages} {222409} (\bibinfo {year} {2017})}\BibitemShut {NoStop}%
\bibitem [{\citenamefont {Bhagat}\ and\ \citenamefont
  {Lubitz}(1974)}]{bhagat_temperature_1974}%
  \BibitemOpen
  \bibfield  {author} {\bibinfo {author} {\bibfnamefont {S.~M.}\ \bibnamefont
  {Bhagat}}\ and\ \bibinfo {author} {\bibfnamefont {P.}~\bibnamefont
  {Lubitz}},\ }\href@noop {} {\bibfield  {journal} {\bibinfo  {journal}
  {PHYSICAL REVIEW B}\ ,\ \bibinfo {pages} {7}} (\bibinfo {year}
  {1974})}\BibitemShut {NoStop}%
\bibitem [{\citenamefont {Gilmore}\ \emph {et~al.}(2007)\citenamefont
  {Gilmore}, \citenamefont {Idzerda},\ and\ \citenamefont
  {Stiles}}]{gilmore_identification_2007}%
  \BibitemOpen
  \bibfield  {author} {\bibinfo {author} {\bibfnamefont {K.}~\bibnamefont
  {Gilmore}}, \bibinfo {author} {\bibfnamefont {Y.~U.}\ \bibnamefont
  {Idzerda}},\ and\ \bibinfo {author} {\bibfnamefont {M.~D.}\ \bibnamefont
  {Stiles}},\ }\href {https://doi.org/10.1103/PhysRevLett.99.027204} {\bibfield
   {journal} {\bibinfo  {journal} {Physical Review Letters}\ }\textbf {\bibinfo
  {volume} {99}},\ \bibinfo {pages} {027204} (\bibinfo {year}
  {2007})}\BibitemShut {NoStop}%
\bibitem [{\citenamefont {Bailey}\ \emph {et~al.}(2001)\citenamefont {Bailey},
  \citenamefont {Kabos}, \citenamefont {Mancoff},\ and\ \citenamefont
  {Russek}}]{bailey_control_2001}%
  \BibitemOpen
  \bibfield  {author} {\bibinfo {author} {\bibfnamefont {W.}~\bibnamefont
  {Bailey}}, \bibinfo {author} {\bibfnamefont {P.}~\bibnamefont {Kabos}},
  \bibinfo {author} {\bibfnamefont {F.}~\bibnamefont {Mancoff}},\ and\ \bibinfo
  {author} {\bibfnamefont {S.}~\bibnamefont {Russek}},\ }\href@noop {}
  {\bibfield  {journal} {\bibinfo  {journal} {IEEE TRANSACTIONS ON MAGNETICS}\
  }\textbf {\bibinfo {volume} {37}},\ \bibinfo {pages} {6} (\bibinfo {year}
  {2001})}\BibitemShut {NoStop}%
\bibitem [{\citenamefont {Klingler}\ \emph {et~al.}(2017)\citenamefont
  {Klingler}, \citenamefont {{Maier-Flaig}},\ and\ \citenamefont
  {Dubs}}]{klingler_gilbert_2017}%
  \BibitemOpen
  \bibfield  {author} {\bibinfo {author} {\bibfnamefont {S.}~\bibnamefont
  {Klingler}}, \bibinfo {author} {\bibfnamefont {H.}~\bibnamefont
  {{Maier-Flaig}}},\ and\ \bibinfo {author} {\bibfnamefont {C.}~\bibnamefont
  {Dubs}},\ }\href@noop {} {\bibfield  {journal} {\bibinfo  {journal} {Appl.
  Phys. Lett.}\ ,\ \bibinfo {pages} {6}} (\bibinfo {year} {2017})}\BibitemShut
  {NoStop}%
\bibitem [{\citenamefont {L{\"o}wen}(2020)}]{lowen_inertial_2020}%
  \BibitemOpen
  \bibfield  {author} {\bibinfo {author} {\bibfnamefont {H.}~\bibnamefont
  {L{\"o}wen}},\ }\href {https://doi.org/10.1063/1.5134455} {\bibfield
  {journal} {\bibinfo  {journal} {The Journal of Chemical Physics}\ }\textbf
  {\bibinfo {volume} {152}},\ \bibinfo {pages} {040901} (\bibinfo {year}
  {2020})}\BibitemShut {NoStop}%
\bibitem [{\citenamefont {Usov}\ \emph {et~al.}(2019)\citenamefont {Usov},
  \citenamefont {Rytov},\ and\ \citenamefont {Bautin}}]{usov_dynamics_2019}%
  \BibitemOpen
  \bibfield  {author} {\bibinfo {author} {\bibfnamefont {N.~A.}\ \bibnamefont
  {Usov}}, \bibinfo {author} {\bibfnamefont {R.~A.}\ \bibnamefont {Rytov}},\
  and\ \bibinfo {author} {\bibfnamefont {V.~A.}\ \bibnamefont {Bautin}},\
  }\href {https://doi.org/10.3762/bjnano.10.221} {\bibfield  {journal}
  {\bibinfo  {journal} {Beilstein Journal of Nanotechnology}\ }\textbf
  {\bibinfo {volume} {10}},\ \bibinfo {pages} {2294} (\bibinfo {year}
  {2019})}\BibitemShut {NoStop}%
\bibitem [{\citenamefont {Coffey}\ and\ \citenamefont
  {Kalmykov}(1996)}]{coffey_inertial_1996}%
  \BibitemOpen
  \bibfield  {author} {\bibinfo {author} {\bibfnamefont {W.}~\bibnamefont
  {Coffey}}\ and\ \bibinfo {author} {\bibfnamefont {{\relax
  Yu.P}.}~\bibnamefont {Kalmykov}},\ }\href
  {https://doi.org/10.1016/S0304-8853(96)00390-3} {\bibfield  {journal}
  {\bibinfo  {journal} {Journal of Magnetism and Magnetic Materials}\ }\textbf
  {\bibinfo {volume} {164}},\ \bibinfo {pages} {133} (\bibinfo {year}
  {1996})}\BibitemShut {NoStop}%
\bibitem [{\citenamefont {Cheng}\ and\ \citenamefont
  {Gupta}(1989)}]{cheng_historical_1989}%
  \BibitemOpen
  \bibfield  {author} {\bibinfo {author} {\bibfnamefont {H.}~\bibnamefont
  {Cheng}}\ and\ \bibinfo {author} {\bibfnamefont {K.~C.}\ \bibnamefont
  {Gupta}},\ }\href {https://doi.org/10.1115/1.3176034} {\bibfield  {journal}
  {\bibinfo  {journal} {Journal of Applied Mechanics}\ }\textbf {\bibinfo
  {volume} {56}},\ \bibinfo {pages} {139} (\bibinfo {year} {1989})}\BibitemShut
  {NoStop}%
\bibitem [{\citenamefont {Young}(2014)}]{young2014leapfrog}%
  \BibitemOpen
  \bibfield  {author} {\bibinfo {author} {\bibfnamefont {P.}~\bibnamefont
  {Young}},\ }\href@noop {} {\bibfield  {journal} {\bibinfo  {journal} {Lecture
  notes in University of california, santa cruz}\ } (\bibinfo {year}
  {2014})}\BibitemShut {NoStop}%
\bibitem [{\citenamefont {Gardiner}(2002)}]{gardiner_handbook_2002}%
  \BibitemOpen
  \bibfield  {author} {\bibinfo {author} {\bibfnamefont {C.~W.}\ \bibnamefont
  {Gardiner}},\ }\href@noop {} {\emph {\bibinfo {title} {Handbook of Stochastic
  Methods: For Physics, Chemistry and the Natural Sciences}}},\ \bibinfo
  {edition} {study ed., 2. ed., 6. print}\ ed.,\ \bibinfo {series} {Springer
  Series in Synergetics}\ No.~\bibinfo {number} {13}\ (\bibinfo  {publisher}
  {{Springer}},\ \bibinfo {address} {{Berlin Heidelberg}},\ \bibinfo {year}
  {2002})\BibitemShut {NoStop}%
\bibitem [{\citenamefont {S{\"a}rkk{\"a}}\ and\ \citenamefont
  {Solin}(2019)}]{sarkka_applied_2019}%
  \BibitemOpen
  \bibfield  {author} {\bibinfo {author} {\bibfnamefont {S.}~\bibnamefont
  {S{\"a}rkk{\"a}}}\ and\ \bibinfo {author} {\bibfnamefont {A.}~\bibnamefont
  {Solin}},\ }\href {https://doi.org/10.1017/9781108186735} {\emph {\bibinfo
  {title} {Applied {{Stochastic Differential Equations}}}}},\ \bibinfo
  {edition} {1st}\ ed.\ (\bibinfo  {publisher} {{Cambridge University Press}},\
  \bibinfo {year} {2019})\BibitemShut {NoStop}%
\bibitem [{\citenamefont {{\O}ksendal}(1998)}]{oksendal_stochastic_1998}%
  \BibitemOpen
  \bibfield  {author} {\bibinfo {author} {\bibfnamefont {B.~K.}\ \bibnamefont
  {{\O}ksendal}},\ }\href@noop {} {\emph {\bibinfo {title} {Stochastic
  Differential Equations: An Introduction with Applications}}},\ \bibinfo
  {edition} {5th}\ ed.,\ Universitext\ (\bibinfo  {publisher} {{Springer}},\
  \bibinfo {address} {{Berlin ; New York}},\ \bibinfo {year}
  {1998})\BibitemShut {NoStop}%
\bibitem [{\citenamefont {Pesce}\ \emph {et~al.}(2013)\citenamefont {Pesce},
  \citenamefont {McDaniel}, \citenamefont {Hottovy}, \citenamefont {Wehr},\
  and\ \citenamefont {Volpe}}]{pesce_stratonovich--ito_2013}%
  \BibitemOpen
  \bibfield  {author} {\bibinfo {author} {\bibfnamefont {G.}~\bibnamefont
  {Pesce}}, \bibinfo {author} {\bibfnamefont {A.}~\bibnamefont {McDaniel}},
  \bibinfo {author} {\bibfnamefont {S.}~\bibnamefont {Hottovy}}, \bibinfo
  {author} {\bibfnamefont {J.}~\bibnamefont {Wehr}},\ and\ \bibinfo {author}
  {\bibfnamefont {G.}~\bibnamefont {Volpe}},\ }\href
  {https://doi.org/10.1038/ncomms3733} {\bibfield  {journal} {\bibinfo
  {journal} {Nature Communications}\ }\textbf {\bibinfo {volume} {4}},\
  \bibinfo {pages} {2733} (\bibinfo {year} {2013})}\BibitemShut {NoStop}%
\bibitem [{\citenamefont {Berkov}\ and\ \citenamefont
  {Gorn}(2002)}]{berkov_thermally_2002}%
  \BibitemOpen
  \bibfield  {author} {\bibinfo {author} {\bibfnamefont {D.~V.}\ \bibnamefont
  {Berkov}}\ and\ \bibinfo {author} {\bibfnamefont {N.~L.}\ \bibnamefont
  {Gorn}},\ }\href {https://doi.org/10.1088/0953-8984/14/13/101} {\bibfield
  {journal} {\bibinfo  {journal} {Journal of Physics: Condensed Matter}\
  }\textbf {\bibinfo {volume} {14}},\ \bibinfo {pages} {L281} (\bibinfo {year}
  {2002})}\BibitemShut {NoStop}%
\bibitem [{\citenamefont {Kraft}\ \emph {et~al.}(2013)\citenamefont {Kraft},
  \citenamefont {Wittkowski}, \citenamefont {{ten Hagen}}, \citenamefont
  {Edmond}, \citenamefont {Pine},\ and\ \citenamefont
  {L{\"o}wen}}]{kraft_brownian_2013}%
  \BibitemOpen
  \bibfield  {author} {\bibinfo {author} {\bibfnamefont {D.~J.}\ \bibnamefont
  {Kraft}}, \bibinfo {author} {\bibfnamefont {R.}~\bibnamefont {Wittkowski}},
  \bibinfo {author} {\bibfnamefont {B.}~\bibnamefont {{ten Hagen}}}, \bibinfo
  {author} {\bibfnamefont {K.~V.}\ \bibnamefont {Edmond}}, \bibinfo {author}
  {\bibfnamefont {D.~J.}\ \bibnamefont {Pine}},\ and\ \bibinfo {author}
  {\bibfnamefont {H.}~\bibnamefont {L{\"o}wen}},\ }\href
  {https://doi.org/10.1103/PhysRevE.88.050301} {\bibfield  {journal} {\bibinfo
  {journal} {Physical Review E}\ }\textbf {\bibinfo {volume} {88}},\ \bibinfo
  {pages} {050301} (\bibinfo {year} {2013})}\BibitemShut {NoStop}%
\bibitem [{\citenamefont {Wittkowski}\ and\ \citenamefont
  {L{\"o}wen}(2012)}]{wittkowski_self-propelled_2012}%
  \BibitemOpen
  \bibfield  {author} {\bibinfo {author} {\bibfnamefont {R.}~\bibnamefont
  {Wittkowski}}\ and\ \bibinfo {author} {\bibfnamefont {H.}~\bibnamefont
  {L{\"o}wen}},\ }\href {https://doi.org/10.1103/PhysRevE.85.021406} {\bibfield
   {journal} {\bibinfo  {journal} {Physical Review E}\ }\textbf {\bibinfo
  {volume} {85}},\ \bibinfo {pages} {021406} (\bibinfo {year}
  {2012})}\BibitemShut {NoStop}%
\bibitem [{\citenamefont {Ermak}\ and\ \citenamefont
  {McCammon}(1978)}]{ermak_brownian_1978}%
  \BibitemOpen
  \bibfield  {author} {\bibinfo {author} {\bibfnamefont {D.~L.}\ \bibnamefont
  {Ermak}}\ and\ \bibinfo {author} {\bibfnamefont {J.~A.}\ \bibnamefont
  {McCammon}},\ }\href {https://doi.org/10.1063/1.436761} {\bibfield  {journal}
  {\bibinfo  {journal} {The Journal of Chemical Physics}\ }\textbf {\bibinfo
  {volume} {69}},\ \bibinfo {pages} {1352} (\bibinfo {year}
  {1978})}\BibitemShut {NoStop}%
\bibitem [{\citenamefont {Beleggia}\ \emph {et~al.}(2004)\citenamefont
  {Beleggia}, \citenamefont {Tandon}, \citenamefont {Zhu},\ and\ \citenamefont
  {De~Graef}}]{beleggia_magnetostatic_2004}%
  \BibitemOpen
  \bibfield  {author} {\bibinfo {author} {\bibfnamefont {M.}~\bibnamefont
  {Beleggia}}, \bibinfo {author} {\bibfnamefont {S.}~\bibnamefont {Tandon}},
  \bibinfo {author} {\bibfnamefont {Y.}~\bibnamefont {Zhu}},\ and\ \bibinfo
  {author} {\bibfnamefont {M.}~\bibnamefont {De~Graef}},\ }\href
  {https://doi.org/10.1016/j.jmmm.2003.12.1314} {\bibfield  {journal} {\bibinfo
   {journal} {Journal of Magnetism and Magnetic Materials}\ }\textbf {\bibinfo
  {volume} {278}},\ \bibinfo {pages} {270} (\bibinfo {year}
  {2004})}\BibitemShut {NoStop}%
\bibitem [{\citenamefont {Cichocki}\ \emph {et~al.}(1994)\citenamefont
  {Cichocki}, \citenamefont {Felderhof}, \citenamefont {Hinsen}, \citenamefont
  {Wajnryb},\ and\ \citenamefont {Bl/awzdziewicz}}]{cichocki_friction_1994}%
  \BibitemOpen
  \bibfield  {author} {\bibinfo {author} {\bibfnamefont {B.}~\bibnamefont
  {Cichocki}}, \bibinfo {author} {\bibfnamefont {B.~U.}\ \bibnamefont
  {Felderhof}}, \bibinfo {author} {\bibfnamefont {K.}~\bibnamefont {Hinsen}},
  \bibinfo {author} {\bibfnamefont {E.}~\bibnamefont {Wajnryb}},\ and\ \bibinfo
  {author} {\bibfnamefont {J.}~\bibnamefont {Bl/awzdziewicz}},\ }\href
  {https://doi.org/10.1063/1.466366} {\bibfield  {journal} {\bibinfo  {journal}
  {The Journal of Chemical Physics}\ }\textbf {\bibinfo {volume} {100}},\
  \bibinfo {pages} {3780} (\bibinfo {year} {1994})}\BibitemShut {NoStop}%
\bibitem [{\citenamefont {Durlofsky}\ \emph {et~al.}(1987)\citenamefont
  {Durlofsky}, \citenamefont {Brady},\ and\ \citenamefont
  {Bossis}}]{durlofsky_dynamic_1987}%
  \BibitemOpen
  \bibfield  {author} {\bibinfo {author} {\bibfnamefont {L.}~\bibnamefont
  {Durlofsky}}, \bibinfo {author} {\bibfnamefont {J.~F.}\ \bibnamefont
  {Brady}},\ and\ \bibinfo {author} {\bibfnamefont {G.}~\bibnamefont
  {Bossis}},\ }\href@noop {} {\bibfield  {journal} {\bibinfo  {journal}
  {Journal of Fluid Mechanics}\ ,\ \bibinfo {pages} {29}} (\bibinfo {year}
  {1987})}\BibitemShut {NoStop}%
\bibitem [{\citenamefont {Banchio}\ and\ \citenamefont
  {Brady}(2003)}]{banchio_accelerated_2003}%
  \BibitemOpen
  \bibfield  {author} {\bibinfo {author} {\bibfnamefont {A.~J.}\ \bibnamefont
  {Banchio}}\ and\ \bibinfo {author} {\bibfnamefont {J.~F.}\ \bibnamefont
  {Brady}},\ }\href@noop {} {\bibfield  {journal} {\bibinfo  {journal} {The
  Journal of Chemical Physics}\ ,\ \bibinfo {pages} {11}} (\bibinfo {year}
  {2003})}\BibitemShut {NoStop}%
\bibitem [{\citenamefont {Brady}\ and\ \citenamefont
  {Bossis}(1988)}]{brady_stokesian_1988}%
  \BibitemOpen
  \bibfield  {author} {\bibinfo {author} {\bibfnamefont {J.~F.}\ \bibnamefont
  {Brady}}\ and\ \bibinfo {author} {\bibfnamefont {G.}~\bibnamefont {Bossis}},\
  }\href {https://doi.org/10.1146/annurev.fl.20.010188.000551} {\bibfield
  {journal} {\bibinfo  {journal} {Annual Review of Fluid Mechanics}\ }\textbf
  {\bibinfo {volume} {20}},\ \bibinfo {pages} {111} (\bibinfo {year}
  {1988})}\BibitemShut {NoStop}%
\bibitem [{\citenamefont {Satoh}\ \emph {et~al.}(1998)\citenamefont {Satoh},
  \citenamefont {Chantrell}, \citenamefont {Coverdale},\ and\ \citenamefont
  {Kamiyama}}]{satoh_stokesian_1998}%
  \BibitemOpen
  \bibfield  {author} {\bibinfo {author} {\bibfnamefont {A.}~\bibnamefont
  {Satoh}}, \bibinfo {author} {\bibfnamefont {R.~W.}\ \bibnamefont
  {Chantrell}}, \bibinfo {author} {\bibfnamefont {G.~N.}\ \bibnamefont
  {Coverdale}},\ and\ \bibinfo {author} {\bibfnamefont {S.-i.}\ \bibnamefont
  {Kamiyama}},\ }\href {https://doi.org/10.1006/jcis.1998.5498} {\bibfield
  {journal} {\bibinfo  {journal} {Journal of Colloid and Interface Science}\
  }\textbf {\bibinfo {volume} {203}},\ \bibinfo {pages} {233} (\bibinfo {year}
  {1998})}\BibitemShut {NoStop}%
\bibitem [{\citenamefont {Sand}\ \emph {et~al.}(2016)\citenamefont {Sand},
  \citenamefont {Stener}, \citenamefont {Toivakka}, \citenamefont {Carlson},\
  and\ \citenamefont {P{\aa}lsson}}]{sand_stokesian_2016}%
  \BibitemOpen
  \bibfield  {author} {\bibinfo {author} {\bibfnamefont {A.}~\bibnamefont
  {Sand}}, \bibinfo {author} {\bibfnamefont {J.~F.}\ \bibnamefont {Stener}},
  \bibinfo {author} {\bibfnamefont {M.~O.}\ \bibnamefont {Toivakka}}, \bibinfo
  {author} {\bibfnamefont {J.~E.}\ \bibnamefont {Carlson}},\ and\ \bibinfo
  {author} {\bibfnamefont {B.~I.}\ \bibnamefont {P{\aa}lsson}},\ }\href
  {https://doi.org/10.1016/j.mineng.2015.10.015} {\bibfield  {journal}
  {\bibinfo  {journal} {Minerals Engineering}\ }\textbf {\bibinfo {volume}
  {90}},\ \bibinfo {pages} {70} (\bibinfo {year} {2016})}\BibitemShut {NoStop}%
\bibitem [{\citenamefont {Goddard}\ \emph {et~al.}(2020)\citenamefont
  {Goddard}, \citenamefont {{Mills-Williams}},\ and\ \citenamefont
  {Sun}}]{goddard_singular_2020}%
  \BibitemOpen
  \bibfield  {author} {\bibinfo {author} {\bibfnamefont {B.~D.}\ \bibnamefont
  {Goddard}}, \bibinfo {author} {\bibfnamefont {R.~D.}\ \bibnamefont
  {{Mills-Williams}}},\ and\ \bibinfo {author} {\bibfnamefont {J.}~\bibnamefont
  {Sun}},\ }\href {https://doi.org/10.1063/5.0009053} {\bibfield  {journal}
  {\bibinfo  {journal} {Physics of Fluids}\ }\textbf {\bibinfo {volume} {32}},\
  \bibinfo {pages} {062001} (\bibinfo {year} {2020})}\BibitemShut {NoStop}%
\bibitem [{\citenamefont {Russel}\ \emph {et~al.}(1989)\citenamefont {Russel},
  \citenamefont {Saville},\ and\ \citenamefont
  {Schowalter}}]{russel_colloidal_1989}%
  \BibitemOpen
  \bibfield  {author} {\bibinfo {author} {\bibfnamefont {W.~B.}\ \bibnamefont
  {Russel}}, \bibinfo {author} {\bibfnamefont {D.~A.}\ \bibnamefont
  {Saville}},\ and\ \bibinfo {author} {\bibfnamefont {W.~R.}\ \bibnamefont
  {Schowalter}},\ }\href@noop {} {\emph {\bibinfo {title} {Colloidal
  Dispersions}}},\ Cambridge Monographs on Mechanics and Applied Mathematics\
  (\bibinfo  {publisher} {{Cambridge University Press}},\ \bibinfo {address}
  {{Cambridge ; New York}},\ \bibinfo {year} {1989})\BibitemShut {NoStop}%
\bibitem [{\citenamefont {Jeffrey}\ and\ \citenamefont
  {Onishi}(1984)}]{jeffrey_calculation_1984}%
  \BibitemOpen
  \bibfield  {author} {\bibinfo {author} {\bibfnamefont {D.~J.}\ \bibnamefont
  {Jeffrey}}\ and\ \bibinfo {author} {\bibfnamefont {Y.}~\bibnamefont
  {Onishi}},\ }\href {https://doi.org/10.1017/S0022112084000355} {\bibfield
  {journal} {\bibinfo  {journal} {Journal of Fluid Mechanics}\ }\textbf
  {\bibinfo {volume} {139}},\ \bibinfo {pages} {261} (\bibinfo {year}
  {1984})}\BibitemShut {NoStop}%
\bibitem [{\citenamefont {Jeffrey}(1992)}]{jeffrey_calculation_1992}%
  \BibitemOpen
  \bibfield  {author} {\bibinfo {author} {\bibfnamefont {D.~J.}\ \bibnamefont
  {Jeffrey}},\ }\href {https://doi.org/10.1063/1.858494} {\bibfield  {journal}
  {\bibinfo  {journal} {Physics of Fluids A: Fluid Dynamics}\ }\textbf
  {\bibinfo {volume} {4}},\ \bibinfo {pages} {16} (\bibinfo {year}
  {1992})}\BibitemShut {NoStop}%
\bibitem [{\citenamefont {Townsend}(2019)}]{townsend_generating_2019}%
  \BibitemOpen
  \bibfield  {author} {\bibinfo {author} {\bibfnamefont {A.~K.}\ \bibnamefont
  {Townsend}},\ }\href {http://arxiv.org/abs/1802.08226} {\bibinfo {title}
  {Generating, from scratch, the near-field asymptotic forms of scalar
  resistance functions for two unequal rigid spheres in low-{{Reynolds-number}}
  flow}} (\bibinfo {year} {2019}),\ \Eprint {https://arxiv.org/abs/1802.08226}
  {arxiv:1802.08226 [physics]} \BibitemShut {NoStop}%
\bibitem [{\citenamefont {Min}\ \emph {et~al.}(2010)\citenamefont {Min},
  \citenamefont {Akbulut}, \citenamefont {Kristiansen}, \citenamefont {Golan},\
  and\ \citenamefont {Israelachvili}}]{min2010role}%
  \BibitemOpen
  \bibfield  {author} {\bibinfo {author} {\bibfnamefont {Y.}~\bibnamefont
  {Min}}, \bibinfo {author} {\bibfnamefont {M.}~\bibnamefont {Akbulut}},
  \bibinfo {author} {\bibfnamefont {K.}~\bibnamefont {Kristiansen}}, \bibinfo
  {author} {\bibfnamefont {Y.}~\bibnamefont {Golan}},\ and\ \bibinfo {author}
  {\bibfnamefont {J.}~\bibnamefont {Israelachvili}},\ }in\ \href@noop {} {\emph
  {\bibinfo {booktitle} {Nanoscience And Technology: A collection of reviews
  from Nature journals}}}\ (\bibinfo  {publisher} {World Scientific},\ \bibinfo
  {year} {2010})\ pp.\ \bibinfo {pages} {38--49}\BibitemShut {NoStop}%
\bibitem [{\citenamefont {Bishop}\ \emph {et~al.}(2009)\citenamefont {Bishop},
  \citenamefont {Wilmer}, \citenamefont {Soh},\ and\ \citenamefont
  {Grzybowski}}]{bishop_nanoscale_2009}%
  \BibitemOpen
  \bibfield  {author} {\bibinfo {author} {\bibfnamefont {K.~J.~M.}\
  \bibnamefont {Bishop}}, \bibinfo {author} {\bibfnamefont {C.~E.}\
  \bibnamefont {Wilmer}}, \bibinfo {author} {\bibfnamefont {S.}~\bibnamefont
  {Soh}},\ and\ \bibinfo {author} {\bibfnamefont {B.~A.}\ \bibnamefont
  {Grzybowski}},\ }\href {https://doi.org/10.1002/smll.200900358} {\bibfield
  {journal} {\bibinfo  {journal} {Small}\ }\textbf {\bibinfo {volume} {5}},\
  \bibinfo {pages} {1600} (\bibinfo {year} {2009})}\BibitemShut {NoStop}%
\bibitem [{\citenamefont
  {Israelachvili}(2011)}]{israelachvili_intermolecular_2011}%
  \BibitemOpen
  \bibfield  {author} {\bibinfo {author} {\bibfnamefont {J.}~\bibnamefont
  {Israelachvili}},\ }\href@noop {} {\emph {\bibinfo {title} {Intermolecular
  and Surface Forces}}},\ \bibinfo {edition} {third edition}\ ed.\ (\bibinfo
  {publisher} {Academic Press},\ \bibinfo {year} {2011})\BibitemShut {NoStop}%
\bibitem [{\citenamefont {M{\o}rup}\ \emph {et~al.}(2010)\citenamefont
  {M{\o}rup}, \citenamefont {Hansen},\ and\ \citenamefont
  {Frandsen}}]{morup_magnetic_2010}%
  \BibitemOpen
  \bibfield  {author} {\bibinfo {author} {\bibfnamefont {S.}~\bibnamefont
  {M{\o}rup}}, \bibinfo {author} {\bibfnamefont {M.~F.}\ \bibnamefont
  {Hansen}},\ and\ \bibinfo {author} {\bibfnamefont {C.}~\bibnamefont
  {Frandsen}},\ }\href {https://doi.org/10.3762/bjnano.1.22} {\bibfield
  {journal} {\bibinfo  {journal} {Beilstein Journal of Nanotechnology}\
  }\textbf {\bibinfo {volume} {1}},\ \bibinfo {pages} {182} (\bibinfo {year}
  {2010})}\BibitemShut {NoStop}%
\bibitem [{\citenamefont {Chuan~Lim}\ and\ \citenamefont
  {Feng}(2012)}]{chuan_lim_agglomeration_2012}%
  \BibitemOpen
  \bibfield  {author} {\bibinfo {author} {\bibfnamefont {E.~W.}\ \bibnamefont
  {Chuan~Lim}}\ and\ \bibinfo {author} {\bibfnamefont {R.}~\bibnamefont
  {Feng}},\ }\href {https://doi.org/10.1063/1.3697865} {\bibfield  {journal}
  {\bibinfo  {journal} {The Journal of Chemical Physics}\ }\textbf {\bibinfo
  {volume} {136}},\ \bibinfo {pages} {124109} (\bibinfo {year}
  {2012})}\BibitemShut {NoStop}%
\end{thebibliography}%


\begin{thebibliography}{6}%
\makeatletter
\providecommand \@ifxundefined [1]{%
 \@ifx{#1\undefined}
}%
\providecommand \@ifnum [1]{%
 \ifnum #1\expandafter \@firstoftwo
 \else \expandafter \@secondoftwo
 \fi
}%
\providecommand \@ifx [1]{%
 \ifx #1\expandafter \@firstoftwo
 \else \expandafter \@secondoftwo
 \fi
}%
\providecommand \natexlab [1]{#1}%
\providecommand \enquote  [1]{``#1''}%
\providecommand \bibnamefont  [1]{#1}%
\providecommand \bibfnamefont [1]{#1}%
\providecommand \citenamefont [1]{#1}%
\providecommand \href@noop [0]{\@secondoftwo}%
\providecommand \href [0]{\begingroup \@sanitize@url \@href}%
\providecommand \@href[1]{\@@startlink{#1}\@@href}%
\providecommand \@@href[1]{\endgroup#1\@@endlink}%
\providecommand \@sanitize@url [0]{\catcode `\\12\catcode `\$12\catcode
  `\&12\catcode `\#12\catcode `\^12\catcode `\_12\catcode `\%12\relax}%
\providecommand \@@startlink[1]{}%
\providecommand \@@endlink[0]{}%
\providecommand \url  [0]{\begingroup\@sanitize@url \@url }%
\providecommand \@url [1]{\endgroup\@href {#1}{\urlprefix }}%
\providecommand \urlprefix  [0]{URL }%
\providecommand \Eprint [0]{\href }%
\providecommand \doibase [0]{https://doi.org/}%
\providecommand \selectlanguage [0]{\@gobble}%
\providecommand \bibinfo  [0]{\@secondoftwo}%
\providecommand \bibfield  [0]{\@secondoftwo}%
\providecommand \translation [1]{[#1]}%
\providecommand \BibitemOpen [0]{}%
\providecommand \bibitemStop [0]{}%
\providecommand \bibitemNoStop [0]{.\EOS\space}%
\providecommand \EOS [0]{\spacefactor3000\relax}%
\providecommand \BibitemShut  [1]{\csname bibitem#1\endcsname}%
\let\auto@bib@innerbib\@empty
\bibitem [{\citenamefont {Happel}\ and\ \citenamefont
  {Brenner}(1981)}]{happel_low_1981}%
  \BibitemOpen
  \bibfield  {author} {\bibinfo {author} {\bibfnamefont {J.}~\bibnamefont
  {Happel}}\ and\ \bibinfo {author} {\bibfnamefont {H.}~\bibnamefont
  {Brenner}},\ }\href {https://doi.org/10.1007/978-94-009-8352-6} {\emph
  {\bibinfo {title} {Low {{Reynolds}} Number Hydrodynamics}}},\ edited by\
  \bibinfo {editor} {\bibfnamefont {R.~J.}\ \bibnamefont {Moreau}},\ \bibinfo
  {series} {Mechanics of Fluids and Transport Processes}, Vol.~\bibinfo
  {volume} {1}\ (\bibinfo  {publisher} {{Springer Netherlands}},\ \bibinfo
  {address} {{Dordrecht}},\ \bibinfo {year} {1981})\BibitemShut {NoStop}%
\bibitem [{\citenamefont {Ermak}\ and\ \citenamefont
  {McCammon}(1978)}]{ermak_brownian_1978}%
  \BibitemOpen
  \bibfield  {author} {\bibinfo {author} {\bibfnamefont {D.~L.}\ \bibnamefont
  {Ermak}}\ and\ \bibinfo {author} {\bibfnamefont {J.~A.}\ \bibnamefont
  {McCammon}},\ }\href {https://doi.org/10.1063/1.436761} {\bibfield  {journal}
  {\bibinfo  {journal} {The Journal of Chemical Physics}\ }\textbf {\bibinfo
  {volume} {69}},\ \bibinfo {pages} {1352} (\bibinfo {year}
  {1978})}\BibitemShut {NoStop}%
\bibitem [{\citenamefont {Brenner}(1967)}]{brenner_coupling_1967}%
  \BibitemOpen
  \bibfield  {author} {\bibinfo {author} {\bibfnamefont {H.}~\bibnamefont
  {Brenner}},\ }\href {https://doi.org/10.1016/0021-9797(67)90185-3} {\bibfield
   {journal} {\bibinfo  {journal} {Journal of Colloid and Interface Science}\
  }\textbf {\bibinfo {volume} {23}},\ \bibinfo {pages} {407} (\bibinfo {year}
  {1967})}\BibitemShut {NoStop}%
\bibitem [{\citenamefont {Sharma}(1988)}]{sharma_field_1988}%
  \BibitemOpen
  \bibfield  {author} {\bibinfo {author} {\bibfnamefont {N.~L.}\ \bibnamefont
  {Sharma}},\ }\href {https://doi.org/10.1119/1.15592} {\bibfield  {journal}
  {\bibinfo  {journal} {American Journal of Physics}\ }\textbf {\bibinfo
  {volume} {56}},\ \bibinfo {pages} {420} (\bibinfo {year} {1988})}\BibitemShut
  {NoStop}%
\bibitem [{\citenamefont {Gardiner}(2002)}]{gardiner_handbook_2002}%
  \BibitemOpen
  \bibfield  {author} {\bibinfo {author} {\bibfnamefont {C.~W.}\ \bibnamefont
  {Gardiner}},\ }\href@noop {} {\emph {\bibinfo {title} {Handbook of Stochastic
  Methods: For Physics, Chemistry and the Natural Sciences}}},\ \bibinfo
  {edition} {study ed., 2. ed., 6. print}\ ed.,\ \bibinfo {series} {Springer
  Series in Synergetics}\ No.~\bibinfo {number} {13}\ (\bibinfo  {publisher}
  {{Springer}},\ \bibinfo {address} {{Berlin Heidelberg}},\ \bibinfo {year}
  {2002})\BibitemShut {NoStop}%
\bibitem [{\citenamefont {S{\"a}rkk{\"a}}\ and\ \citenamefont
  {Solin}(2019)}]{sarkka_applied_2019}%
  \BibitemOpen
  \bibfield  {author} {\bibinfo {author} {\bibfnamefont {S.}~\bibnamefont
  {S{\"a}rkk{\"a}}}\ and\ \bibinfo {author} {\bibfnamefont {A.}~\bibnamefont
  {Solin}},\ }\href {https://doi.org/10.1017/9781108186735} {\emph {\bibinfo
  {title} {Applied {{Stochastic Differential Equations}}}}},\ \bibinfo
  {edition} {1st}\ ed.\ (\bibinfo  {publisher} {{Cambridge University Press}},\
  \bibinfo {year} {2019})\BibitemShut {NoStop}%
\end{thebibliography}%

\end{document}


\title{Supplementary material for the article \\ ``Conservation laws for interacting magnetic nanoparticles at finite temperature''}
\author{Frederik L. Durhuus}
\affiliation{NEXMAP, Department of Physics, Technical University of Denmark, 2800 Kgs.\ Lyngby, Denmark}
\author{Marco Beleggia}
\affiliation{Department of Physics, University of Modena and Reggio Emilia, 41125 Modena, Italy}
\affiliation{DTU Nanolab, Technical University of Denmark, 2800 Kgs. Lyngby, Denmark}
\author{Cathrine Frandsen}
\affiliation{NEXMAP, Department of Physics, Technical University of Denmark, 2800 Kgs.\ Lyngby, Denmark}
\email{fraca@fysik.dtu.dk}    

\onecolumngrid

\maketitle

\section{Generalisation to non-spherical particles}

In the main text, we derive a general model for the time-evolution of a collection of single-domain, \textit{spherical} magnetic nanoparticles (MNPs). We derive expressions for the transfer of energy, linear- and angular momentum between MNPs and environment within this model, and verify said expressions numerically. We also discuss a generalisation of the model to non-spherical particles and give the corresponding generalised transfer rates. The derivations of the generalised rates are highly similar to those for the spherical case, but for ease of reference we present them here in full. Note that we expect the reader to have already read the main text, so only some of the symbols are reintroduced.

\subsection{Langevin dynamics model}

We consider a collection of uniformly magnetised MNPs of arbitrary shape. 
Here we present the general equations of motion in our Langevin dynamics model for interacting, uniformly magnetised MNPs of arbitrary shape.

The system energy is, up to a constant,
\begin{align}
    E &= \frac{1}{2} \sum_i \bs{\omega}_i \vdot \vb{L}_i + \frac{1}{2} \mathfrak{m} \sum_i v_i^2 - \frac{1}{2} \sum_i \bs{\mu}_i \vdot \vb{B}^\text{int}_i
    \notag\\
    &\quad - \sum_i \bs{\mu}_i \vdot \vb{B}^\text{ext} + E^\text{ani}_i. \label{eq:E_multi_MNP}
\end{align}
where $\vb{L}_i = \mathrm{I} \bs{\omega}_i$, $\mathrm{I}$ is the moment of inertia tensor, $\vb{B}^\text{int}_i$ is the average magnetic field over MNP $i$ from all the other MNPs and $E^\text{ani}$ is the energy from shape- and magnetocrystalline anisotropy. Unlike the main paper the moment of inertia is a tensor, $\vb{B}^\text{int}$ is not dipolar and we do not assume uniaxial anisotropy. Since the particles no longer have a rotational symmetry axis, more information is required to specify their orientation. To MNP $i$ we assign the 3 orientation vectors $\vb{u}_{\text{a}i}, \vb{u}_{\text{b}i}, \vb{u}_{\text{c}i}$, which are orthogonal unit vectors fixed relative to the atomic lattice. We then have
\begin{align}
    \vb{\dot{L}}_{0,i} = \sum_{\text{k} = \text{a,b,c}} \pdv{E}{\vb{u}_{\text{k}i}} \cross \vb{u}_{\text{k}i}   \label{eq:L0_dot}
\end{align}
where $\vb{\dot{L}}_{0,i}$ is the derivative without damping, hence also without thermal agitation.

The resulting mechanical equations of motion are
\begin{align}
    \vb{\dot{L}}_i &= \mu \gamma^{-1} \vb{\dot{m}}_i + \mu\vb{m}_i \cross \vb{B}_i+ \bs{\tau}_i^\text{visc} + \bs{\tau}^\text{th}_i \label{eq:omega_dot}
    \\
    \vb{\dot{p}}_i &= \vb{F}^\text{int}_i + \vb{F}^\text{visc}_i + \vb{F}^\text{th}_i \qq{,} \vb{\dot{r}}_i = \vb{v}_i \label{eq:v_dot}
    \\
    \vb{\dot{u}}_{\text{a}i} &= \bs{\omega}_i \cross \vb{u}_{\text{a}i} \:\: , \:\: \vb{\dot{u}}_{\text{b}i} = \bs{\omega}_i \cross \vb{u}_{\text{b}i} \:\: , \:\: \vb{\dot{u}}_{\text{c}i} = \bs{\omega}_i \cross \vb{u}_{\text{c}i} \label{eq:u_dot}
\end{align}
Here $\vb{p}_i = \mathfrak{m} \vb{\dot{v}}_i$ is the linear momentum, $\vb{F}^\text{int}_i$ is the force from gradients in $\vb{B}^\text{int}$, the magnetic field from all sources outside MNP $i$ is $\vb{B}_i = \vb{B}^\text{ext} + \vb{B}_i^\text{int}$ and
\begin{align}
\mqty(\vb{F}^\text{visc}_i \\ \bs{\tau}^\text{visc}_i) = -\mathrm{Z} \mqty(\vb{v}_i \\ \bs{\omega}_i) = \mqty(- \mathrm{Z}^\text{tt} \vb{v}_i - \mathrm{Z}^\text{tr} \bs{\omega}_i \\ - \mathrm{Z}^\text{rr} \bs{\omega}_i - \mathrm{Z}^\text{rt} \vb{v}_i),    \label{eq:viscous_drag_non_spherical}
\end{align}
where
\begin{align*}
    \mathrm{Z} = \mqty(\mathrm{Z}^\text{tt} & \mathrm{Z}^\text{tr} \\ \mathrm{Z}^\text{rt} & \mathrm{Z}^\text{rr})
\end{align*}
is the 6-by-6 friction tensor\cite{happel_low_1981}. Using the compact notation $\vb{f}^\text{visc}_i = (\vb{F}^\text{visc}_i, \bs{\omega}_i^\text{visc})$ and $\vb{V}_i = (\vb{v}_i, \bs{\omega}_i)$ this can be written as $\vb{f}^\text{visc}_i = \mathrm{Z} \vb{V}_i$. Using $\vb{f}_i^\text{th} = (\vb{F}_i^\text{th}, \bs{\tau}_i^\text{th})$ the thermal force and torque are given by\cite{ermak_brownian_1978}
\begin{align*}
    \expval{\vb{f}^\text{th}} = 0 \qq{,}
    \expval{f^\text{th}_\alpha(t), f^\text{th}_\beta(t')} = 2k_B T Z^{-1}_{\alpha \beta} \delta(t - t')
\end{align*}
That is, the thermal force and torque are still Gaussian distributed with zero mean and no autocorrelation, but there are correlations between different components. Interestingly for a shape with at least 3 perpendicular planes of mirror symmetry\cite{happel_low_1981,brenner_coupling_1967} $\mathrm{Z}^\text{tr} = \mathrm{Z}^\text{rt} = 0$. Example shapes include rectangular boxes, ellipsoids and cylinders.

The LLG equation can be written in the equivalent forms
\begin{align}
    \vb{\dot{m}}_i &= -\gamma \vb{m}_i \cross \vb{B}^\text{eff}_i + \alpha \vb{m}_i \cross \dot{\vb{m}}_i,  \label{eq:mu_dot_mid}
\\
\dot{\vb{m}}_i &= -\gamma' \vb{m}_i \cross \vb{B}^\text{eff}_i - \alpha \gamma' \vb{m}_i \cross \left[\vb{m}_i \cross \vb{B}^\text{eff}_i\right] \label{eq:mu_dot}
\\
\vb{\dot{m}}_i &= \bs{\Omega}_i \cross \vb{m}_i \qq{,} \bs{\Omega}_i = \gamma' \vb{B}_i^\text{eff} + \alpha \gamma' \vb{m}_i \cross \vb{B}_i^\text{eff},     \label{eq:Omega}
\end{align}
all of which are useful for analytical derivations. The effective magnetic field is
\begin{align}
    \vb{B}^\text{eff}_i = \vb{B}^\text{ext}_i + \vb{B}^\text{int}_i + \vb{B}^\text{ani}_i + \vb{B}^\text{th}_i - \alpha\vb{m}_i \cross \vb{B}^\text{bar}_i,   \label{eq:B_eff}
\end{align}
where $\vb{B}^\text{ani}_i = - \frac{1}{\mu}\partial_{\vb{m}_i} E^\text{ani}_i$ is the effective field from anisotropy and 
\begin{align}
 \expval{B^\text{th}_{\alpha}(t)B^\text{th}_{\beta}(t')} &= \frac{2k_BT\alpha}{\gamma m} \delta_{\alpha\beta} \delta(t-t')  \label{eq:B_th}
\end{align}
just like the spherical case.

For reference, we recall the time-derivative of the moment at zero damping
\begin{align}
    \dot{\bs{\mu}}_{0,i} &= - \gamma \bs{\mu}_i \cross (\vb{B}_i + \vb{B}^\text{ani}_i) \label{eq:mu_dot_no_damping}
\end{align}
and the definition of the Barnett field
\begin{align}
    \vb{B}^\text{bar}_i = -\gamma^{-1} \bs{\omega}_i. \label{eq:B_bar}
\end{align}

\subsection{Transfer of momenta}

The momenta have the same form as the spherical case. Using the equations of motion \cref{eq:mu_dot,eq:v_dot,eq:omega_dot,eq:viscous_drag_non_spherical,eq:u_dot,eq:mu_dot_mid,eq:Omega} the time derivatives of the angular momentum components for the $i$'th MNP are
\begin{align}
    \vb{\dot{S}}_i &= -\gamma^{-1} \dot{\bs{\mu}}_i \label{eq:Si_dot}
    \\
    \vb{\dot{J}}_i &=\bs{\tau}^\text{visc}_i + \bs{\mu}_i \cross \vb{B}_i + \bs{\tau}^\text{th}_i   \label{eq:Ji_dot}
\end{align}
with $\vb{\dot{L}}_i$ given by \cref{eq:omega_dot}. The change in linear momentum is given by \cref{eq:v_dot}.

The change in total angular momentum is
\begin{align*}
    \vb{\dot{J}} &= \sum_i \left(\dot{\vb{J}}_i 
    + \vb{r}_i \cross \dot{\vb{p}}_i \right) 
    \\
    &= \eval{\vb{\dot{J}}}_{\text{Isolated}} + \sum_i \bigg(\bs{\mu}_i \cross \vb{B}^\text{ext}_i + \bs{\tau}^\text{th}_i
    \\
    &\hspace{1cm} \bs{\tau}^\text{visc}_i  - \vb{r}_i \cross [\vb{F}^\text{visc}_i + \vb{F}^\text{th}_i]\bigg)
\end{align*}
The first term is for a system of MNPs at $T=0$ in vacuum, with no external fields. This is not in general zero because as the MNPs interact and move, they induce E-fields which leads to angular momentum in the EM-fields\cite{sharma_field_1988} and radiative losses. However, as is most common, we use the fields of magnetostatics for modelling purposes, in which case there is no field momentum and $\eval{\vb{\dot{J}}}_{\text{Isolated}} = 0$. For a formal proof see Appendix B of the main article.

\subsection{Energy transfer \label{subsec:Energy_transfer}}

The system energy, $E$, is a function of the independent variables $\vb{m}_i, \vb{u}_i, \bs{\omega}_i, \vb{r}_i, \vb{v}_i, t$, so the time derivative is
\begin{align}
    \dv{t}E = &\sum_i\bigg[\pdv{E}{\vb{m}_i} \vdot \dot{\vb{m}}_i + \sum_{\text{k}=\text{a},\text{b},\text{c}} \pdv{E}{\vb{u}_{\text{k}i}} \vdot \dot{\vb{u}}_{\text{k}i}
    \notag\\
    &+ \pdv{E}{\vb{L}_i} \vdot \dot{\vb{L}}_i  +\pdv{E}{\vb{r}_i} \vdot \vb{v}_i + \pdv{E}{\vb{v}_i} \vdot \dot{\vb{v}}_i\bigg] + \pdv{E}{t} \label{appeq:dE_dt}
\end{align}
Here we assume the Stratonovich interpretation, which means the ordinary rules of calculus apply despite the stochastic terms\cite{gardiner_handbook_2002,sarkka_applied_2019}.

We consider a single term and suppress the $i$-subscript to avoid notational clutter.
Let 
$\dot{\vb{m}}_0,\dot{\vb{L}}_0, \dot{\vb{p}}_0$ be the derivatives without damping. The derivatives with damping are given in \cref{eq:mu_dot_mid,eq:omega_dot,eq:v_dot} 
and $\dot{\vb{u}}_\text{k} = \bs{\omega} \cross \vb{u}_\text{k}$ applies in both cases. For reference we recall the general vector identity 
\begin{align*}
\vb{a} \vdot (\vb{b} \cross \vb{c}) = \vb{b} \vdot (\vb{c} \cross \vb{a}) = \vb{c} \vdot (\vb{a} \cross \vb{b})
\end{align*}

The only explicit time-dependence is in the external field, so
\begin{align}
    \pdv{E}{t} = -\mu \vb{m} \vdot \dot{\vb{B}}^\text{ext} \label{appeq:pdv_E_t}
\end{align}
By definition $\partial_{\vb{r}} E = -\vb{\dot{p}}_0$, hence
\begin{align*}
    \pdv{E}{\vb{r}} \vdot \vb{v} &= -\dot{\vb{p}}_0 \vdot \vb{v} = -(\vb{\dot{p}} - \vb{F}^\text{visc}_i - \vb{F}^\text{th}) \vdot \vb{v} 
    \\
    &= - \vb{p} \vdot \dot{\vb{v}} + \vb{F}^\text{visc}_i \vdot \vb{v}_i + \vb{F}^\text{th} \vdot \vb{v}
\end{align*}
Using \cref{eq:E_multi_MNP} $\partial_{\vb{v}}E = \vb{p}$, thus
\begin{align}
    \pdv{E}{\vb{r}} \vdot \vb{v} + \pdv{E}{\vb{v}} \vdot \dot{\vb{v}} = \vb{v} \vdot (\vb{F}^\text{visc} + \vb{F}^\text{th})  \label{appeq:diss_trans}
\end{align}
\Cref{appeq:diss_trans} is the final result for the translational degrees of freedom.


For mechanical rotation we find, using eq. \eqref{eq:u_dot} that
\begin{align*}
    \pdv{E}{\vb{u}_\text{a}} \vdot \dot{\vb{u}}_\text{a} &= \pdv{E}{\vb{u}_\text{a}} \vdot (\bs{\omega} \cross \vb{u}_\text{a}) = -\bs{\omega} \vdot \left(\pdv{E}{\vb{u}_\text{a}} \cross \vb{u}_\text{a} \right)
\end{align*}
and similarly for $\vb{u}_\text{b}, \vb{u}_\text{c}$. Referring to \cref{eq:L0_dot} it follows that 
\begin{align*}
    \sum_{\text{k} = \text{a,b,c}} \pdv{E}{\vb{u}_\text{k}} \vdot \dot{\vb{u}}_\text{k} = - \bs{\omega} \vdot \vb{\dot{L}}_0
\end{align*}
Next we need the gradient 
\begin{align*}
    \pdv{E}{\vb{L}} = \frac{1}{2} \pdv{\vb{L}} (\bs{\omega} \vdot \mathrm{I} \bs{\omega}) = \frac{1}{2} \pdv{\vb{L}} (\vb{L} \vdot \mathrm{I}^{-1} \bs{\omega})
\end{align*}
To evaluate this, we pick a coordinate system where $\mathrm{I}^{-1}$ is diagonal, so that
\begin{align*}
     \pdv{E}{\vb{L}} = \frac{1}{2} \pdv{\vb{L}} \sum_{k=1,2,3} (\mathrm{I}^{-1})_k L_k^2 = \sum_k (\mathrm{I}^{-1})_k \omega_k \vb{e}_k = \mathrm{I}^{-1} \vb{L} = \bs{\omega}
\end{align*}
It follows that
\begin{align}
    \pdv{E}{\vb{u}} \vdot \dot{\vb{u}} + \pdv{E}{\vb{L}} \vdot \vb{\dot{L}} &= \bs{\omega} \vdot (\vb{\dot{L}} - \vb{\dot{L}}_0) 
    \label{appeq:mech_rot_pdv}\\
    &= \frac{\mu}{\gamma}\bs{\omega} \vdot (\dot{\vb{m}} - \dot{\vb{m}}_0) + \bs{\omega} \vdot (\bs{\tau}^\text{visc} + \bs{\tau}^\text{th}) \notag
\end{align}

For magnetic rotation, we find analogously that
\begin{align*}
    \pdv{E}{\vb{m}} \vdot \vb{\dot{m}} &= \pdv{E}{\vb{m}} \vdot (\bs{\Omega} \cross \vb{m}) = -\bs{\Omega} \vdot \dot{\vb{S}}_0 = \frac{\mu}{\gamma} \bs{\Omega} \vdot \vb{\dot{m}}_0
\end{align*}
where $\bs{\Omega}$ is defined in \cref{eq:Omega} and $\vb{\dot{S}}_0 = \partial_{\vb{m}} E \cross \vb{m}_0$. We note that $\bs{\Omega} \vdot \vb{\dot{m}} = \bs{\Omega} \vdot (\bs{\Omega} \cross \vb{m}) = 0$, so
\begin{align}
    \pdv{E}{\vb{m}} \vdot \vb{\dot{m}} = \frac{\mu}{\gamma} \bs{\Omega} \vdot (\vb{\dot{m}}_0 - \vb{\dot{m}})   \label{appeq:mag_rot_pdv}
\end{align}

Now, adding \cref{appeq:mech_rot_pdv,appeq:mag_rot_pdv} we get 
\begin{align}
    \pdv{E}{\vb{u}} \vdot \vb{\dot{u}} + &\pdv{E}{\bs{\omega}} \vdot \bs{\dot{\omega}} + \pdv{E}{\vb{m}} \vdot \vb{\dot{m}} 
    \label{appeq:rot_pdv}\\
    &= \frac{\mu}{\gamma}(\bs{\Omega} - \bs{\omega}) \vdot (\vb{\dot{m}}_0 - \vb{\dot{m}}) + \bs{\omega} \vdot (\bs{\tau}^\text{visc} + \bs{\tau}^\text{th}) \notag
\end{align}
It follows from \cref{eq:mu_dot_no_damping,eq:mu_dot_mid,eq:B_eff} that
\begin{align*}
    \vb{\dot{m}}_0 - \vb{\dot{m}} = \gamma \vb{m} \cross \vb{B}^\text{th} - \alpha \vb{m} \cross (\bs{\Omega} \cross \vb{m} - \gamma\vb{B}^\text{bar}).
\end{align*}
Inserting the definition of the Barnett field, \cref{eq:B_bar}, this can be written
\begin{align*}
    \vb{\dot{m}}_0 - \vb{\dot{m}} = \gamma \vb{m} \cross \left[\vb{B}^\text{th} - \alpha (\bs{\Omega} - \bs{\omega}) \cross \vb{m}\right].
\end{align*}
Hence
\begin{align}
\frac{\mu}{\gamma}(\bs{\Omega} &- \bs{\omega}) \vdot (\vb{\dot{m}}_0 - \vb{\dot{m}}) \label{appeq:mag_mech_rot_identity}\\
&= \mu \left[(\bs{\Omega} - \bs{\omega}) \cross \vb{m} \right] \vdot \vb{B}^\text{th} - \alpha \mu \gamma^{-1} ([\bs{\Omega} - \bs{\omega}] \cross \vb{m})^2 \notag
\end{align}
Inserting \cref{appeq:mag_mech_rot_identity} in \cref{appeq:rot_pdv} then adding \cref{appeq:pdv_E_t,appeq:diss_trans} gives the total energy transfer
\begin{align*}
    \dot{E} = P^\text{hyst} + P^\text{mag} + P^\text{visc}
\end{align*}
where $P^\text{hyst}, P^\text{mag}$ are identical to the spherical case and
\begin{align*}
    P^\text{visc} &= \vb{v} \vdot (\vb{F}^\text{visc} + \vb{F}^\text{th}) + \bs{\omega} \vdot (\bs{\tau}^\text{visc} + \bs{\tau}^\text{th})
    \\
    &= \vb{V} \vdot (\vb{f}^\text{visc} + \vb{f}^\text{th})
    \\
    &= - \vb{V} \vdot \mathrm{Z}\vb{V} + \vb{V} \vdot \vb{f}^\text{th}
\end{align*}

\bibliography{references.bib}